\makeatletter
\def\cl@chapter{}
\makeatother
\documentclass[]{article}       
\usepackage{fancyhdr}
\usepackage[margin=1in]{geometry}
\usepackage{AAS_packages}

\DeclareMathOperator*{\argmin}{arg\,min}

\title{Bayesian Shape Reconstruction and Optimal Guidance for Autonomous Landing on Asteroids}

\author{
    Shankar Kulumani \and Taeyoung Lee
}

\begin{document}

\maketitle

\begin{abstract}
    Construction of the precise shape of an asteroid is critical for spacecraft operations as the gravitational potential is determined by spatial mass distribution.
    The typical approach to shape determination requires a prolonged ``mapping'' phase of the mission over which extensive measurements are collected and transmitted for Earth-based processing.
    This paper presents a set of approaches to explore an unknown asteroid with onboard calculations, and to land on its surface area selected in an optimal fashion. 
    The main motivation is to avoid the extended period of mapping or preliminary ground observations that are commonly required in spacecraft missions around asteroids. 
    First, range measurements from the spacecraft to the surface are used to incrementally correct an initial shape estimate according to the Bayesian framework. 
    Then, an optimal guidance scheme is proposed to control the vantage point of the range sensor to construct a complete 3D model of the asteroid shape. 
    This shape model is then used in a nonlinear controller to track a desired trajectory about the asteroid.
    Finally, a multi resolution approach is presented to construct a higher fidelity shape representation in a specified location while avoiding the inherent burdens of a uniformly high resolution mesh. 
    This approach enables for an accurate shape determination around a potential landing site.
    We demonstrate this approach using several radar shape models of asteroids and provide a full dynamical simulation about asteroid 4769 Castalia.
\end{abstract}



\section{Introduction}
Small solar system bodies, such as asteroids and comets, continue to remain a focus of scientific study.
The small size of these bodies prevents the formation of large internal pressures and temperatures which helps to preserve the early chemistry of the solar system.
This insight offers additional detail into the formation of the Earth and also of the probable formation of other extrasolar planetary bodies.
Of particular interest are those near-Earth asteroids (NEA) which inhabit heliocentric orbits in the vicinity of the Earth. 
These easily accessible bodies provide attractive targets to support space industrialization, mining operations, and scientific missions.
In spite of the significant interest, and the extensive research by the community, the operation of spacecraft near small bodies remains a challenging problem.

The dynamical environment around asteroids is strongly perturbed and challenging for analysis and mission operations~\cite{scheeres2012}.
Due to their low mass, which in turn causes a low gravitational attraction, asteroids may have irregular shapes.
Furthermore, asteroids may also have a chaotic spin state due to the absorption and emittance of solar radiation~\cite{rubincam2000}.
As a result, approaches utilizing an inverse square gravitational model do not capture the  true dynamical environment.
In addition, the vast majority of asteroids are difficult to track and characterize using ground based sensors.
Due to their small size, frequently with a maximum radius less than \SI{1}{\kilo\meter}, and low albedo, the reflected energy of these asteroids is insufficient for reliable detection or tracking.
Therefore, the dynamical model of the asteroid is relatively coarse prior to in situ measurements from a dedicated spacecraft.
As a result, any spacecraft mission to an asteroid must include the ability to update the dynamical model given in situ measurements and remain robust to unmodelled forces.

Another key dynamical consideration is the coupling between rotational and translational states around the asteroid.
The coupling is induced due to the different gravitational forces experienced on various portions of the spacecraft. 
The effect of the gravitational coupling is related to the ratio of the spacecraft size and orbital radius~\cite{hughes2004}.
For operations around asteroids, the ratio is relatively large which causes a much larger coupling between the translational and rotational states.
In~\cite{elmasri2005} and \cite{sanyal2004a}, the coupling of an elastic dumbbell spacecraft in orbit about a central body is considered, assuming that the central body is spherically symmetric.
Furthermore, the spacecraft model is assumed to remain in a planar orbit.
As a result, these developments are not directly applicable to motion about an asteroid, which experiences highly non-Keplerian motion.
In~\cite{misra2015b}, the effect of coupled motion is investigated for long term trajectories around asteroids.
However, the analysis only considered a second order spherical harmonic gravitational potential model. 
Therefore, these results are only valid when far from the asteroid surface and will diverge when used within the Brillouin sphere.

An accurate gravitational potential model is critical for performing low altitude and/or surface operations around asteroids.
Due to the irregular shape, trajectories will pass within the Brillouin sphere, where the typical spherical harmonic model diverges from the true gravitational potential.
The standard approach for asteroid missions is to compute the gravitational potential using a polyhedron potential model~\cite{werner1996}.
The polyhedron potential model provides the exact gravitational potential, and subsequently the gravitational acceleration, for a given triangular faceted shape model of an asteroid.
The method provides the exact potential at any point outside the body for a given shape model.
As a result, the accuracy of the gravitational potential is primarily dependent on the accuracy with which the shape model represents the true surface.
A high fidelity shape, which necessarily has many vertices and faces, is required for an accurate computation of the gravitational acceleration and enabling low altitude operations.

Prior to the arrival of a spacecraft at an asteroid, Earth-based sensors are used to characterize the body.
Using both optical and radar sensors allows for the precise orbit of the asteroid to be determined.
Another vital task is the determination of the asteroid shape from radar data~\cite{hudson1994,busch2011}.
This is a challenging problem as it requires the simultaneous estimation of the asteroid spin state and shape.
Furthermore, determining the shape from radar is currently the only Earth-based technique that can produce detailed three-dimensional shape information of near-Earth objects~\cite{greenberg2015}.
The current approach is based on an estimation scheme which iteratively perturbs a shape to match given radar data.
This computationally intensive approach is only able to capture the gross size and shape and is unable to capture the small surface features of the asteroid.
Frequently, only a coarse model is possible from the ground and an accurate shape must be determined only after a spacecraft has rendezvoused with the asteroid.
As a result, upon arrival the gravitational environment near the asteroid is poorly modeled as the shape of the asteroid is not accurate.
Therefore, the polyhedron potential model is not appropriate immediately upon arrival but rather only after the shape has been determined.

On approach to an asteroid, spacecraft navigation and guidance is primarily based on ground measurements.
After arrival, spacecraft will generally spend upwards of several months in a mapping mission phase~\cite{williams2018,kubota2003,cole1998}.
For example, the recent OSIRIS-REx mission used global imagery to create digital terrain maps over the span of approximately seven months~\cite{williams2018}.
During this period, spacecraft sensors, such as on board optical telescopes or Light radio Detection and Ranging (LIDAR), are used to characterize the asteroid.
The resulting imagery and range data is transmitted to the ground and the resulting asteroid shape and motion is estimated. 
During this mapping phase the spacecraft must remain in a quiescent state devoted entirely to mapping the surface.
Depending on the mission type, this long period of mapping is crucial to the mission, such as sample collection~\cite{williams2018}. 
For example, the OSIRIS-REx mission spent approximately seven months in mapping phase where digitial terrain maps and a shape model were constructed in support of the main sample collection mission.
However, other missions, such as asteroid mitigation, may be severely limited by the time and ground resources required to generate a surface shape.
Furthermore, the long distances involved necessitate on-board autonomy to enable to spacecraft to operate without ground communications.
Similarly, during landing the spacecraft will require the ability to sense and model the surface topography in order to safely land in an unknown environment.
Onboard shape reconstruction techniques enable autonomous operations which allow for a greater range of spacecraft operations.
Mitigation of this ground based surface modeling will greatly expand the range of missions possible.

This paper presents three techniques to explore an asteroid without need for an extended period of mapping or reliance on ground based measurements. 
First, we develop a method to compute the surface shape of an asteroid from onboard range measurements.
Assuming that a meshed model of an initial estimate, such as an ellipsoidal shape, is available, the radius of any vertex in the measured area is adjusted according to the measurements. 
This is completed in a stochastic fashion where the degree of confidence in the current shape is compared with the level of measurement noise. 
Our approach is able to operate in real time and incrementally update the shape model of an asteroid as new range measurements are collected.
This approach allows for the shape to be continually updated as range measurements are used to locally modify the shape estimate.
This technique is verified by numerical examples of shape construction for asteroids Geographos and Golevka. 

Next, we present an optimal guidance scheme to construct a complete and accurate shape model of an asteroid. 
The motivation is to actively control the relative position of the spacecraft such that the range sensor covers the surface area from every perspective. 
Utilizing the stochastic formulation of the above shape construction method, this is addressed by an optimization to find the pose of the spacecraft that minimizes the degree of uncertainties in the shape model. 

Finally, we consider the scenario of landing on the surface of an asteroid. 
Once a complete shape model is constructed, we study a landing site selection problem. 
For safety concern, the area with higher slope is excluded. 
Within the low slope region, a landing site is selected to minimize a control cost subtracted by a measure of scientific interests. 
Once the landing site is selected, the nearby area is mapped with a higher resolution to create a detailed topological map.  
The proposed multi resolution mapping yields a high fidelity map without the computational burden of a uniformly high resolution mesh.  

The optimal guidance and multi resolution mapping schemes are illustrated by dynamical simulation of landing on the asteroid Castalia. 
During this simulation, this updated shape model is then used in a nonlinear controller to track a desired state trajectory for the dynamics of a rigid body spacecraft.
The dynamics are developed on the nonlinear manifold of rigid body motions, namely the special Euclidean group.
This formulation is based on an intrinsic geometric description of the motion and accurately captures the coupling between orbit and attitude dynamics. 


In short, this paper presents a method to incrementally update the shape  model of an asteroid from range measurements. 
Our approach alleviates the need for a dedicated mapping phase as the spacecraft is able to update its shape model in real time and without expensive computations.
This type of approach allows for the spacecraft to maneuver and land on the asteroid immediately upon arrival rather than spending several months mapping the surface.

\section{Problem Formulation}\label{sec:problem}

\begin{figure}[htbp]
    \centering

\newcommand\pgfmathsinandcos[3]{%
  \pgfmathsetmacro#1{sin(#3)}%
  \pgfmathsetmacro#2{cos(#3)}%
}
\newcommand\LongitudePlane[3][current plane]{%
  \pgfmathsinandcos\sinEl\cosEl{#2} 
  \pgfmathsinandcos\sint\cost{#3} 
  \tikzset{#1/.style={cm={\cost,\sint*\sinEl,0,\cosEl,(0,0)}}}
}
\newcommand\LatitudePlane[3][current plane]{%
  \pgfmathsinandcos\sinEl\cosEl{#2} 
  \pgfmathsinandcos\sint\cost{#3} 
  \pgfmathsetmacro\yshift{\cosEl*\sint}
  \tikzset{#1/.style={cm={\cost,0,0,\cost*\sinEl,(0,\yshift)}}} %
}

\newcommand\DrawLongitudeCircle[4][1]{
\LongitudePlane{\angEl}{#2}
\tikzset{current plane/.prefix style={scale=#1}}
\pgfmathsetmacro\angVis{
atan(sin(#2)*cos(\angEl)/sin(\angEl))} %
\draw[shift={(#3, #4)}][current plane]
(\angVis:1) arc (\angVis:\angVis+180:1);
\draw[shift={(#3, #4)}][current plane,dashed]
(\angVis-180:1)arc(\angVis-180:\angVis:1);
}
\newcommand\DrawLatitudeCircle[4][1]{
\LatitudePlane{\angEl}{#2}
\tikzset{current plane/.prefix style={scale=#1}}
\pgfmathsetmacro\sinVis{
sin(#2)/cos(#2)*sin(\angEl)/cos(\angEl)}
\pgfmathsetmacro\angVis{
asin(min(1,max(\sinVis,-1)))}
\draw[shift={(#3, #4)}][current plane]
(\angVis:1) arc (\angVis:-\angVis-180:1);
\draw[shift={(#3, #4)}][current plane,dashed]
(180-\angVis:1)arc(180-\angVis:\angVis:1);
}


\tikzset{%
  >=latex, 
  inner sep=0pt,%
  outer sep=2pt,%
  mark coordinate/.style={inner sep=0pt,outer sep=0pt,minimum size=3pt,
    fill=black,circle}%
}

\tikzsetnextfilename{dumbbell}
\begin{tikzpicture}[scale=0.8] 

\def\R{2} 
\def\angEl{35} 
\def\angAz{-105} 
\def\length{4} 
\def\opacity{0.2}

\coordinate (m1) at (\length, 0);
\coordinate (m2) at (-\length, 0);

\draw [ thick, dashed, -Latex] (m2) -- ($ (m1) + (3, 0) $) node [below right] {$b_1$};
\draw [thick, dashed, -Latex] (0, 0) -- (0, \R) node [above right] {$b_2$};
\filldraw[ball color=blue, opacity=\opacity] (m1) circle (\R);
\foreach \t in {-45, 0, 45} { \DrawLatitudeCircle[\R]{\t}{\length}{0} }
\foreach \t in {-30, -60,...,-150} { \DrawLongitudeCircle[\R]{\t}{\length}{0} }

\filldraw[ball color=blue, opacity=\opacity] (m2) circle (\R);
\foreach \t in {-60,-30,...,60} { \DrawLatitudeCircle[\R]{\t}{-\length}{0} }
\foreach \t in {-5,-35,...,-175} { \DrawLongitudeCircle[\R]{\t}{-\length}{0} }

\draw[thick,-Latex] node [below right] {$\zeta$} (0, 0) --  (m1);
\draw[thick,-Latex] (m1) -- ++(30:\R) node [above right] {$\eta$};


\end{tikzpicture}
    \caption{Dumbbell Model of Rigid Spacecraft\label{fig:dumbbell_sc}}
\end{figure}
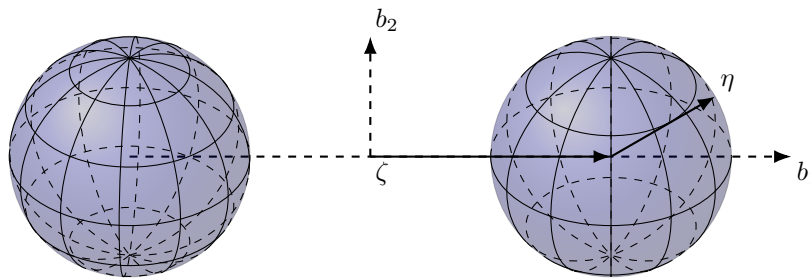

In this paper, we consider the motion of a dumbbell model of spacecraft around an asteroid.
The dumbbell is defined by two spherical masses of radius \( r_1, r_2 \in \R \) with masses \( m_1, m_2 \in  \R\).
The masses are separated by a massless rod of length \( l \) and attached to the centers of each mass.
\Cref{fig:dumbbell_sc} shows the model and associated parameters.
The dumbbell model is in some sense the simplest approximation of an extended rigid body.
This model captures the important dynamics of an extended rigid body with a minmum of additional complexity and is prevelant the literature~\cite{kulumani2017b,lee2006a,lee2007a,elmasri2005}.
In addition, the developments of the dumbbell model can be trivially extended to include a larger number of discrete masses to better approximate a more complicated spacecraft model.

Without loss of generality, we define body fixed frames for both the spacecraft and asteroid, which are aligned with the principle axes of each body and originate at their respective center of mass. 
The spacecraft body fixed frame is centered at the center of mass of the vehicle.
The \( b_1 \)  axis is aligned with the connecting rod and directed along the axis of symmetry.
The \( b_2, b_3 \) axes span the plane orthogonal to the axis of symmetry of the dumbbell.
The distance from the center of mass to each spherical mass is defined as
\begin{align}\label{eq:dumbbell_mass_distances}
    l_1 &= \frac{m_2}{m_1 + m_2} l, \\
    l_2 &= l - l_1.
\end{align}
The asteroid is modeled as a constant density polyhedron with constant, and known, spin about the axis of its maximum moment of inertia. 
The axes of the body fixed frame for the asteroid are denoted by $\{f_1,f_2,f_3\}$. 
We also define the inertial frame, whose axes are denoted by $\{e_1,e_2,e_3\}$. 

The kinematics of the dumbbell and asteroid are described in the inertial frame by
\begin{itemize}
    \item \( \vc{x}\in\R^3 \) - the position of the center of mass of the  spacecraft represented in the inertial frame, \( \vc{e}_i\),
    \item \( R\in\SO \) - the rotation matrix which transforms the representation of vectors defined in the spacecraft fixed frame, \( \vc{b}_i \), to the inertial frame, \( \vc{e}_i \),
    \item \( \vc{\Omega}\in\R^3 \) - the angular velocity of the spacecraft body fixed frame relative to the inertial frame and represented in the dumbbell body fixed frame, \( \vc{b}_i \), and
    \item \( R_A\in \SO \) - the rotation matrix which transforms the representation of vectors defined in the asteroid fixed frame, \( \vc{f}_i \), to the inertial frame, \( \vc{e}_i \).
\end{itemize}
In this work, we assume that the asteroid is much more massive than the spacecraft and its motion is not affected by that of the spacecraft.
This assumption allows us to treat the motion of the vehicle independently from the dynamics of the asteroid, which is assumed to spin at a fixed rate.  

\subsection{Spacecraft Dynamical Model}

Using Hamilton's principle one can derive the inertial equations of motion of the dumbbell spacecraft~\cite{kulumani2017b} as
\begin{align}
    \dot{\vc{x}} &= \vc{v}, \label{eq:x_dot}\\
    \parenth{m_1 + m_2} \dot{\vc{v}} &= m_1 R_A \deriv{U}{\vc{z}_1} + m_2 R_A \deriv{U}{\vc{z}_2} + u_f,\label{eq:v_dot} \\
    \dot{R} &= R S(\vc{\Omega}) , \label{eq:R_dot}\\
    J \dot{\vc{\Omega}} + \vc{\Omega} \times J \vc{\Omega} &= \vc{M}_1 + \vc{M}_2 + u_m. \label{eq:omega_dot}
\end{align}
The vectors \( \vc{z}_1 \) and \( \vc{z}_2\in\R^3\) define the position of the dumbbell masses in the asteroid fixed frame and are defined as
\begin{align}
    \vc{z}_1 &= R_A^T \parenth{\vc{x} + R \vc{\rho}_1} , \\
    \vc{z}_2 &= R_A^T \parenth{\vc{x} + R \vc{\rho}_2},
\end{align}
where \( \rho_i \in\R^3 \) defines the position of each mass in the spacecraft fixed body frame.
The control inputs to the spacecraft are defined by \( u_f, u_m \in \R^3 \) which define the control force represented in the inertial frame and the control moment represented in the spacecraft frame, respectively.
The standard moment of inertia of the dumbbell model is 
\begin{align}\label{eq:dumbbell_moment_of_inertia}
    J_I = \sum_i^n J_i + m_i \parenth{\vc{\zeta}_i^T \vc{\zeta}_i I_{3\times 3} - \vc{\zeta}_i \vc{\zeta}_i^T} , 
\end{align}
where \( \vc{\zeta}_i \) is the position of \( m_i \) in the spacecraft fixed frame and the moment of inertia of each sphere is
\begin{align}\label{eq:sphere_moment_of_inertia}
    J_i = \begin{bmatrix} 
        \frac{2}{5} m_i r_i^2 & 0 & 0 \\
        0 & \frac{2}{5} m_i r_i^2 & 0 \\
        0 & 0 & \frac{2}{5} m_i r_i^2 
    \end{bmatrix}.
\end{align}
\Cref{eq:dumbbell_moment_of_inertia} is consistent with the well-known parallel-axis theorem~\cite{greenwood1988}.
The gravitational moment on the dumbbell \( \vc{M}_i\) is defined as
\begin{align}
    \vc{M}_i = m_i \parenth{S(R_A^T \vc{\rho}_i) R^T \deriv{U}{\vb{z}_i}}.
\end{align}
In the above equations, the gravitational potential is denoted by $U\in\R$, which is computed by a polyhedron model as described in the subsequent section. 


\subsection{Polyhedron Potential Model}\label{sec:polyhedron_potential}

An accurate gravitational potential model is necessary for the operation of spacecraft about asteroids.
Additionally, a detailed shape model of the asteroid is needed for trajectories passing close to the body.
The classic approach is to expand the gravitational potential into a harmonic series and compute the series coefficients.
However, the harmonic expansion is always an approximation as a result of the infinite order series used in the representation.
Additionally, the harmonic model used outside of the circumscribing sphere is not guaranteed to converge inside the sphere, which makes it unsuitable for trajectories near the surface.

We represent the gravitational potential of the asteroid using a polyhedron gravitation model.
This model is composed of a polyhedron, which is a three-dimensional solid body, that is defined by a series of vectors in the body-fixed frame.
The vectors define vertices in the body-fixed frame as well as planar faces which compose the surface of the asteroid.
We assume that each face is a triangle composed of three vertices and three edges.
As a result, only two faces meet at each edge while three faces meet at each vertex.
Only the body-fixed vectors, and their associated topology, is required to define the exterior gravitational model.
References~\cite{werner1994} and~\cite{werner1996} give a detailed derivation of the polyhedron model.

Here, we summarize the key equations required for implementation.
The polyhedron potential is defined as 
\begin{align}
    U(\vc{r}) &= \frac{1}{2} G \sigma \sum_{e \in \text{edges}} \vc{r}_e \cdot \vc{E}_e \cdot \vc{r}_e \cdot L_e - \frac{1}{2}G \sigma \sum_{f \in \text{faces}} \vc{r}_f \cdot \vc{F}_f \cdot \vc{r}_f \cdot \omega_f,
\end{align}
and \( \vc{r}_e\) and \(\vc{r}_f \) are the vectors from the spacecraft to any point on the respective edge or face, \( G\in\R\) is the universal gravitational constant, and \( \sigma\in\R \) is the constant density of the asteroid.
We define the attraction, gravity gradient matrix, and Laplacian as
\begin{align}
    \nabla U ( \vc{r} ) &= -G \sigma \sum_{e \in \text{edges}} \vc{E}_e \cdot \vc{r}_e \cdot L_e + G \sigma \sum_{f \in \text{faces}} \vc{F}_f \cdot \vc{r}_f \cdot \omega_f \in \R^{3 \times 1} , \label{eq:attraction}\\
    \nabla \nabla U ( \vc{r} ) &= G \sigma \sum_{e \in \text{edges}} \vc{E}_e  \cdot L_e - G \sigma \sum_{f \in \text{faces}} \vc{F}_f \cdot \omega_f \in \R^{3 \times 3}, \label{eq:gradient_matrix}\\
    \nabla^2 U &= -G \sigma \sum_{f \in \text{faces}}  \omega_f \in \R^1 .\label{eq:laplacian}
\end{align}
The Laplacian, defined in~\cref{eq:laplacian}, gives a simple method to determine if the spacecraft has collided with the body~\cite{werner1996}. 
The sum \( \sum \omega_f \) vanishes when outside the body and equals \( -4 \pi \) inside.
These equations are utilized to compute the gravitational potential of  the presented dumbbell spacecraft model.

\section{Bayesian Shape Reconstruction}\label{sec:radius_update}

One of the first tasks for any spacecraft mission to a small body is to generate an estimate of the shape.
We assume that upon arrival at a target body, the spacecraft contains an initial estimate for the shape of the small body.
This shape can be a coarse estimate computed from ground measurements or it can be a triaxial ellipsoid based on the semimajor axes of the asteroid.
Additionally, we assume that the shape estimate is a triangular faceted, closed surface mesh, emulating those used in practice to represent asteroids.
Furthermore, the number of vertices in the estimate can be scaled according to the desired final accuracy or computational capabilities.
In this section, we present a stochastic formulation to update the initial estimate of the shape incrementally based on range measurements from spacecraft in real time. 

\subsection{Range Measurements}

We assume the spacecraft contains a range sensor, such as LIDAR, that allows for the accurate measurement of the relative distance between the spacecraft and asteroid~\cite{zuber1997,zuber2000}.
Recent missions such as OSIRIS-REx have featured a scanning LIDAR system which allows for much higher measurement rates compared to previous missions~\cite{lauretta2017,daly2017}.
In these mission, ranging measurements are combined with imagery to derive a shape model using a ground based method~\cite{gaskell2008}.
In constrast to previous missions, the scanning LIDAR system on OSIRIS-REx enables for high surface sample rates in spite of the low ground-track velocities~\cite{daly2017}.
This type of sensor measures the round-trip time for a pulse of energy to leave the spacecraft, reflect off the surface, and return to a collector on board.
Given the time total time of flight ($\Delta TOF$) the distance can be accurately computed using \( d = \frac{c \Delta TOF }{2} \) where \( c = \SI{2.998e8}{\meter\per\second}\) is the constant speed of light.
Assuming accurate knowledge of the pointing direction of the spacecraft, in the form of the rotation matrix \( R \in \SO \), we can compute a direction from the spacecraft to the measurement location on the surface.
The output of this sensor is a vector, \( \vc{d}_i\in\R^3 \), defined in the spacecraft fixed frame which gives the direction to a measurement point on the surface. 
Using the state of the asteroid, we can transform this measurement to the asteroid fixed frame using the simple transformation
\begin{align*}
    \vc{p}_i = R_A^T (x+R \vc{d}_i), 
\end{align*}
which is the vector from the origin of the asteroid fixed frame to the surface point of the measurement.

\begin{figure}
    \centering
    \includegraphics[width=0.75\textwidth]{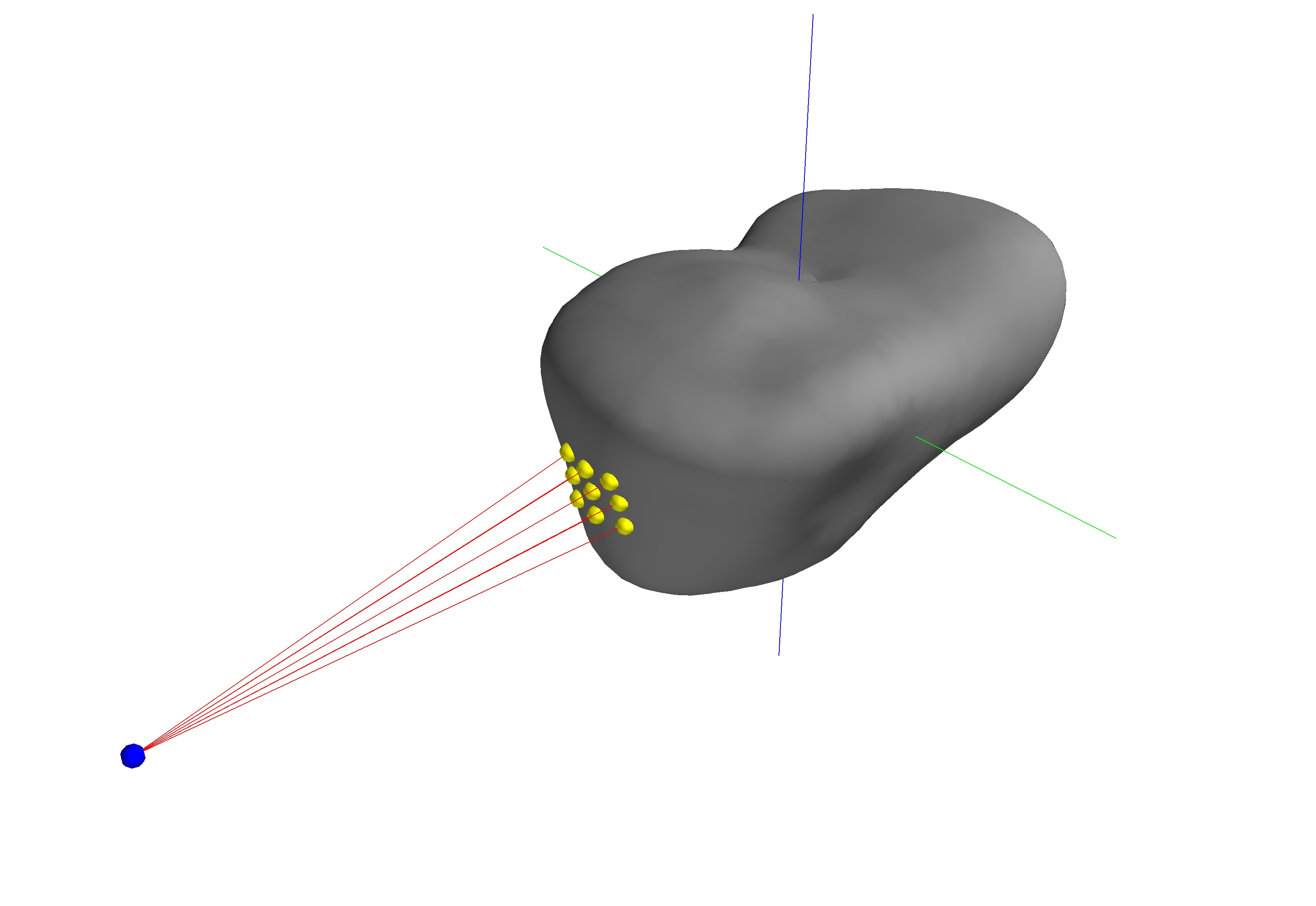}
    \caption{Simulated LIDAR measurements of asteroid Castalia~\label{fig:lidar_example}}
\end{figure}

Given many measurements, \( \vc{p}_i \in \R^3 \), of the asteroid surface we can efficiently update our initial shape estimate to that of the true surface.
\Cref{fig:lidar_example} shows asteroid 4769 Castalia and a representation of several LIDAR measurements. 
The spacecraft measures the range between itself and the asteroid surface to several points within the field of view of the sensor. 
These measurements provide a collection of points to form so called \textit{point cloud}, which allows us to reconstruct the shape.

\subsection{Bayesian Shape Update}

Our algorithm applies a probabilistic framework to radially modify each vertex \( \vc{v}_i \in \R^3\) of the shape estimate based on measurement \( \vc{p}_i \in \R^3 \). 
In other words, each vertex from the initial estimate of shape is either stretched or shrunk along its radial direction, without any rotation. 
We assume that the initial intertial state of the vehicle is available. 
This data is typically available from a ground based orbit determination scheme which enabled the original orbital rendezvous. 
The approach presented here is reasonable assuming that both the initial estimate and the actual shape of the body does not have any hole. 
This approach alleviates much of complexity of incorporating new vertices or surface triangulation common in surface reconstruction methods~\cite{berg2008}.
This implies that the total number of vertices of the shape model is fixed.
However, additional detail, in the form of additional vertices, is possible by using standard mesh subdivision algorithms~\cite{orourke1998}, which is discussed in the subsequent section for multi resolution mapping.

The proposed scheme follows a Bayesian estimation scheme, where the degree of confidence in the current estimate is compared with that of new measurements. 
More specifically, the radial distance of each vertex, \( v_i = \norm{\vc{v}_i}\), is assumed to be distributed according to the Gaussian distribution
\begin{align*}
    v_i \sim \mathcal{N}(r_i, w_i^2)
\end{align*}
where \( r_i \in \R \) is the initial estimate of the radial distance of the $i$-th vertex \( \vc{v}_i\in\R\), and \( w_i\in\R \) is the initial variance, or confidence, in the radial distance.

In order to reduce the computational demands, which are typical in point cloud applications, we do not update the complete shape model for each measurement.
Instead we define an area of interest, \( \Delta S\in\R \), about each measurement which defines the surface area over which the measurement will affect the mesh estimate.
We relate \( \Delta S \) to an equivalent angular constraint using
\begin{align}\label{eq:region_of_interest}
    \Delta \sigma_{max} = \sqrt \frac{\Delta S}{r_b^2}
\end{align}
where \( r_b \in \R \) defines the Brillouin  sphere radius, or the radius of the circumscribing sphere of the asteroid.
Only vertices which satisfy \( \Delta \sigma_i \leq \Delta \sigma_{max} \) are considered in the Bayesian update defined as follows.

Each measurement is defined by the index \( j \) while the associated vertex satisfying \cref{eq:region_of_interest} is defined by \( i \). 
As a result, the measurement \( p_{j, i} \) defines the distribution of measurement \( j \) with respect to vertex \( i \). 
The radial distance of each measurement, \( p_{j,i} = \norm{\vc{p}_j}\), is also assumed to be distributed according to the Gaussian distribution
\begin{align*}
    p_{j,i} \sim \mathcal{N}(r_{j,i}, w_{j,i}^2)
\end{align*}
where \( r_{j,i} = \norm{\vc{p}_{j,i}} \) defines the radial distance of the surface vector measurement and \( w_{j, i} \in\R\) defines the variance of the measurement with respect to vertex \( \vc{v}_i\).

The variance for each measurement vector is assumed to be related to the ``distance'' from the measurement to vertex \( \vc{v}_i \).
Here, we use the geodesic distance to parameterize the difference, and hence  uncertainty, of associating the measurement with a given vertex.
More explicitly, the variance of measurement \( \vc{p}_i \) with respect to vertex \( \vc{v}_i \) is then defined by the geodesic distance as
\begin{align}
    w_{j, i} = c \norm{\vc{p}_j} \Delta \sigma_{j,i} ,
\end{align}
where \( c \) is an additional scaling constant.
This approach relates the uncertainty of the measurement \( \vc{p}_j \) with the geodesic distance to a given vertex, \( \vc{v}_i \).
As a result, measurements which are far from a vertex, i.e.\ \( \Delta \sigma \) is large, will tend to have a larger variance and hence more uncertainty. 
This approach can be considered as a form of a correlation based sensor model~\cite{thrun2005}.
The main benefit of a correlation based approach, in contrast to feature extraction is the relative simplicity of implementation.
However, the resulting correlation values do not precisely represent the noise or uncertainty characteristics of the sensor in a quantitative manner.


From spherical trigonometry~\cite{gade2010}, the central angle between measurement \( \vc{p}_j \) and vertex \( \vc{v}_i \) of the shape estimate is given by
\begin{align}\label{eq:geodesic_distance}
    \Delta \sigma_{j,i} = \arctan \parenth{\frac{\norm{\vc{p}_j \times \vc{v}_i}}{\vc{p}_j \cdot \vc{v}_i }}.
\end{align}
The parameters \( c \) and \( \Delta \sigma_{max} \) are used to define the region of impact of each measurement. 
The angular constraint \( \Delta \sigma_{max} \) may be used to adjust the computational requirements of the shape refinement process.
Larger values of \( \Delta \sigma_{max} \) result in a greater computational demand as a given measurement ray is used to update a larger number of vertices.
\Cref{sec:landing_refinement} presents an approach to allow for a higher resolution mesh in certain regions to capture finer surface details without excessive resources. 
The scalar \( c \) may then be used to adjust the variance \( w_{j,i} \) as a function of the distance between measurement \( j \) and vertex \( i \).
These two parameters may be used to minimize steep gradients that may occur during the shape reconstruction process.

From Bayes' theorem, the a posteriori probability of the vertex radius is given by
\begin{align}
    p(v_i | p_{j, i}) = \frac{p(p_{j, i} | v_i) p(v_i)}{p( p_{j, i})} \propto p(p_{j,i} | v_i) p(v_i).
\end{align}
From the properties of Gaussian distributions, the posterior probability given a measurement is also distributed according to a Gaussian distribution~\cite{bishop2006} and given by
\begin{align}\label{eq:posterior_probability}
    \mathcal{N} \parenth{\frac{w_{j, i}^2 r_i + w_i^2 r_{j, i}}{w_i^2 + w_{j, i}^2} , \frac{w_i^2  w_{j, i}^2}{w_i^2 +  w_{j, i}^2}} .
\end{align}
From~\cref{eq:posterior_probability}, the a posterior mean conditioned on the measurement is the average of the prior knowledge and the measurement weighted by the reciprocal of the variance. 
As such, it will be closer to the value with a smaller variance. 
For example, measurements that are far from the vertex will have a high uncertainty or variance and will have a reduced impact on the radial position of the vertex.

The approach presented in this section allows one to update the shape of small body given a single range measurement of the surface.
A sequential process can be used to iteratively update the shape estimate given many measurements of the surface. 

\subsection{Numerical Examples}\label{sec:kinematic_exploration}

In this section, we demonstrate the use of the incremental shape reconstruction algorithm with asteroids 1620 Geographos and 6489 Golevka.
Their properties are listed at \Cref{tab:kinematic_asteroids}.
Truth shape models for Geographos and Golevka are computed based on Earth based radar measurements~\cite{neese2004}.
\begin{table}[htbp]
    \centering
    \begin{tabular}{lccc}
        \toprule
        Asteroid & Semi-major axes (\si{\kilo\meter}) & Vertices & Faces\\
        \midrule
        \num{1620} Geographos & \( 2.5 \times 1.0 \times 1.05 \) & \num{8192} & \num{16380}  \\
        \num{6489} Golevka & \( 0.53 \times 0.53 \times 0.53 \)  & \num{2048} & \num{4092} \\
        \bottomrule
    \end{tabular} 
    \caption{Asteroid properties for kinematic only exploration~\label{tab:kinematic_asteroids}}
\end{table}

The results in this section utilize a kinematics only model of the spacecraft instead of the full dynamical simulation. 
We ignore the dynamics of the asteroid and spacecraft and instead focus solely on the shape reconstruction, and the spacecraft is assumed to be able to arbitrarily move around the asteroid and collect measurements.
The spacecraft remains at a fixed orbital radius defined as twice the maximum semi-major axis of the asteroid.
The simulations begin with an triaxial ellipsoid mesh that is sized to match the semi-major axes of each asteroid.
With this estimate, measurements are made of the surface and used to reconstruct the true shape following the process in~\cref{sec:radius_update}.
LIDAR measurements are generated until the total uncertainty of the model
\begin{align*}
    \sum_i w_i,
\end{align*}
is sufficiently small.
This uncertainty metric is the summation of the uncertainty of each vertex of the shape estimate.
As a result, a small total uncertainty is used to indicate that additional measurements would not substantially modify the shape.
In addition, we compute the volume of the estimated shape and compare it against the volume of the truth shape model~\cite{neese2004}.

\begin{figure}[htbp]
    \centering
    \subcaptionbox{Initial Shape Estimate\label{fig:geographos_partial_0}}{\includegraphics[trim={20cm 15cm 20cm 15cm}, clip,width=0.5\textwidth,keepaspectratio]{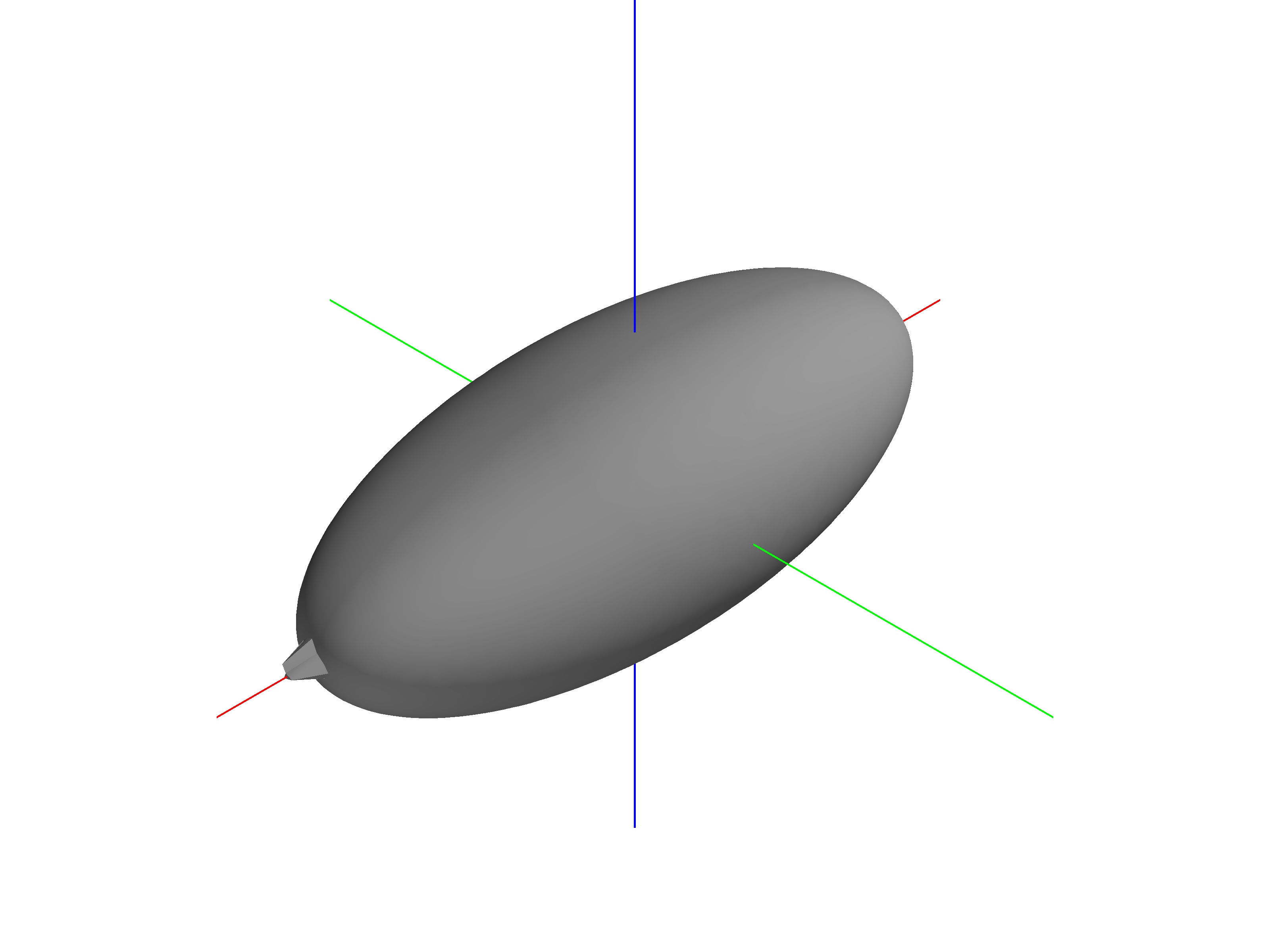}}%
    \subcaptionbox{\SI{25}{\percent} of measurements added\label{fig:geographos_partial_25}}{\includegraphics[trim={20cm 15cm 20cm 15cm},clip,keepaspectratio,width=0.5\textwidth]{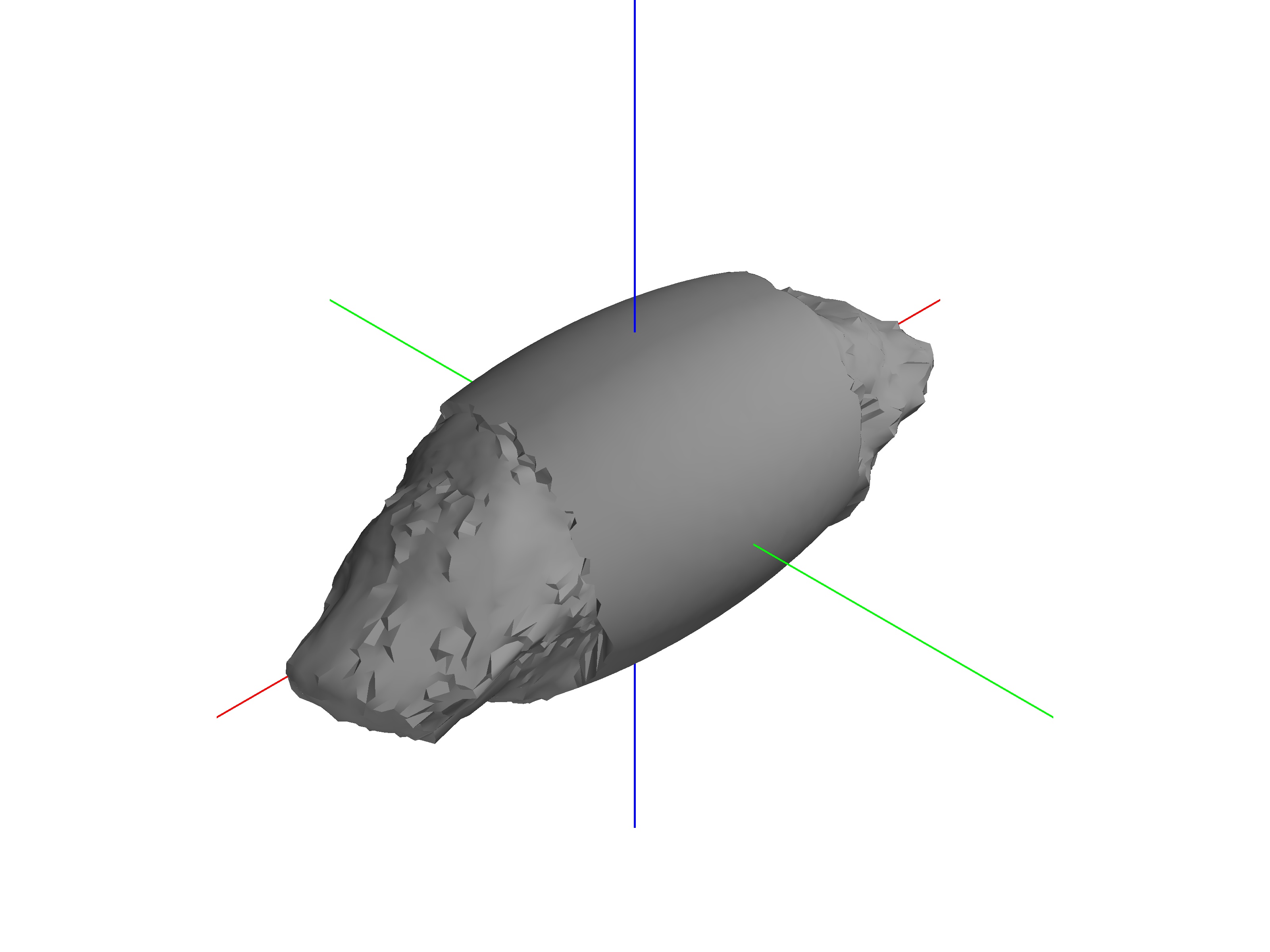}}

    \subcaptionbox{\SI{50}{\percent} of measurements added\label{fig:geographos_partial_50}}{\includegraphics[trim={20cm 15cm 20cm 15cm},clip,keepaspectratio,width=0.5\textwidth]{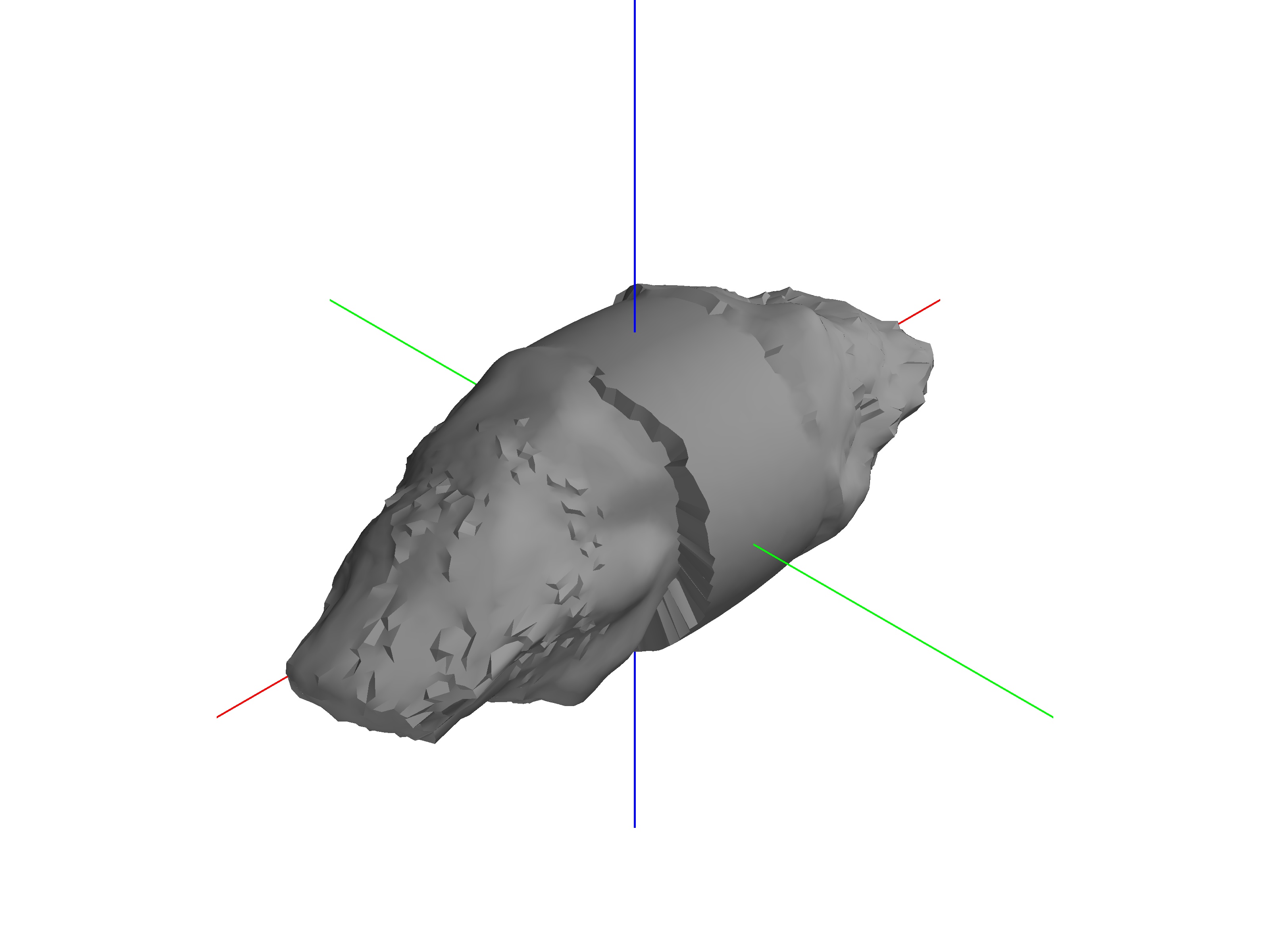}}%
    \subcaptionbox{\SI{75}{\percent} of measurements added\label{fig:geographos_partial_75}}{\includegraphics[trim={20cm 15cm 20cm 15cm},clip,keepaspectratio,width=0.5\textwidth]{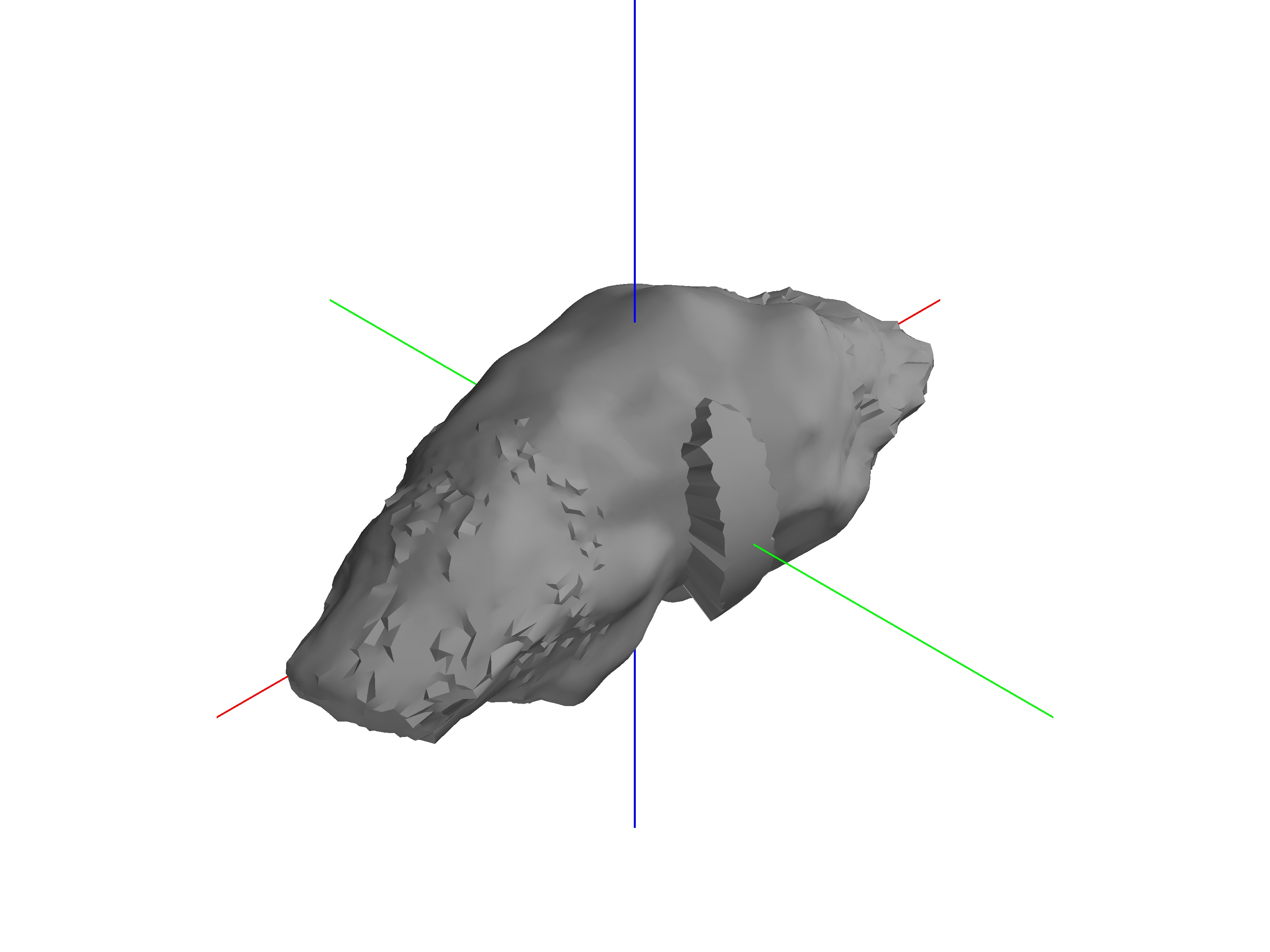}}

    \subcaptionbox{\SI{100}{\percent} of measurements added\label{fig:geographos_partial_100}}{\includegraphics[trim={20cm 15cm 20cm 15cm},clip,keepaspectratio,width=0.5\textwidth]{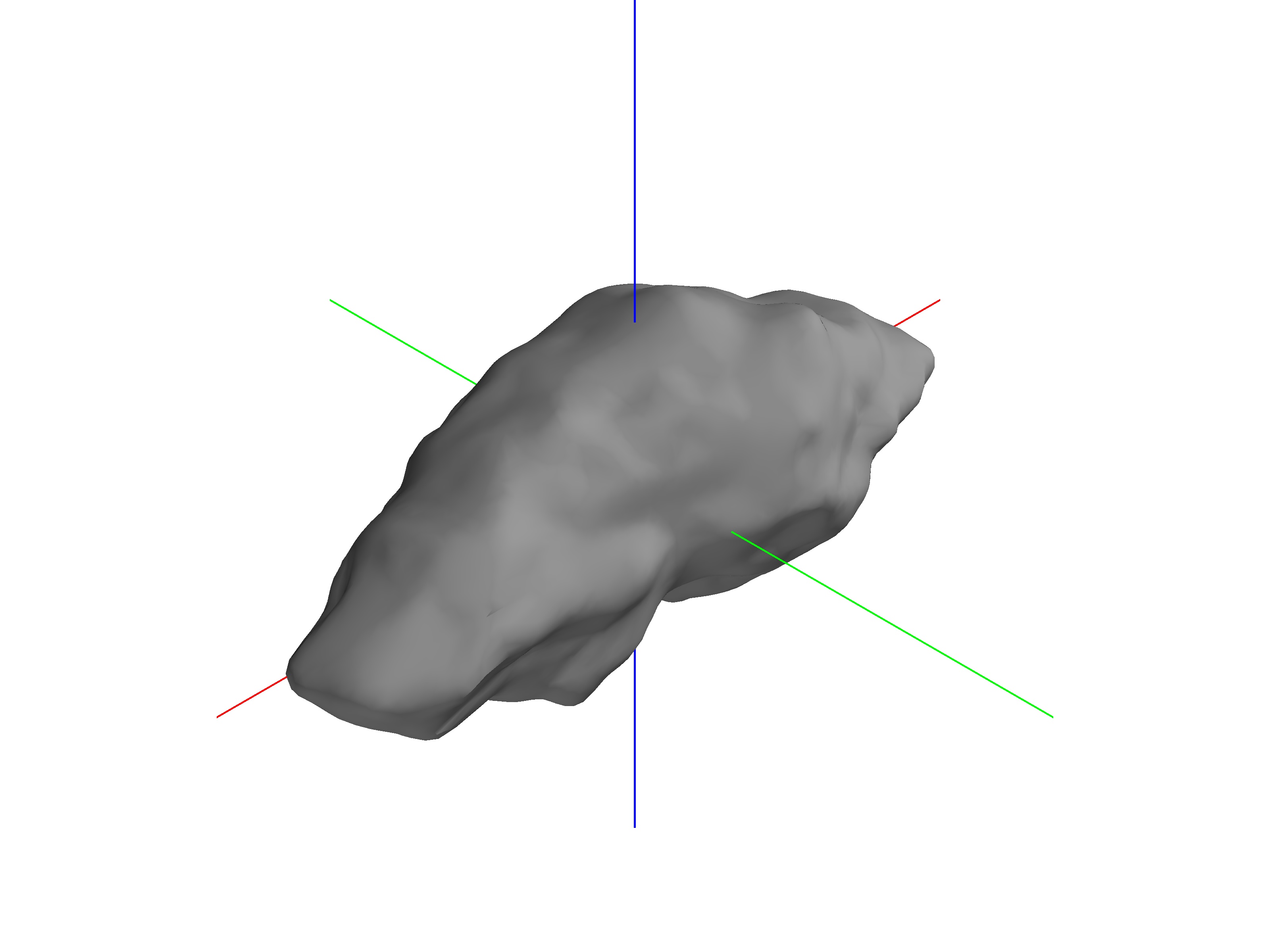}}%
    \subcaptionbox{True Shape Model\label{fig:geographos_truth}}{\includegraphics[trim={20cm 15cm 20cm 15cm},clip,keepaspectratio,width=0.5\textwidth]{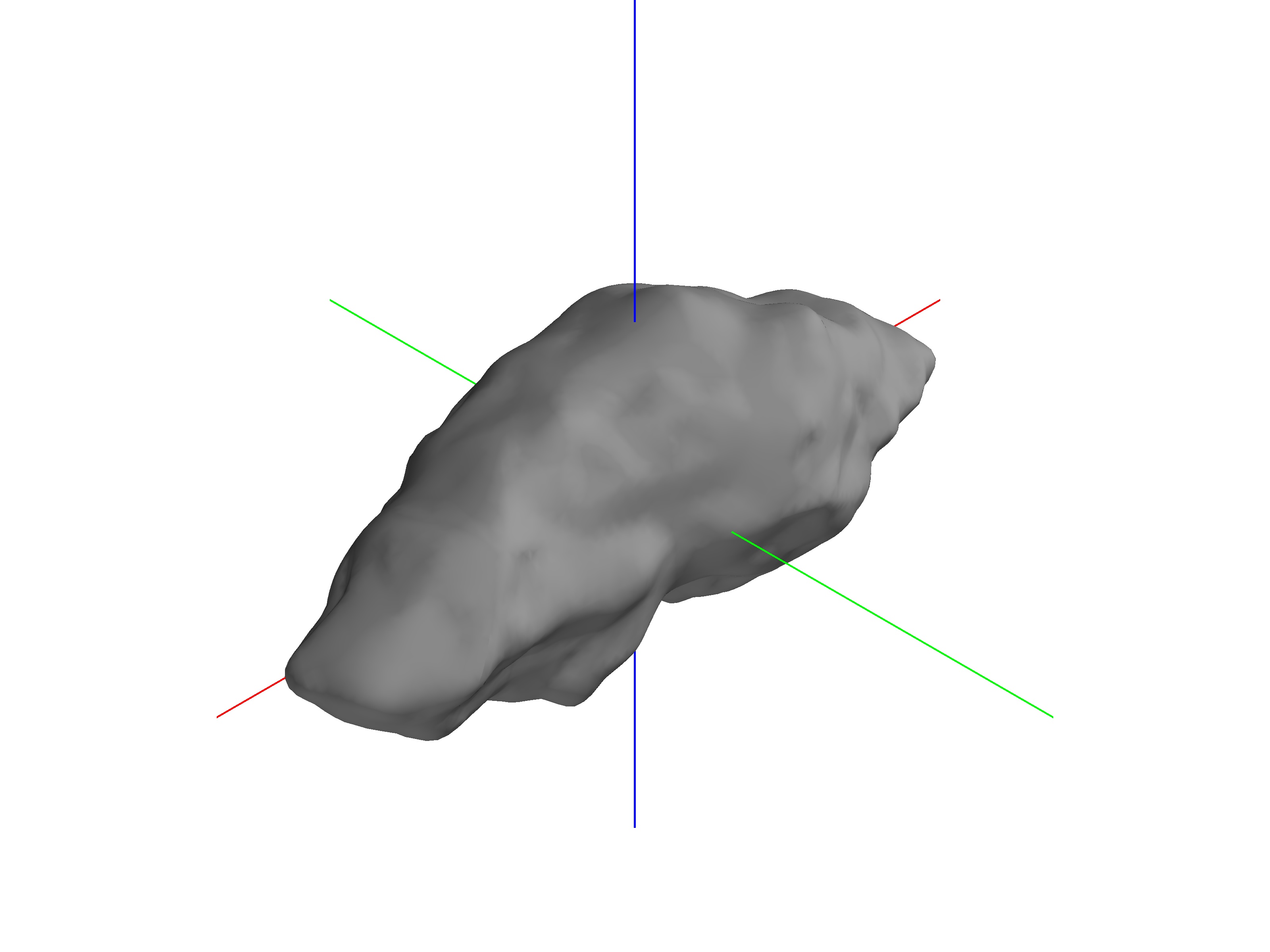}}
    \caption[Asteroid Geographos incremental reconstruction]{Incremental reconstruction of asteroid Geographos~\label{fig:geographos_reconstruction}}
\end{figure}

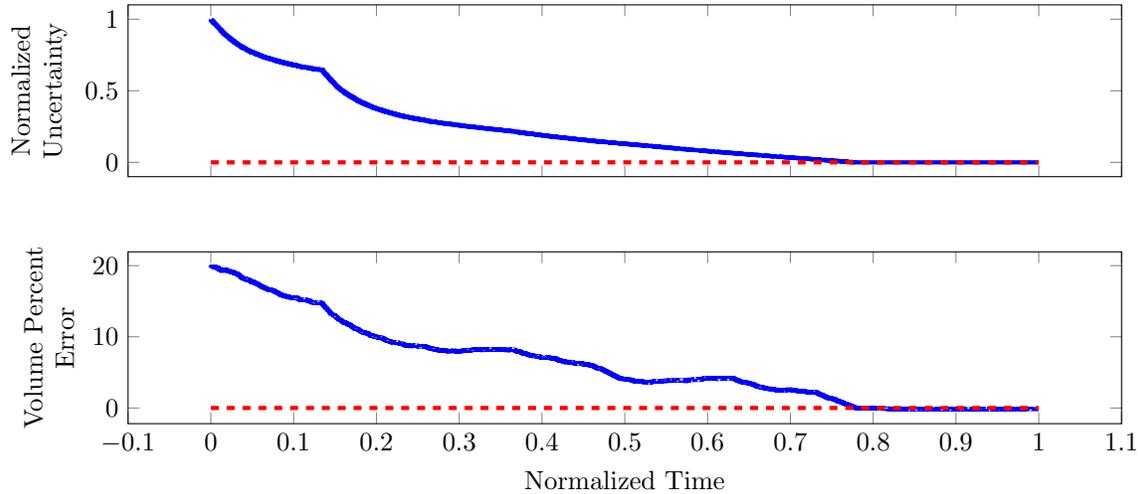
\begin{figure}[htbp]
    \centering
    \tikzsetnextfilename{geographos_metrics}
\begin{tikzpicture}[baseline]
    \begin{groupplot}[
        group style={
            group name={geographos_metrics},
            group size=1 by 2,
            xlabels at=edge bottom,
            ylabels at=edge left,
            xticklabels at=edge bottom,
        },
        xlabel={Normalized Time},
        scale only axis,
        width=0.8\textwidth,
        height=0.1\textheight,
        ylabel style={align=center},
    ]
    \nextgroupplot[ylabel={Normalized\\Uncertainty}]
    \addplot [ultra thick, color=blue, mark=none] table [x=NORMALIZED_TIME, y=NORMALIZED_UNCERTAINTY, col sep=comma] {mesh_update_geographos_uncertainty.csv};
    \addplot [ultra thick,red, mark=none, dashed] coordinates {
        (0.0, 0.0) (1.0, 0.0) 
    };

    \nextgroupplot[ylabel={Volume Percent\\Error}]
    \addplot [ultra thick, blue, mark=none] table [x=NORMALIZED_TIME, y=VOLUME_PERCENT_ERROR, col sep=comma] {mesh_update_geographos_volume.csv};
        \addplot [ultra thick,red, mark=none, dashed] coordinates {
            (0.0, 0.0) (1.0, 0.0) 
        };
\end{groupplot}
\end{tikzpicture}
    \caption{Normalized uncertainty and volume percent error for Geographos\label{fig:geographos_metrics}}
\end{figure}

\paragraph{Asteroid \num{1620} Geographos Reconstruction} 
\num{1620} Geographos is a highly elongated stony asteroid of the Apollo group.
Discovered in \num{1951}, Geographos is a potentially hazardous asteroid which passes sufficiently close to the Earth.
\Cref{fig:geographos_reconstruction} shows the shape reconstruction for asteroid Geographos at several distinct points during the process.
Comparing~\cref{fig:geographos_partial_100,fig:geographos_truth} shows that the final shape closely matches the true radar model.
In~\cref{fig:geographos_metrics} we display the vertex uncertainty and mesh volume as a function of time.
The plots show that the reconstruction achieves an accurate shape estimate with a total volume which closely matches the true volume.

\begin{figure}[htbp]
    \centering
    \subcaptionbox{Initial Shape Estimate\label{fig:golevka_partial_0}}{\includegraphics[trim={20cm 5cm 20cm 5cm},clip,keepaspectratio,width=0.5\textwidth,height=0.25\textheight]{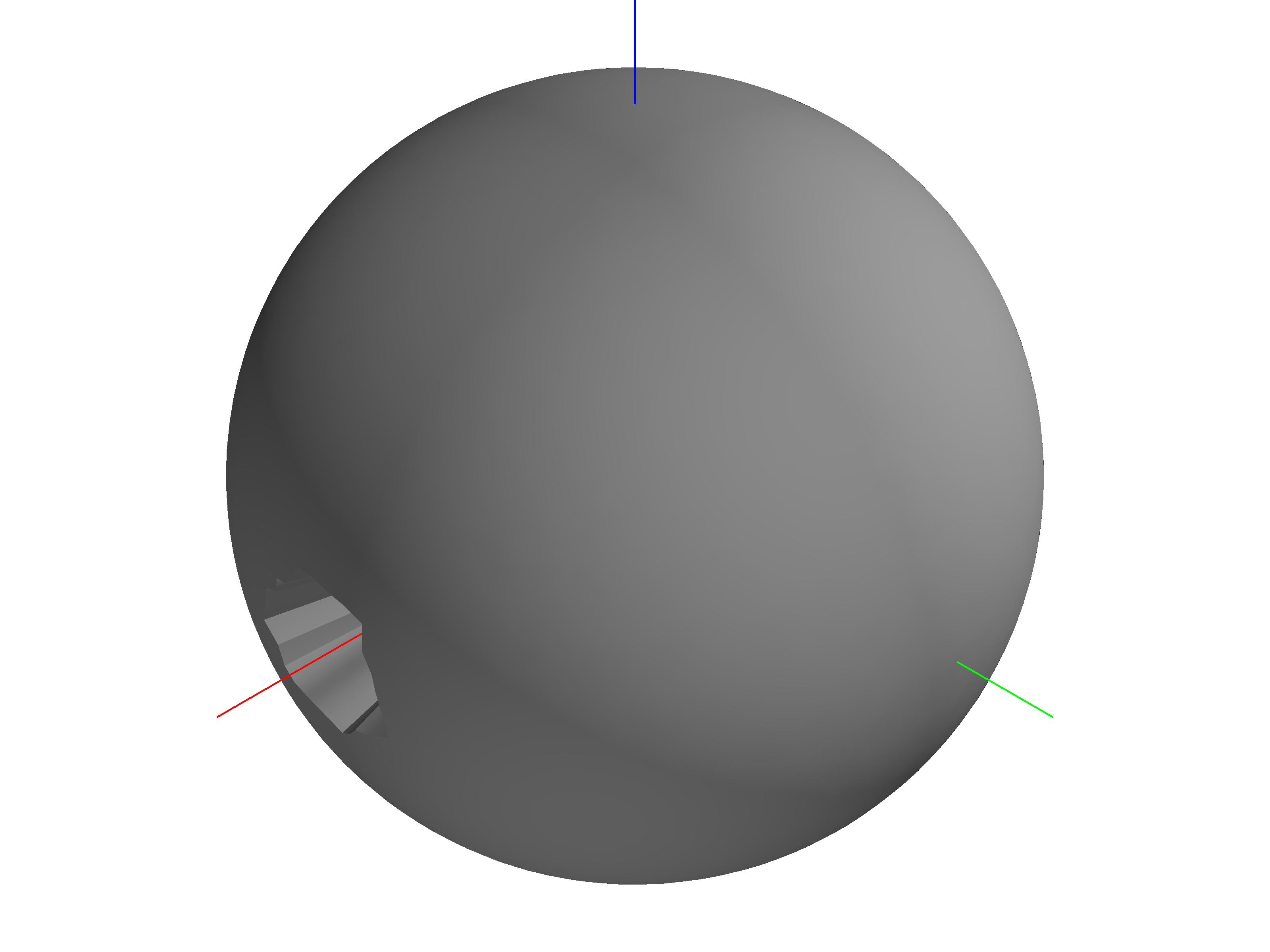}}%
    \subcaptionbox{\SI{25}{\percent} of measurements added\label{fig:golevka_partial_25}}{\includegraphics[trim={20cm 10cm 20cm 10cm},clip,keepaspectratio,width=0.5\textwidth,height=0.25\textheight]{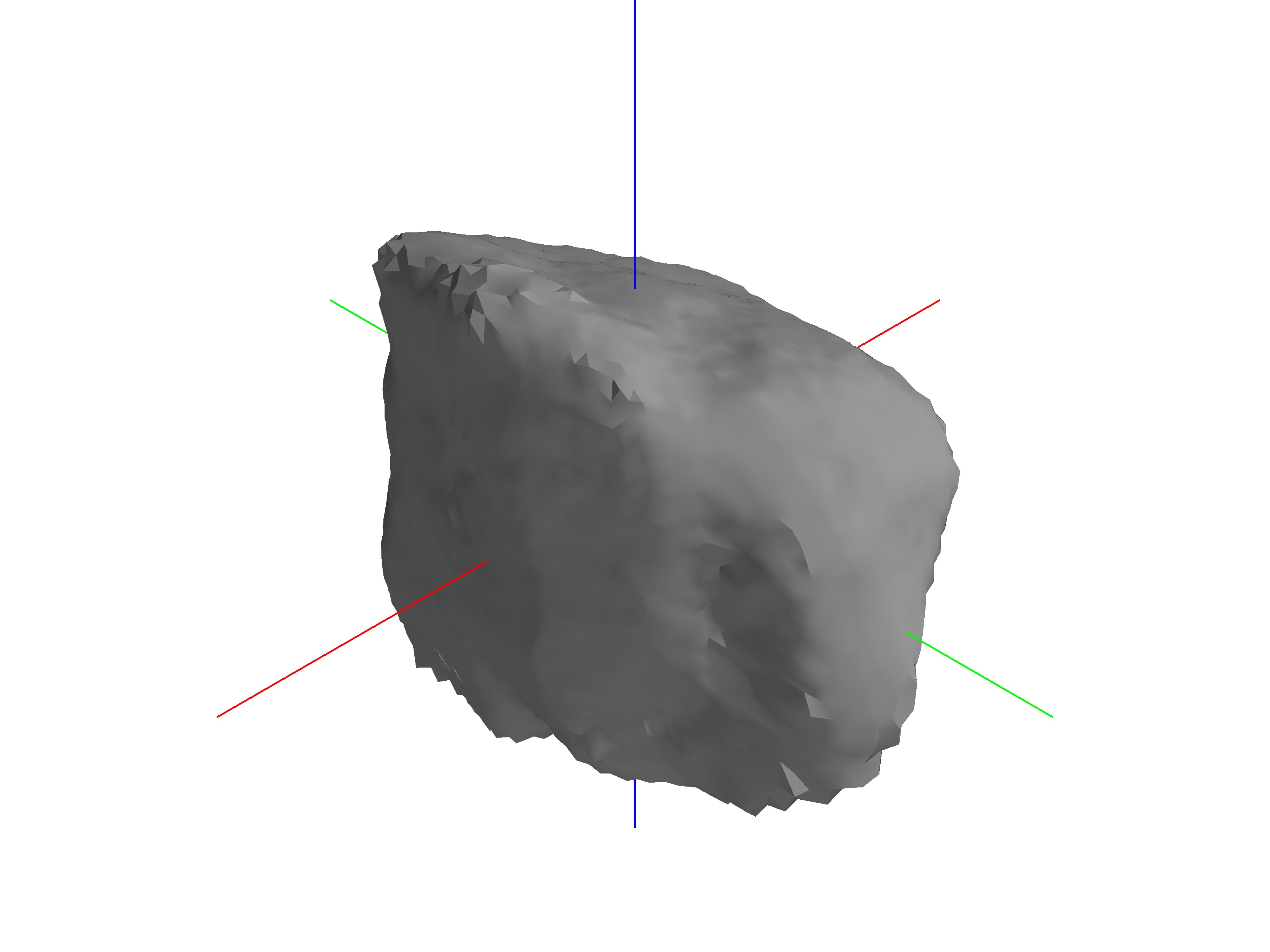}}\\%

    \subcaptionbox{\SI{50}{\percent} of measurements added\label{fig:golevka_partial_50}}{\includegraphics[trim={20cm 10cm 20cm 10cm},clip,keepaspectratio,width=0.5\textwidth,height=0.25\textheight]{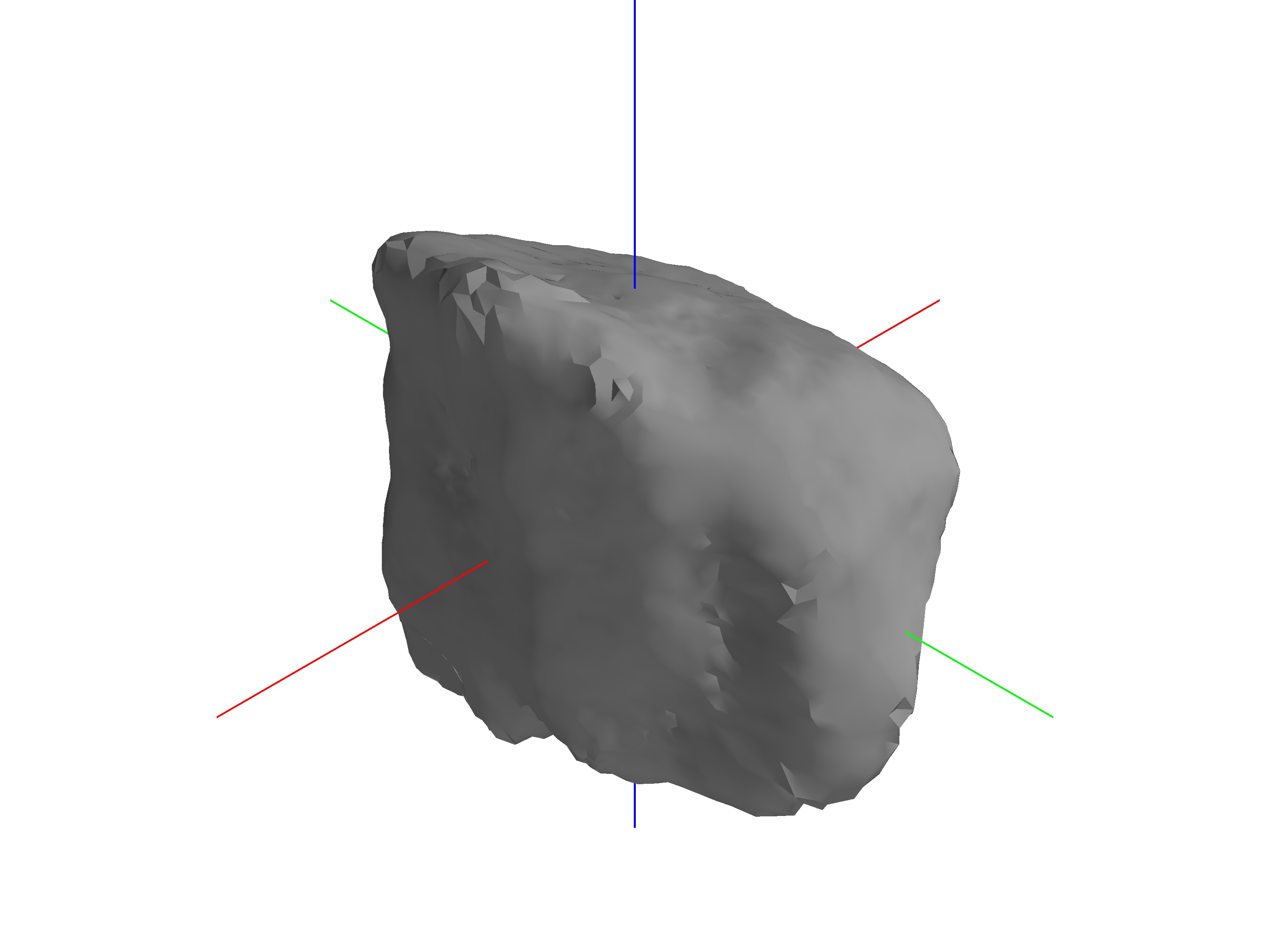}}%
    \subcaptionbox{\SI{75}{\percent} of measurements added\label{fig:golevka_partial_75}}{\includegraphics[trim={20cm 10cm 20cm 10cm},clip,keepaspectratio,width=0.5\textwidth,height=0.25\textheight]{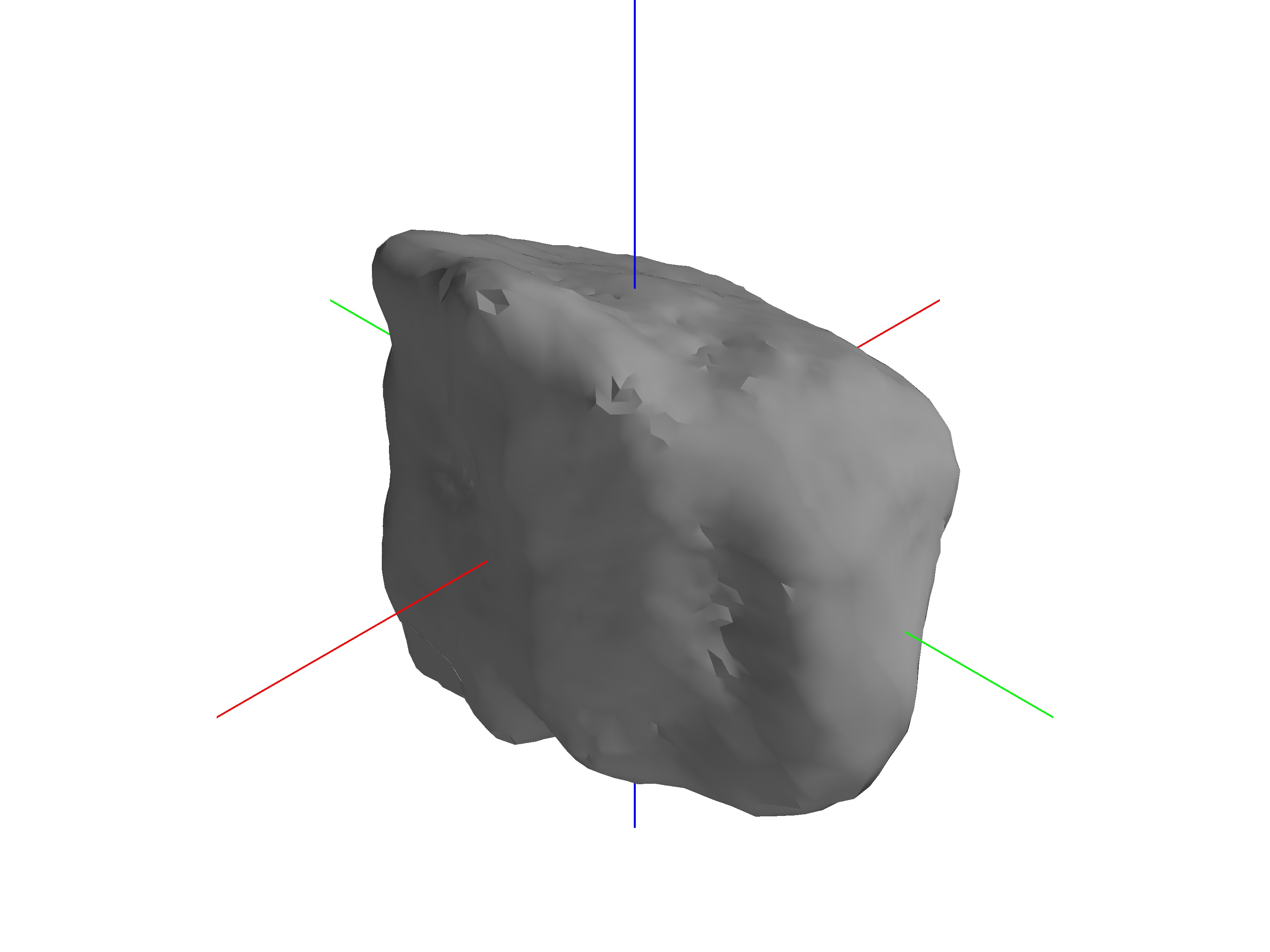}}\\%

    \subcaptionbox{\SI{100}{\percent} of measurements added\label{fig:golevka_partial_100}}{\includegraphics[trim={20cm 10cm 20cm 10cm},clip,keepaspectratio,width=0.5\textwidth,height=0.25\textheight]{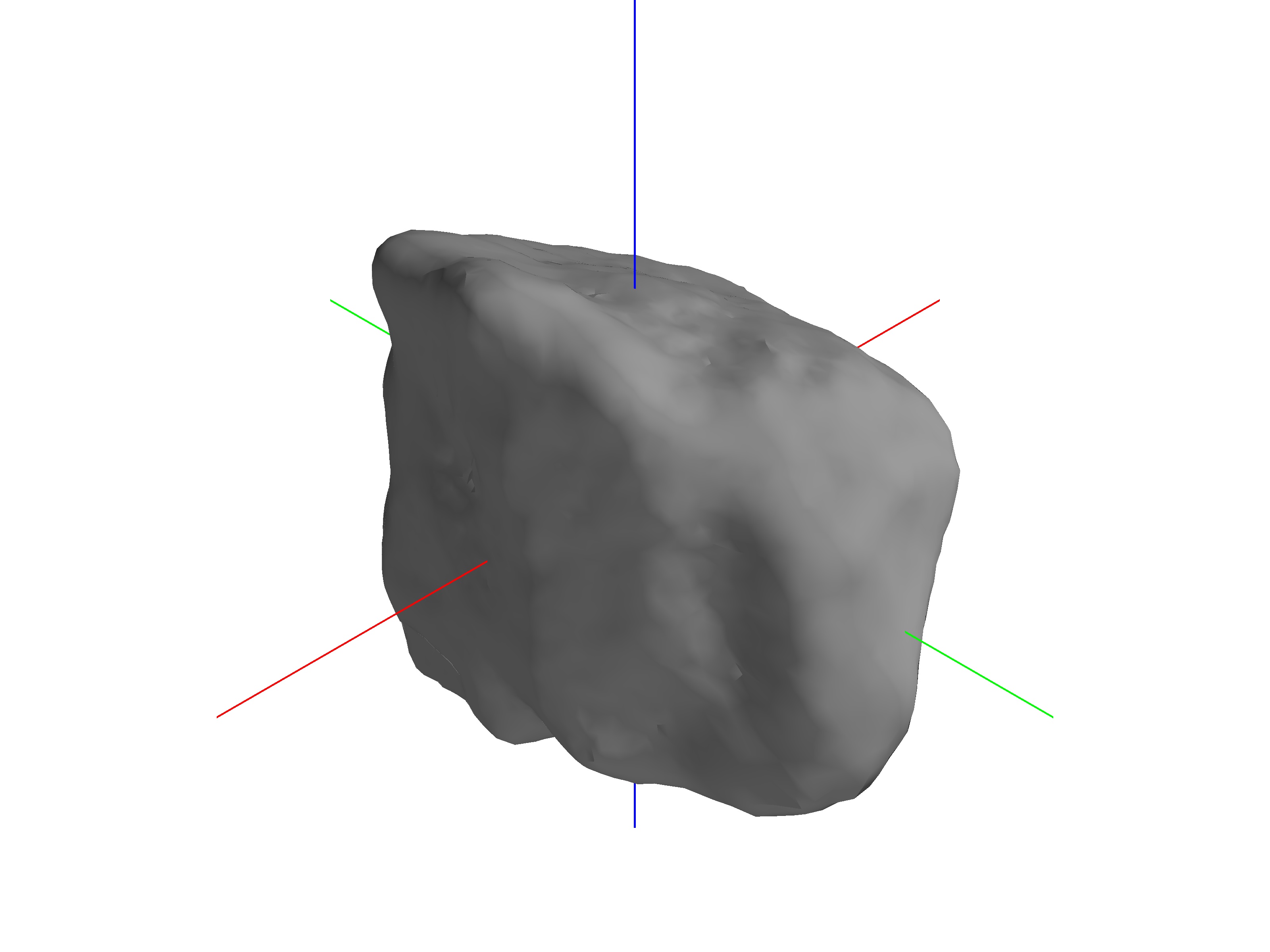}}%
    \subcaptionbox{True Shape Model\label{fig:golevka_truth}}{\includegraphics[trim={20cm 10cm 20cm 10cm},clip,keepaspectratio,width=0.5\textwidth,height=0.25\textheight]{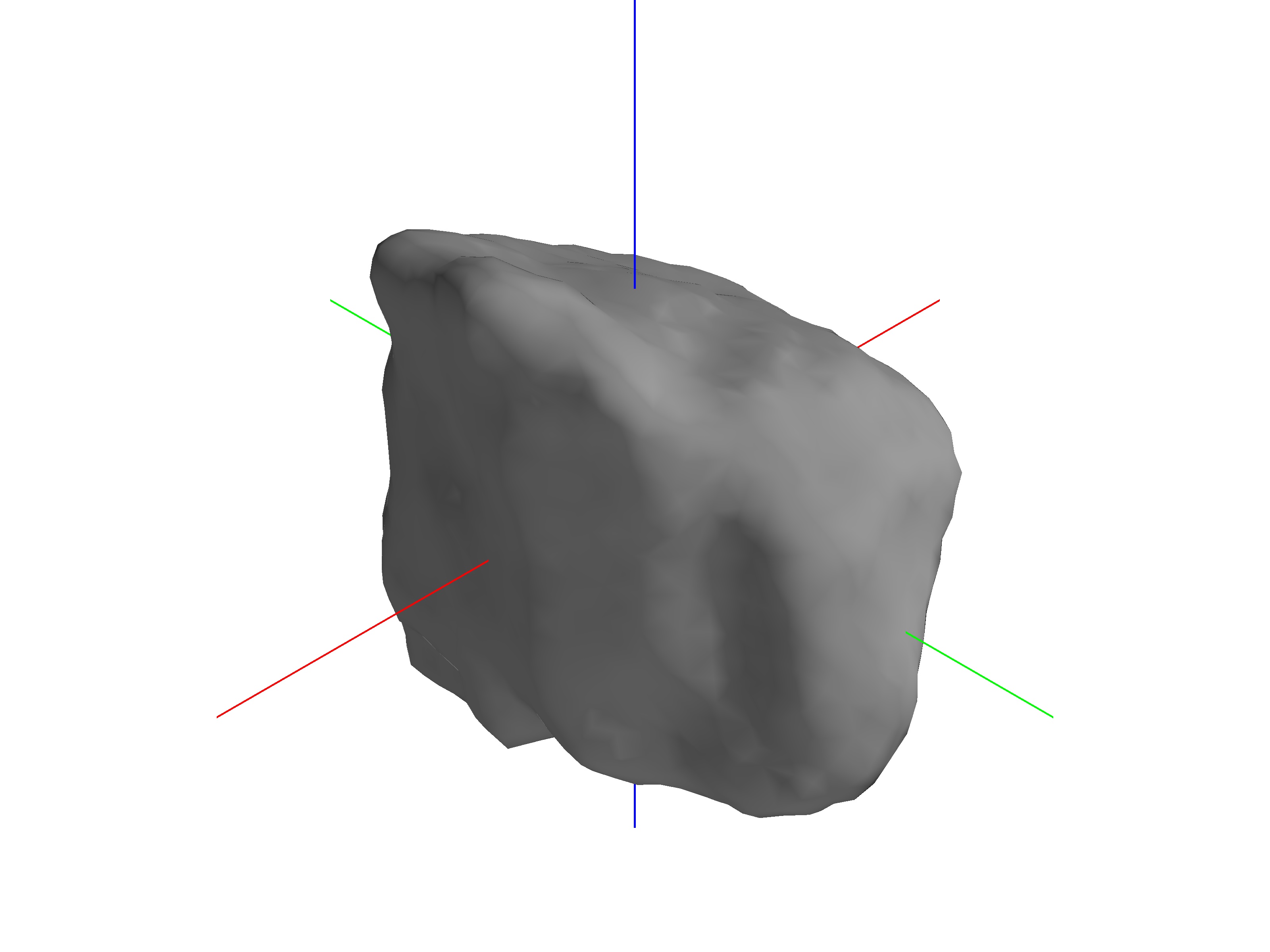}}
    \caption[Asteroid Golevka incremental reconstruction]{Incremental reconstruction of asteroid Golevka~\label{fig:golevka_reconstruction}}
\end{figure}

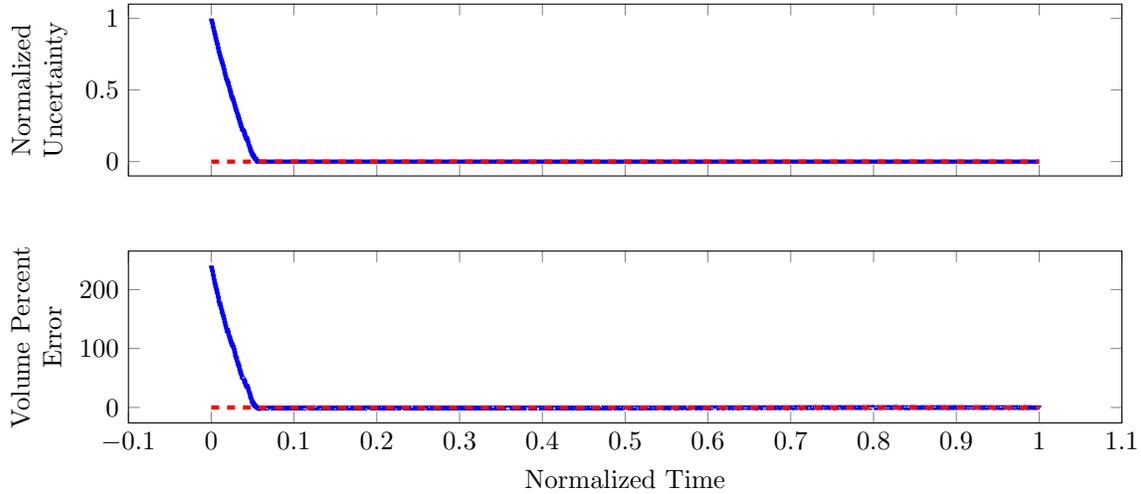
\begin{figure}[htbp]
    \centering
    \tikzsetnextfilename{golevka_metrics}
\begin{tikzpicture}[baseline]
    \begin{groupplot}[
        group style={
            group name={golevka_metrics},
            group size=1 by 2,
            xlabels at=edge bottom,
            ylabels at=edge left,
            xticklabels at=edge bottom,
        },
        xlabel={Normalized Time},
        scale only axis,
        width=0.8\textwidth,
        height=0.1\textheight,
        ylabel style={align=center},
    ]
    \nextgroupplot[ylabel={Normalized\\Uncertainty}]
    \addplot [ultra thick, color=blue, mark=none] table [x=NORMALIZED_TIME, y=NORMALIZED_UNCERTAINTY, col sep=comma] {mesh_update_golevka_uncertainty.csv};
    \addplot [ultra thick,red, mark=none, dashed] coordinates {
        (0.0, 0.0) (1.0, 0.0) 
    };

    \nextgroupplot[ylabel={Volume Percent\\Error}]
    \addplot [ultra thick, blue, mark=none] table [x=NORMALIZED_TIME, y=VOLUME_PERCENT_ERROR, col sep=comma] {mesh_update_golevka_volume.csv};
        \addplot [ultra thick,red, mark=none, dashed] coordinates {
            (0.0, 0.0) (1.0, 0.0) 
        };
\end{groupplot}
\end{tikzpicture}
    \caption{Normalized uncertainty and volume percent error for Golevka\label{fig:golevka_metrics}}
\end{figure}

\paragraph{Asteroid \num{6489} Golevka Reconstruction} 
Next, \num{6489} Golevka is a small angular shaped asteroid of the Apollo group.
Discovered in \num{1995}, Golevka is another potentially hazardous asteroid which passes close to the Earth.
\Cref{fig:golevka_reconstruction} shows the shape reconstruction for asteroid Golevka at several distinct points during the process.
Comparing~\cref{fig:golevka_partial_100,fig:golevka_truth} shows that the final shape closely matches the true radar model.
In~\cref{fig:golevka_metrics} we display the vertex uncertainty and mesh volume as a function of time.

The plots show that the reconstruction achieves an accurate shape estimate with a total volume which closely matches the true volume.
It is interesting to note that the reconstruction of Golevka achieves an accurate shape reconstruction in a much smaller amount of time as compared to Geographos.
This is primarily due to the smaller size of Golevka and the relatively spherical shape of the body in contrast to the highly elliptical shape of Geographos.

\section{Optimal Guidance for Shape Reconstruction}\label{sec:explore_asteroid}

The shape reconstruction algorithm presented in the preceding section does not offer a method to determine which portion of the surface needs to be measured. 
In this section, we present a guidance scheme or a motion planning scheme in order to guide the spacecraft into the most uncertain region. 
This is to reduce the shape uncertainty in an optimal fashion while considering the control cost to change the orbital properties of the spacecraft. 
Then, a nonlinear geometric controller is utilized which allows the spacecraft to maneuver to the optimized location that will update the shape estimate~\cite{kulumani2017b}.

\subsection{Optimal Guidance}

We define a cost associated with each vertex \( \vc{v}_i \) of the shape estimate as
\begin{align}\label{eq:explore_cost}
    \mathcal{J}_i (x, R, R_A) = \alpha_w \mathcal{J}_{w_i} + \alpha_d \mathcal{J}_{d_i}(x_r) + \alpha_c \mathcal{J}_{c_i}(x_r)
\end{align}
where the weighting factors \( \alpha_w, \alpha_d, \alpha_c \in \R \) are chosen such that \( \alpha_w + \alpha_d + \alpha_c = 1 \).
The cost function is defined as a function of the current inertial position, \( x \in \R^3 \), and the attitude, \( R \in \SO \) of the spacecraft.
Furthermore, the knowledge of the small body rotation is required in order to determine the position of the spacecraft in the small body fixed frame, \( x_r = R_A^T x\).

The term \( \mathcal{J}_{w_i} \in \R^1 \) represents the cost associated with the uncertainty of vertex \( i \) as
\begin{align}\label{eq:weight_cost}
    \mathcal{J}_{w_i} &= - \frac{w_i}{w_m}
\end{align}
where \( w_i \) is the uncertainty of vertex \( i \), which is defined as the variance of the radius in the preceding section, and \( w_m \in\R \) is a maximum uncertainty used to scale the values.
The term \( \mathcal{J}_{d_i} \) represents the scaled geodesic distance between the current state of the spacecraft and vertex \( i \),
\begin{align}\label{eq:distance_cost}
    \mathcal{J}_{d_i}(\rpos) &= \frac{1}{\pi} \arctan \parenth{ \frac{\norm{\rpos \times \vc{v}_i}}{\rpos \cdot \vc{v}_i}}.
\end{align}

Finally, a control component is included in the cost function which penalizes vertices that are difficult to reach.
Consider, the current position of the spacecraft in the small body fixed frame as \( \rpos\) and a desired vertex \( \vc{v}_i \) of the shape estimate.
We can define a normal vector to the plane spanned by \( \rpos, \vc{v}_i \) as
\begin{align}\label{eq:normal_to_plane}
    \vc{n}_i = \frac{\rpos \times \vc{v}_i}{\norm{\rpos} \norm{\vc{v}_i}}.
\end{align}
Then a desired trajectory \( x_d(\theta) \) as
\begin{align}\label{eq:spherical_waypoint}
    x_d(\theta) = r_d \exp{\parenth{\theta \hat{\vc{n}_i}} } \frac{\rpos}{\norm{\rpos}},
\end{align}
where \( \theta : \bracket{0, \frac{\rpos \cdot \vc{v}_i}{\norm{\rpos}\norm{\vc{v}_i}}} \to \R\) parameterizes the desired trajectory.
\Cref{eq:spherical_waypoint} simply describes a portion of a great circle trajectory between the current state, \( \rpos \), and the desired vertex \( \vc{v}_i \)~\cite{chen2016}, with a desired radius $r_d$.
The radius of the spacecraft, \( r_d \in \R \), can be chosen based on sensor characteristics of safety concerns.
For example, \( r_d \) can be chosen as the distance of the Biroullin sphere with an additional safety margin to mitigate any surface collision~\cite{scheeres2012a}.

We assume $\theta$ varies linearly with respect to $t$, and substitute the desired trajectory into \Cref{eq:v_dot} to obtain the control force required to follow the desired trajectory in the absence of any tracking error:
\begin{align}\label{eq:tracking_control_cost}
    u_f(\theta) = -F_{ext}(x_d(\theta)), 
\end{align}
which is computed by the polyhedron potential model given in~\cref{eq:attraction}.
The control cost is then defined as the integral over the desired trajectory~\cref{eq:spherical_waypoint} between the current state and the desired vertex as
\begin{align}\label{eq:control_cost}
    \mathcal{J}_{c_i}(\rpos) = \frac{1}{u_m} \int_{\theta_0}^{\theta_f} u_f(\theta)^T R u_f(\theta) d\theta,
\end{align}
where \( u_m \) is used to normalize and scale \( J_{c_i} \).
\Cref{eq:control_cost} is numerically integrated over the trajectory \( x_d(t) \) and used to penalize vertices which have a larger cost.

The vertex which minimizes~\cref{eq:explore_cost} 
\begin{align*}
    \vc{v}_{min} = \argmin_{i} \{\mathcal{J}_i(x, R, R_A)\},
\end{align*}
is determined.
This vertex is then used to determine the desired trajectory of the spacecraft in order to collect a measurement as in~\Cref{eq:spherical_waypoint}.

Next, the desired attitude command, \( R_d\), is chosen such that the spacecraft camera axis, \( \vc{b}_1 \), is directed along the nadir towards the asteroid.
It is sufficient to define two orthogonal vectors to uniquely determine the attitude of the spacecraft.
The \( \vc{b}_{3d} \) vector is chosen to lie in the plane spanned by \(\vc{b}_{1d} \) and \( \vc{e}_3 = \vc{f}_3 \).
The desired attitude command is defined as
\begin{align}
    \vc{b}_{1d} &= - \frac{\vc{x}}{\norm{\vc{x}}} , \\
    \vc{b}_{3d} &= \frac{\vc{f}_3 - \parenth{\vc{f}_3 \cdot \vc{b}_{1d}} \vc{b}_{1d}}{\norm{\vc{f}_3 - \parenth{\vc{f}_3 \cdot \vc{b}_{1d}} \vc{b}_{1d}}}, \\
    \vc{b}_{2d} &= \vc{b}_{3d} \times \vc{b}_{1d} , \\
    R_d &= \begin{bmatrix} \vc{b}_{1d} & \vc{b}_{2d} & \vc{b}_{3d} \end{bmatrix} .
\end{align}
This form of \( R_d \) will direct the \( b_1 \) axis towards the small body, and can be modified for a different camera orientations~\cite{kulumani2017a}.

\subsection{Geometric Tracking Control}

We utilize a geometric tracking control system to follow the desired trajectory for the position and the attitude defined above. 
The spacecraft is assumed to be fully controllable and as a result the translational and rotational dynamics can be decoupled.
We first present an geometric attitude controller used to track the desired attitude command \( R_d \).
This is followed by a translational controller to track the desired position \( x_d \).

First, an attitude error function \(A : \SO \times \SO  \to \R \), an attitude error vector \( e_R : \SO \times \SO \to \R \), and an angular velocity error \( e_\Omega : \SO \times \R^3 \times \SO \times \R^3 \) are defined as
\begin{subequations}\label{eq:attitude_error_function}
\begin{align}
    A(R, R_d) &= \frac{1}{2}  \tr{G\parenth{I - R_d^T R}}, \label{eq:A} \\
    e_R &= \frac{1}{2} G \parenth{R_d^T R - R^T R_d^\vee}, \label{eq:eRA}\\
    e_\Omega &= \Omega - R^T R_d \Omega_d \label{eq:eWA}.
\end{align}
\end{subequations}
Then the following properties hold:
\begin{enumerate}
    \item \label{item:prop_A_psd} \( A \) is  positive definite about \( R = R_d \) on \( \SO \).
    \item \label{item:prop_eRA} The variation of \( A \) with respect to a variation of \( \delta R = R \hat{\eta} \) for \( \eta \in \R^3 \) is given by
	\begin{align}\label{eq:dirDiff_A}
		\dirDiff{A}{R} &= \eta \cdot e_{R} ,
	\end{align}
	where the notation \( \dirDiff{A}{R} \) represents the directional derivative of $A$ with respect to $R$ along the direction $\delta R$.
\item \label{item:prop_A_critical_points} The critical points of \( A \), where \( e_R = 0 \) are \( \braces{R_d} \cup \braces{R_d \exp(\pi \hat s) } \) for \( s \in \braces{e_1, e_2, e_3}\)
    \item \label{item:prop_A_quadratic} \( \Psi \) is a locally quadratic function, which means there exist constants \( 0 < n_1 \leq n_2 \) such that
    \begin{align}\label{eq:A_bound}
        n_1 \norm{e_R}^2 \leq \Psi(R) \leq n_2 \norm{e_R}^2 ,
    \end{align}
    for $0<\psi < h_1 $ where the constants \( n_1 = \frac{h_1}{h_2 + h_3} \) and \( n_2 = \frac{h_1 h_4}{h_5 \parenth{h1 - \psi} }\) for
	\begin{align*}
		h_1 &= \min\braces{g_1 + g_2, g_2 + g_3 , g_3 + g_1} ,\\
		h_2 &= \max\braces{\parenth{g_1 -g_2}^2,\parenth{g_2 -g_3}^2 , \parenth{g_3 -g_1}^2} ,\\
		h_3 &= \max\braces{\parenth{g_1 + g_2}^2, \parenth{g_2 + g_3}^2 , \parenth{g_3 + g_1}^2}, \\		
        h_4 &= \max\braces{g_1 + g_, g_2 +g_3, g_3 + g_1}, \\
        h_5 &= \max\braces{\parenth{g_1 + g_2}^2, \parenth{g_2 + g_3}^2, \parenth{g_3 + g_1}^2}.
	\end{align*}
\end{enumerate}
The proof of these properties is available in Reference~\cite{kulumani2017a}.

With the appropriate attitude configuration error we now present the error dynamics, which are used in the subsequent development of the nonlinear control system.
The attitude error dynamics for \( A(R, R_d), e_\Omega \) satisfy
\begin{gather}
    \diff{}{t} \parenth{A(R)} = e_{R_A} \cdot e_\Omega , \label{eq:A_dot} \\
    \diff{}{t} \parenth{e_{R_A}} = E(R, R_d) e_\Omega , \label{eq:eRA_dot} \\
    \diff{}{t} \parenth{e_\Omega} = J^{-1} \parenth{-\hat{\Omega} J \Omega + u_m + M_{ext}} , \label{eq:eW_dot}
\end{gather}
where the matrix \(E(R,R_d)\in \R^{3\times3} \) is given by
\begin{align}
    E(R,R_d) = &\frac{1}{2} \parenth{\tr{R^T R_d G}I - R^T R_d G} , \label{eq:E}
\end{align}
The proof is available in~\cite{kulumani2017a}.

Using these properties we can now define an attitude control input to allow the vehicle to track attitude errors.
Given a desired attitude command \( \parenth{R_d, \Omega_d = 0} \) and positive constants \( k_R, k_\Omega \in \R \) we define a control input \( u \in \R^3 \) as follows
\begin{gather}
    u = -k_R e_R - k_\Omega e_\Omega + \Omega \times J \Omega - M_{ext}. \label{eqn:nodist_control}
\end{gather}
Then the zero equilibrium of the attitude error is asymptotically stable.
The proof is show in Reference~\cite{kulumani2017a}.

The required translational control are defined in terms of the position and velocity tracking errors.
The tracking error vectors are easier to define as they evolve on a Euclidean space rather than a nonlinear manifold and are given by
\begin{subequations}\label{eq:translation_error_variables}
\begin{align}
    e_x &= x - x_d ,\\
    e_v &= v - \dot{x}_d.
\end{align}
\end{subequations}
The error dynamics are given by
\begin{subequations}\label{eq:translation_error_dynamics}
\begin{align}
    \dot{e}_x &= v - \dot{x}_d, \\
    \dot{e}_v &= \dot{v} - \ddot{x}_d .
\end{align}
\end{subequations}
A control input, \( u_f \), is derived to ensure asymptotic trajectory tracking for the translational dynamics.
    Given a desired trajectory \( \vc{x}_d \) and positive constants \( k_x, k_v, c \in \R \) we define a control input \( \vc{u}_f \in \R \) as follows:
    \begin{align}\label{eq:translational_control}
        \vc{u}_f = -k_x e_x - k_v e_v - F_{ext} + m \ddot{x}_d .
    \end{align}
    If \( c \) is chosen such that
    \begin{align}\label{eq:translational_control_gain_bound}
        0 < c < \min \braces{\sqrt{k_x m}, \frac{4 k_x k_v m}{ k_v \parenth{4 k_x m - k_v} }   }
    \end{align}
    then the zero equilibrium of the error vectors~\cref{eq:translation_error_variables} is stable in the sense of Lyapunov.
    Furthermore, \( e_x, e_v \to 0 \) as \( t \to \infty\).
    The proof is shown in~\cite{kulumani2017a}.

The control gains are chosen based on the desired closed-loop system response. 
A variety of techniques are available to choose these gains, but a simple linear analysis offers a straightforward and systematic approach to choosing suitable values. 
We use the control inputs defined in~\cref{eq:translational_control,eqn:nodist_control} and substitute them into the dynamical equations of motion in~\cref{eq:x_dot,eq:v_dot,eq:R_dot,eq:omega_dot}.

\subsection{Numerical Examples}

We utilize radar shape models of asteroid \num{4769} Castalia and (\num{52760}) \num{1998} \(\text{ML}_{14}\)~\cite{neese2004}.
The examples demonstrate the full dynamical simulation of a rigid spacecraft with an autonomous closed loop control scheme to both reconstruct the asteroid shape.
The proposed optimal guidance scheme is executed in the outer-loop to generate the desired trajectory, which is followed by the nonlinear control scheme in the inner-loop. 
More specifically, the nonlinear controllers described previously are used to control both the translational and rotational states of the vehicle, where the control inputs are computed using the current shape estimate of the asteroid. 
As measurements are collected, the spacecraft autonomously updates its shape estimate and uses this estimate to compute the control inputs and desired future states.
As such, both the shape model and the gravitational potential available to the controller will be gradually refined. 
However, throughout the simulation, the actual spacecraft dynamics are computed with the gravity computed by the full shape model unknown to the controller.
These demonstrate the ability of a spacecraft to autonomously explore and maneuver around an initially poorly modeled asteroid while incrementally updating the shape model.

In this numerical example we have assumed the perfect knowledge of the spacecraft state. 
This is a strong assumption that would not exist in any realistic situation. 
However, the measurement model defined in the previous section may be extended to consider the uncertainty in both the state \( x \) and the range measurement \( d \) rather than combining them into a single random variable \( p \).
In addition, the dynamics defined by~\crefrange{eq:x_dot}{eq:omega_dot} do not consider any non-gravitational accelerations such as solar radiation pressure on the vehicle.
In general, these types of external forces will primarily serve to increase state uncertainty and cause large errors in the closed loop control system.
Adaptive control techniques have been demonstrated in the past to address both fixed and time varying disturbances of this kind~\cite{kulumani2017a}.
However, these topics are beyond the scope of this paper focusing on optimal guidance for Bayesian shape reconstruction, and relegated to future works. 

The asteroids are assumed to constantly rotate about the \( \vc{f}_3 = \vc{e}_3\) axis according to the parameters given in~\cref{tab:dynamic_asteroids}.
Furthermore, the state of the asteroid, namely the rotation matrix \( R_A \), is assumed to be known based on ground measurements or previous data.
\begin{table}[htbp]
    \centering
    \begin{tabular}{lcc}
        \toprule
        Property & \num{4769} Castalia & (\num{52760}) \num{1998} \(\text{ML}_{14}\) \\
        \midrule
        Semi-major axes(\si{\kilo\meter}) & \( 0.8065 \times 0.4905 \times 0.413 \) & \( 1.1 \times 1.1 \times 1.1 \) \\
        Rotational Period (\si{\hour}) & \num{4.095} & \num{14.98} \\
        Density (\si{\gram\per\centi\meter^3}) & \num{2.1} & \num{2.1} \\
        Vertices & \num{2048}  & \num{8192} \\
        Faces & \num{4092} & \num{16320} \\
        \bottomrule
    \end{tabular}
    \caption{Asteroid properties for dynamical exploration~\label{tab:dynamic_asteroids}}
\end{table}
At the beginning of the simulation the spacecraft is assumed to lie on the inertial \( e_1 \) axis, i.e.\ \( x_0 = [ x_0,\, 0,\,  0 ]\, \si{\kilo\meter} \).
In addition, at the initial state the spacecraft is orientated such that the \( b_1 \) axis is aligned with the inertial \( e_2 \) axis.
In other words the initial orientation is given by \( R_0 = \exp(\frac{\pi}{2} \hat{e}_3)\).
The shape reconstruction phase of the simulation is performed over \SI{15000}{\second}, over which time the spacecraft will take LIDAR measurements of the surface at \SI{1}{\hertz}.
Once the total uncertainty has been reduced sufficiently the spacecraft maneuvers to a ``home'' position aligned with the \( f_1 \) axis of the asteroid.

\paragraph{Asteroid 52760 Reconstruction}

\begin{figure}[htbp]
    \centering
    \subcaptionbox{Initial Shape Estimate\label{fig:52760_partial_weights_0}}{\includegraphics[trim={30cm 15cm 30cm 15cm},clip,height=0.25\textheight,width=0.5\textwidth,keepaspectratio]{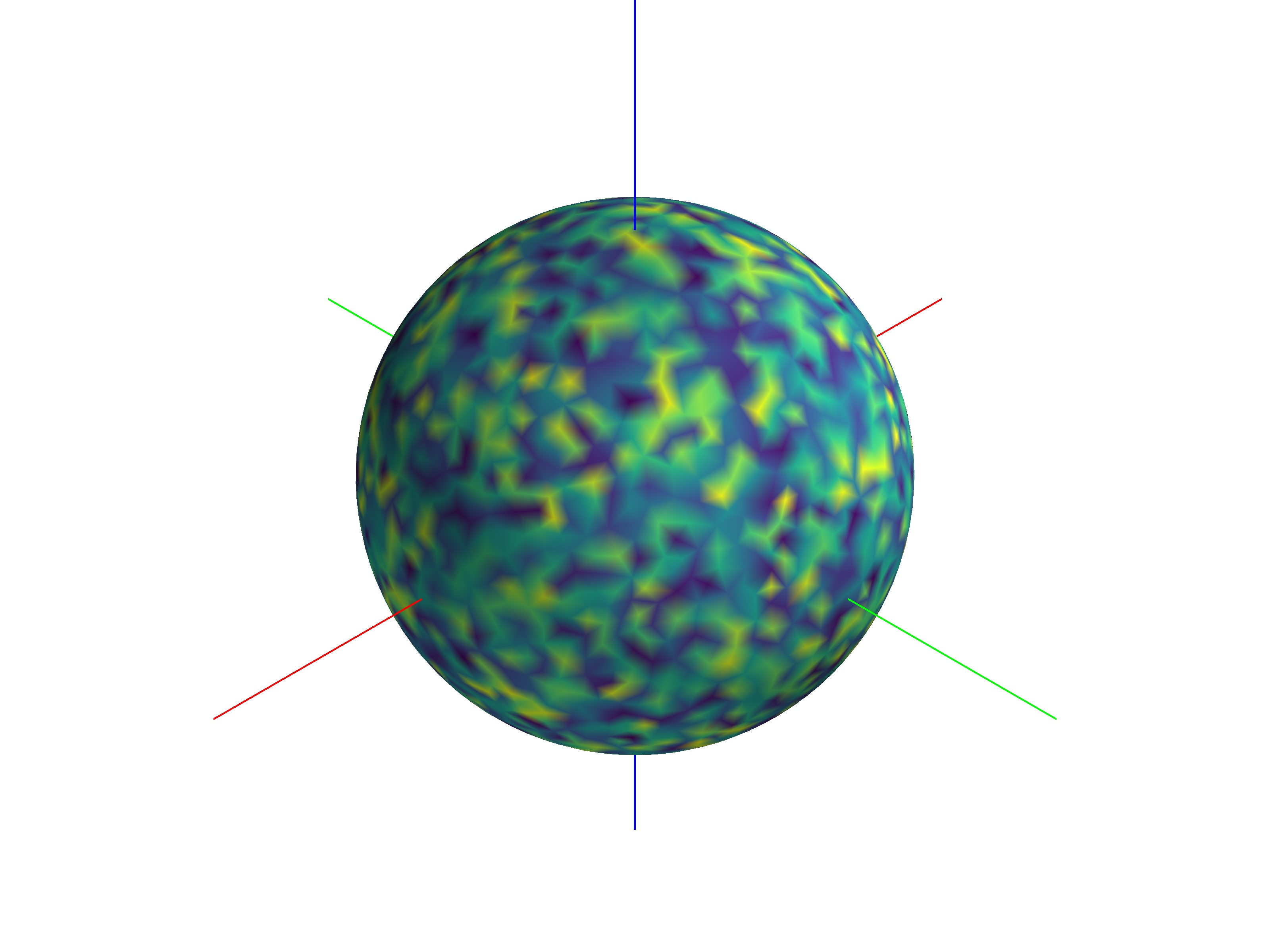}}%
    \subcaptionbox{\SI{25}{\percent} of measurements\label{fig:52760_partial_weights_25}}{\includegraphics[trim={30cm 15cm 30cm 15cm},clip,height=0.25\textheight,width=0.5\textwidth,keepaspectratio]{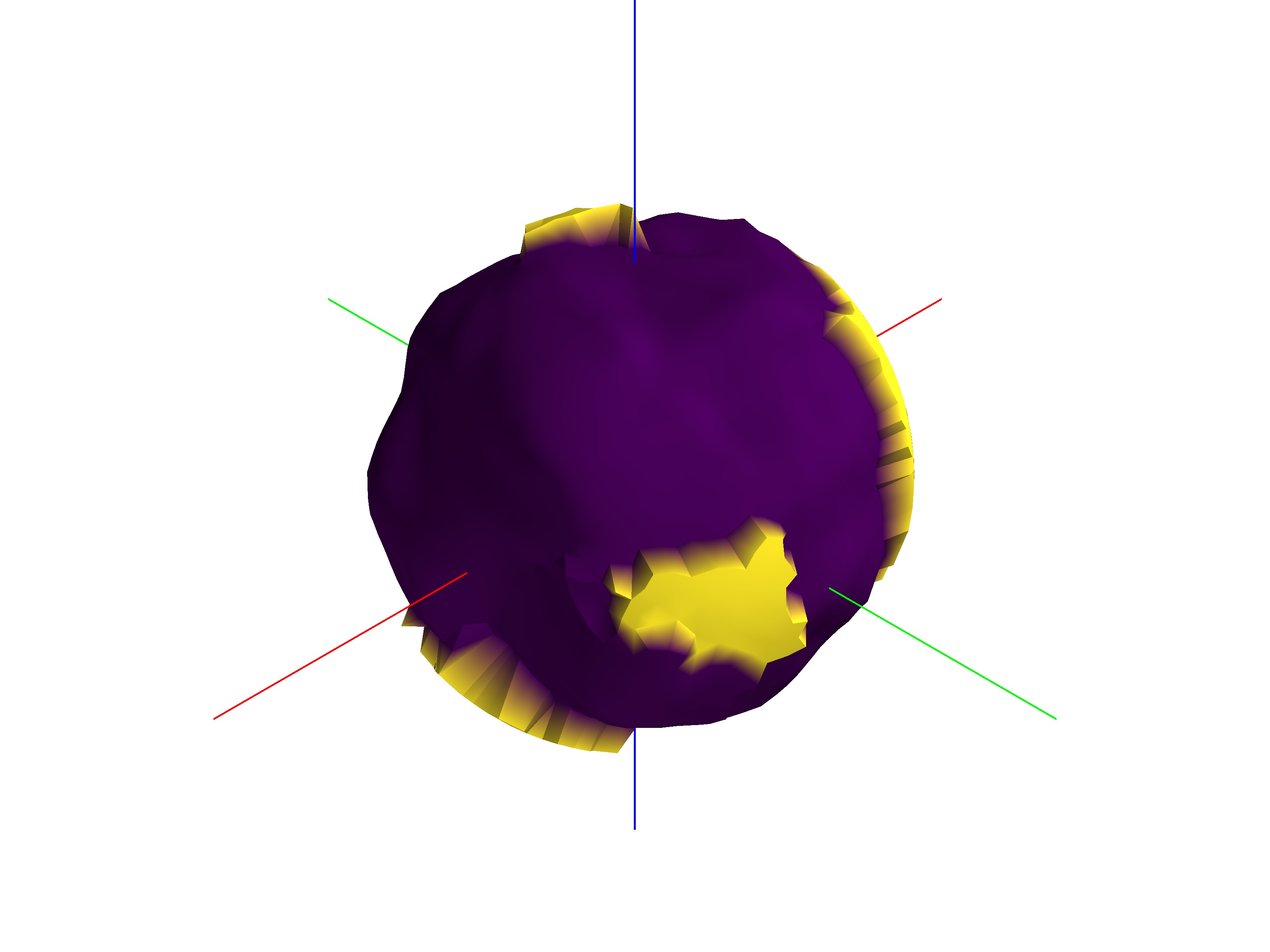}}%

    \subcaptionbox{\SI{50}{\percent} of measurements\label{fig:52760_partial_weights_50}}{\includegraphics[trim={30cm 15cm 30cm 15cm},clip,height=0.25\textheight,width=0.5\textwidth,keepaspectratio]{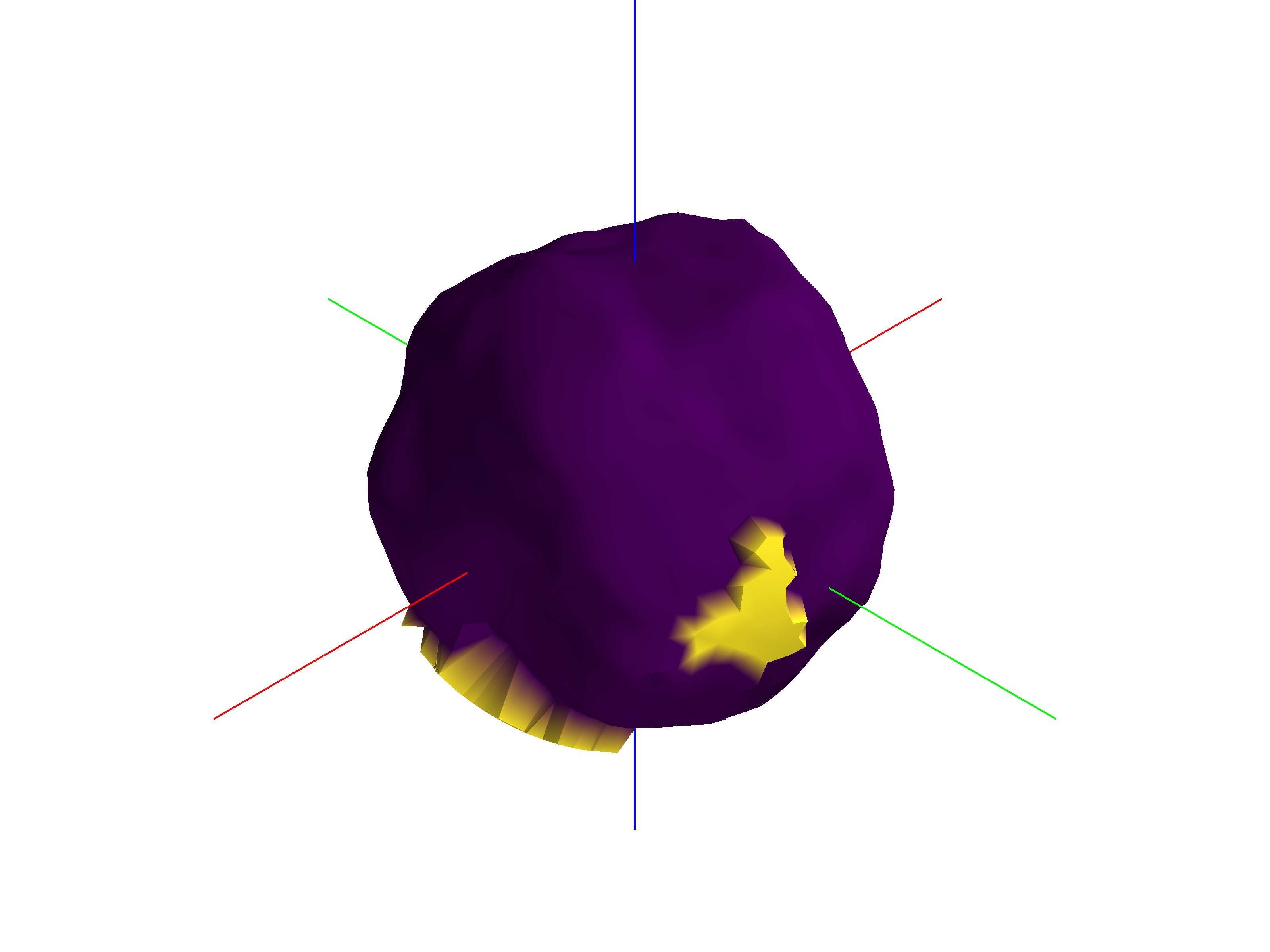}}
    \subcaptionbox{\SI{75}{\percent} of measurements\label{fig:52760_partial_weights_75}}{\includegraphics[trim={30cm 15cm 30cm 15cm},clip,height=0.25\textheight,width=0.5\textwidth,keepaspectratio]{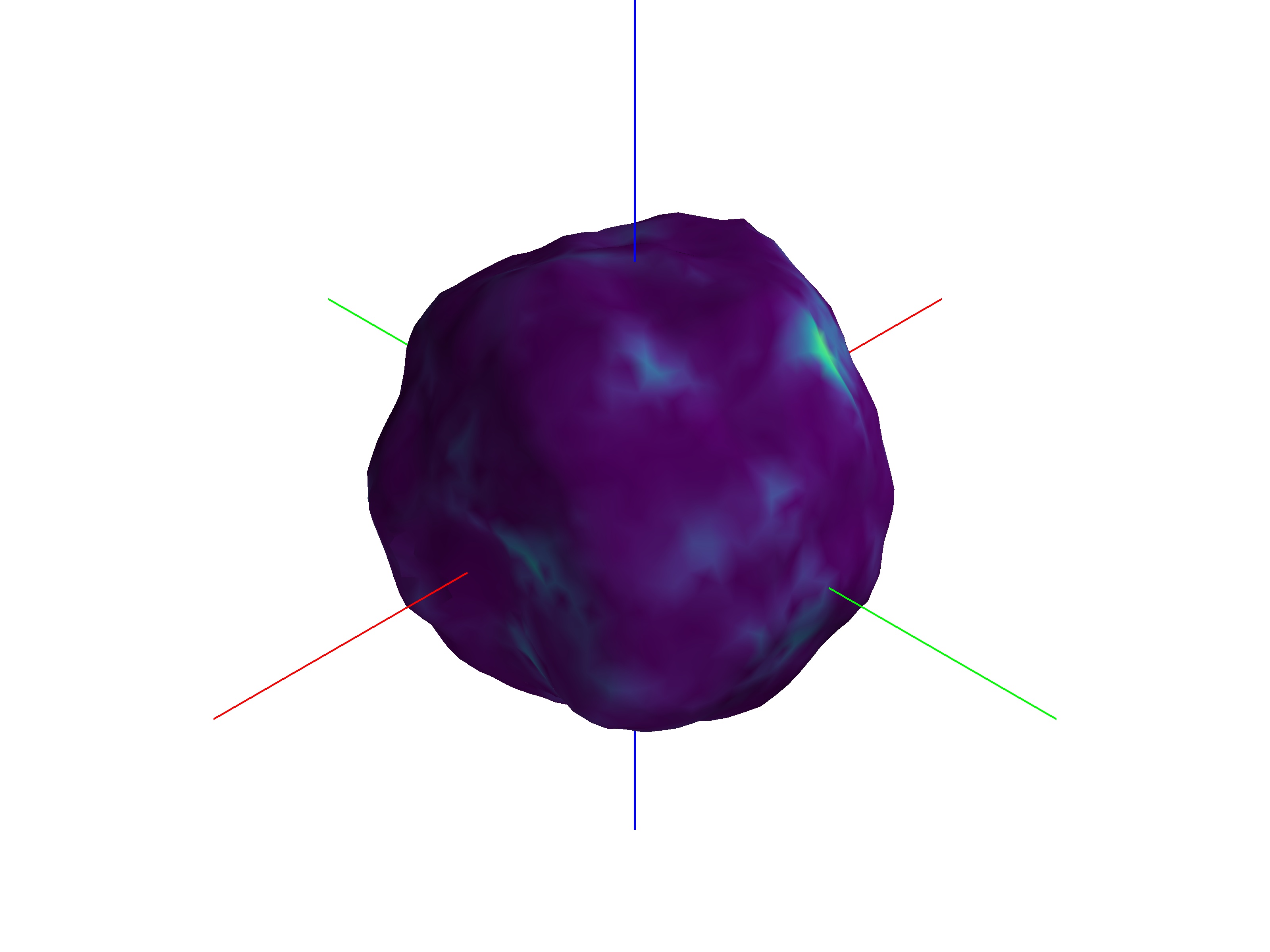}}%

    \subcaptionbox{\SI{100}{\percent} of measurements\label{fig:52760_partial_weights_100}}{\includegraphics[trim={30cm 15cm 30cm 15cm},clip,height=0.25\textheight,width=0.5\textwidth,keepaspectratio]{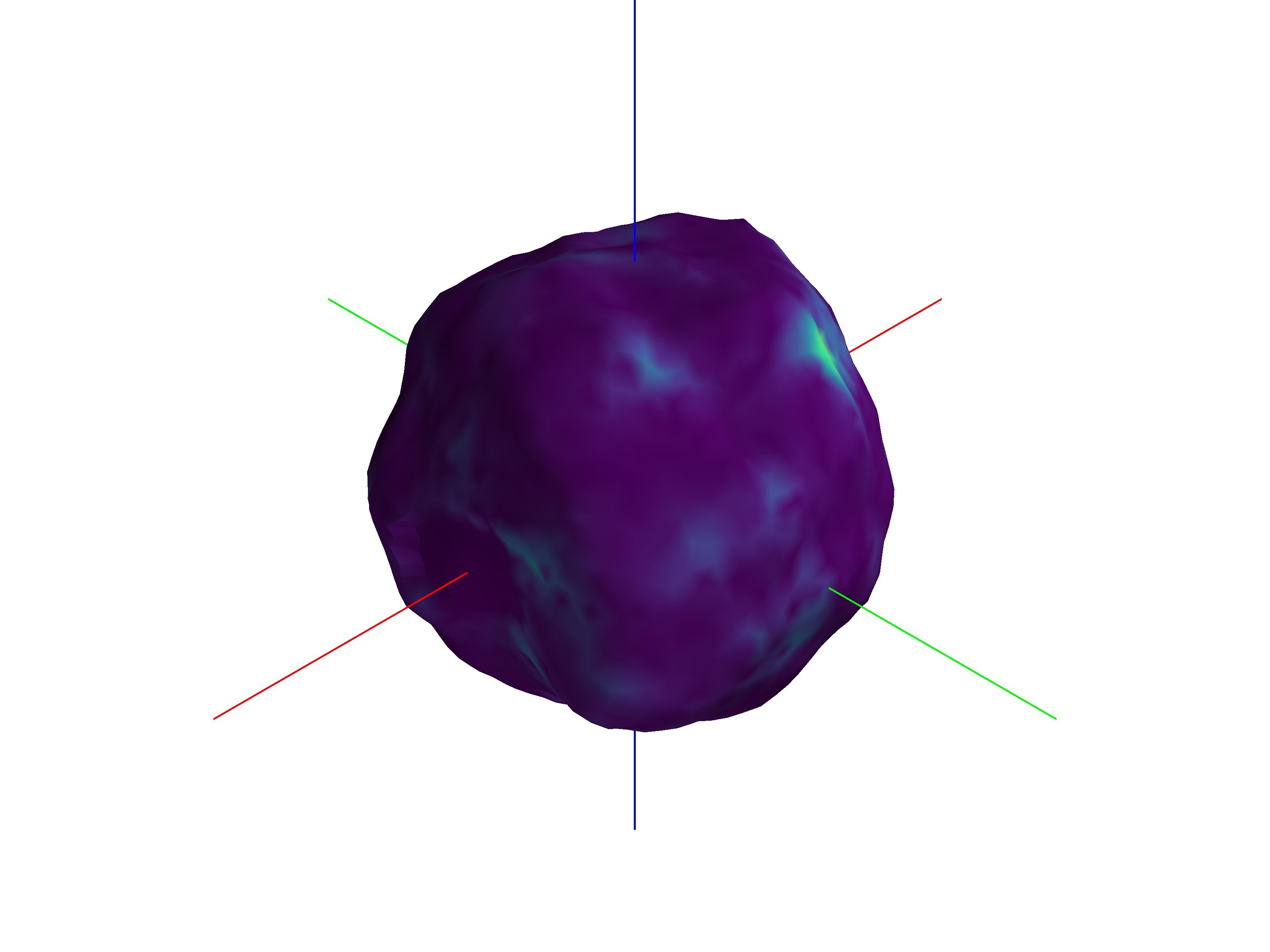}}%
    \subcaptionbox{True Shape Model\label{fig:52760_weights_truth}}{\includegraphics[trim={15cm 0cm 15cm 5cm},clip,height=0.25\textheight,width=0.5\textwidth,keepaspectratio]{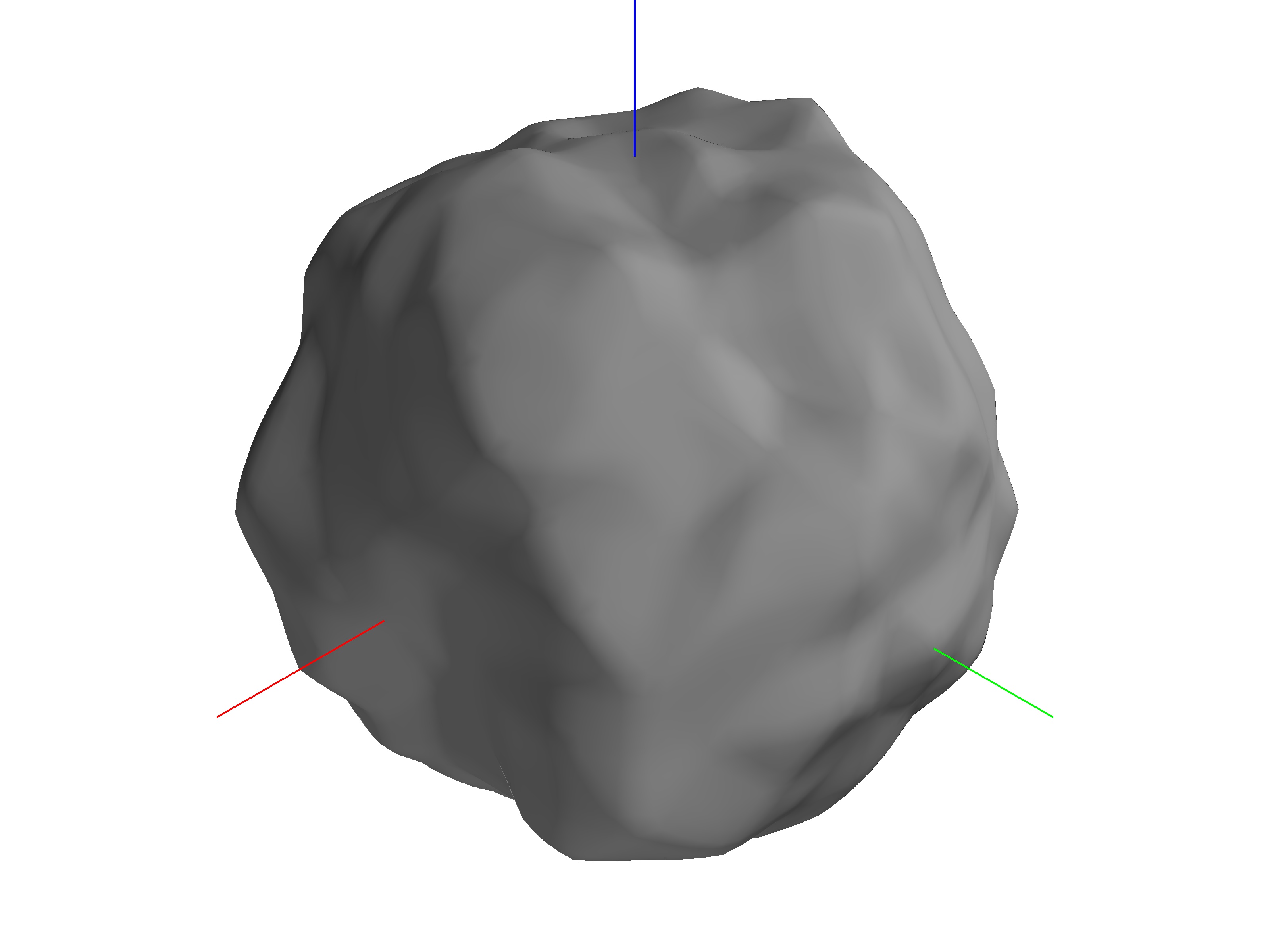}}
    \caption[Asteroid 52760 shape reconstruction with uncertainty]{Incremental reconstruction of asteroid 52760. The images colored according to the shape uncertainty. Areas of high uncertainty are in yellow while ares of low uncertainty are in purple.
    The guidance scheme continually selects regions of high uncertainty until the total shape uncertainty is sufficiently small.~\label{fig:52760_weights_reconstruction}}
\end{figure}

Asteroid (\num{52760}) \num{1998} \(\text{ML}_{14}\) was discovered in \num{1998} and is near Earth asteroid of the Apollo group and classified as a potentially hazardous body.
The asteroid is roughly spherical with a mean radius of approximately \SI{1}{\kilo\meter}.
The initial shape estimate is assumed to be spherical with approximately the same number of vertices as the truth model.
\Cref{fig:52760_weights_reconstruction} show the shape reconstruction at several discrete points during the simulation.
Due to the roughly spherical shape of the asteroid large portions of the surface are quickly modified to match the measurements.
In addition~\cref{fig:52760_weights_reconstruction} displays the vertex uncertainty \( w_i \) as a colormap on the surface. 
Areas of high uncertainty are denoted in yellow while areas of low uncertainty are in purple/blue.

\paragraph{Asteroid 4769 Castalia Reconstruction}

\begin{figure}[htbp]
    \centering
    \subcaptionbox{Initial Shape Estimate\label{fig:castalia_partial_weights_0}}{\includegraphics[trim={15cm 10cm 15cm 10cm},clip,height=0.5\textheight,width=0.3\textwidth,keepaspectratio]{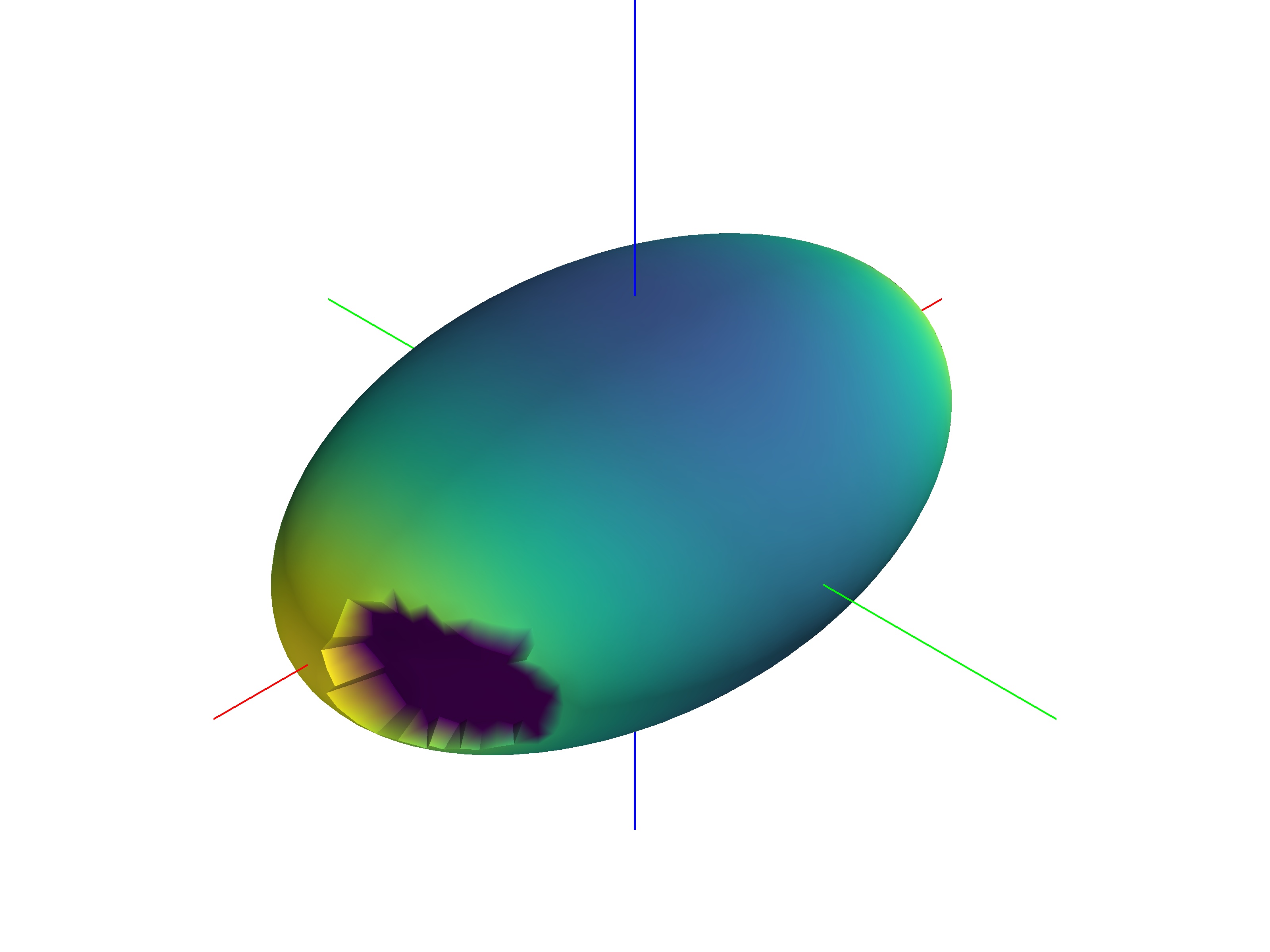}}%
    \subcaptionbox{\SI{25}{\percent} of measurements added\label{fig:castalia_partial_weights_25}}{\includegraphics[trim={15cm 10cm 15cm 10cm},clip,height=0.5\textheight,width=0.3\textwidth,keepaspectratio]{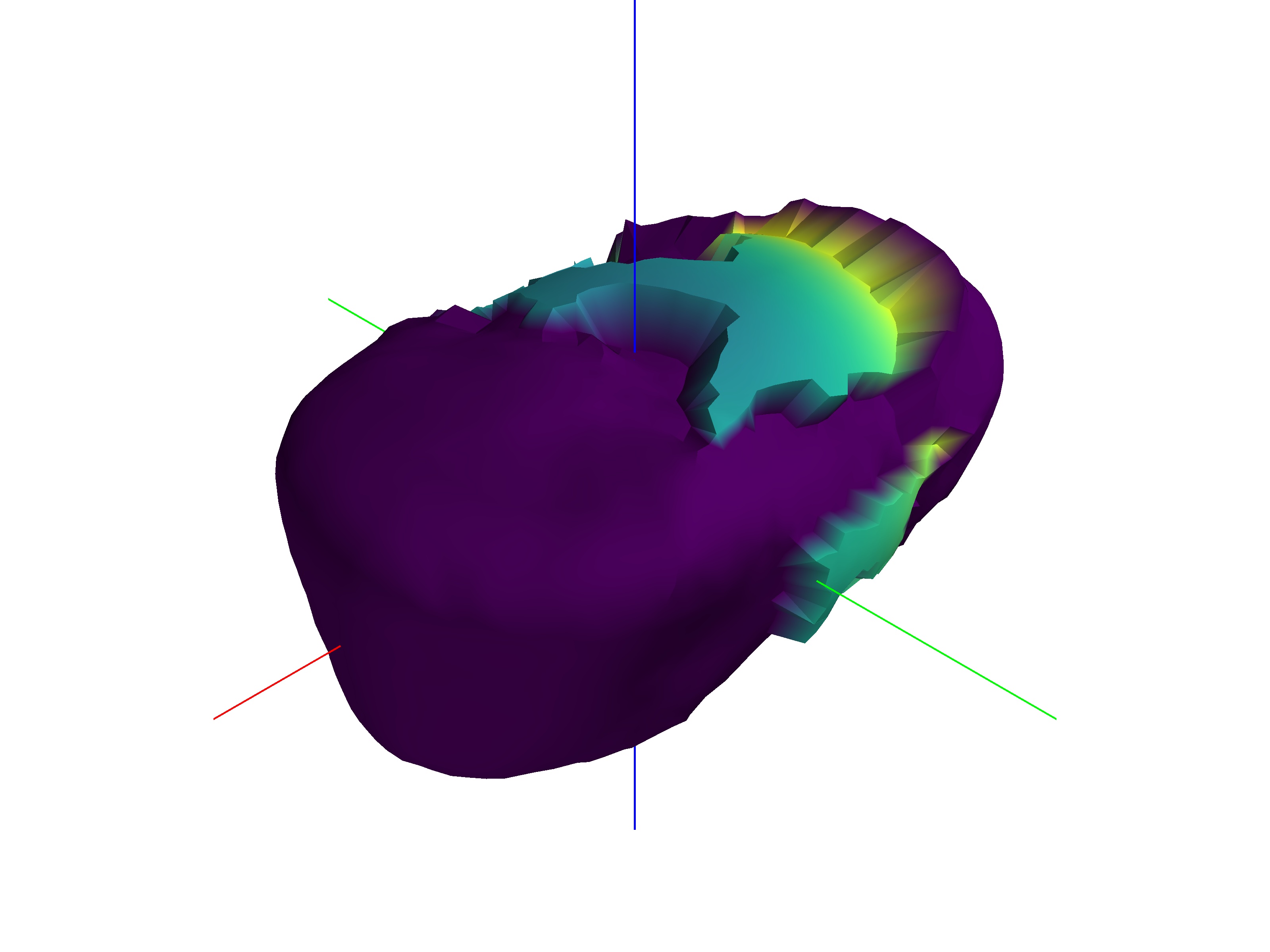}}%
    \subcaptionbox{\SI{50}{\percent} of measurements added\label{fig:castalia_partial_weights_50}}{\includegraphics[trim={15cm 10cm 15cm 10cm},clip,height=0.5\textheight,width=0.3\textwidth,keepaspectratio]{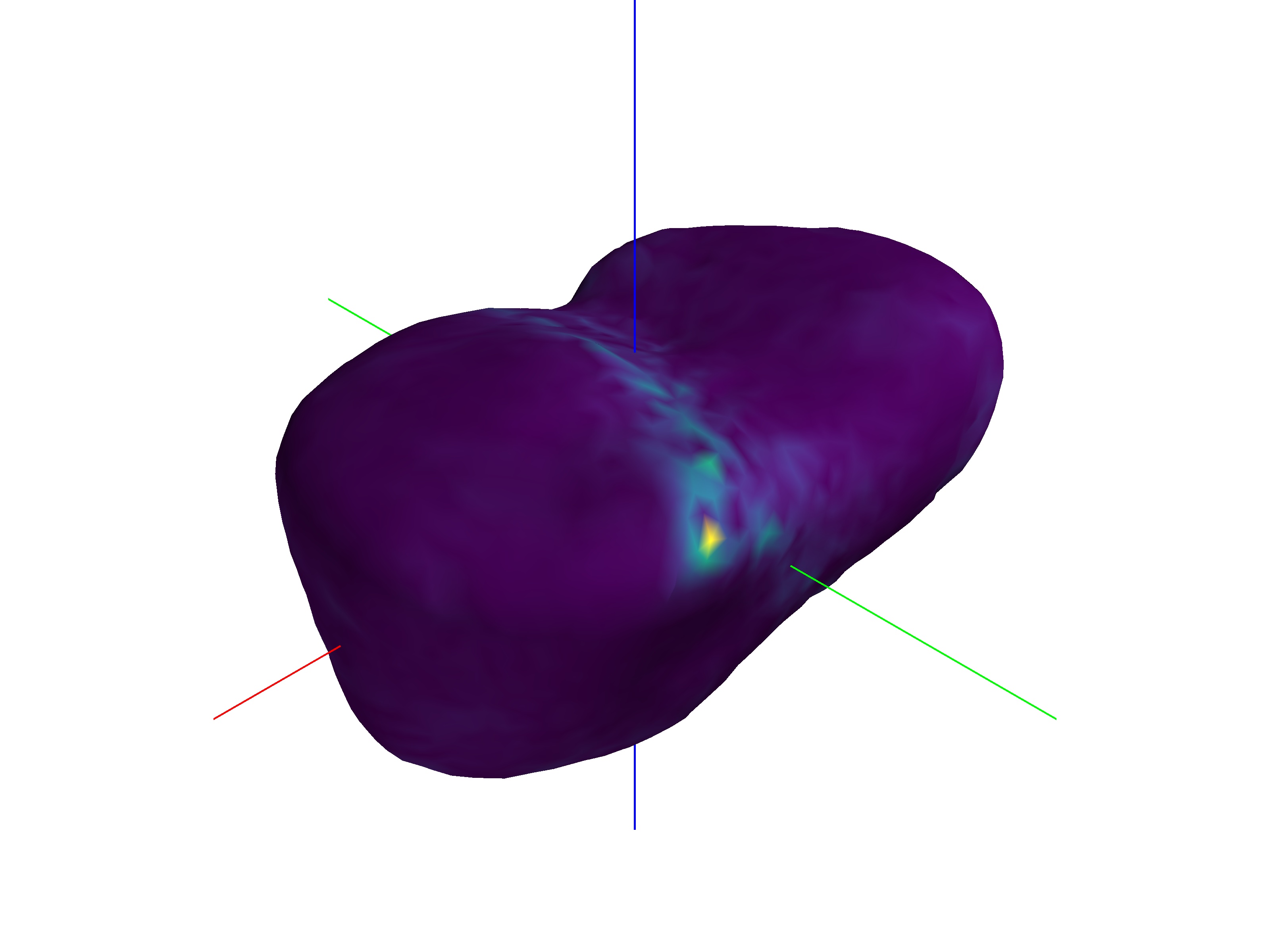}}

    \subcaptionbox{\SI{75}{\percent} of measurements added\label{fig:castalia_partial_weights_75}}{\includegraphics[trim={15cm 10cm 15cm 10cm},clip,height=0.5\textheight,width=0.3\textwidth,keepaspectratio]{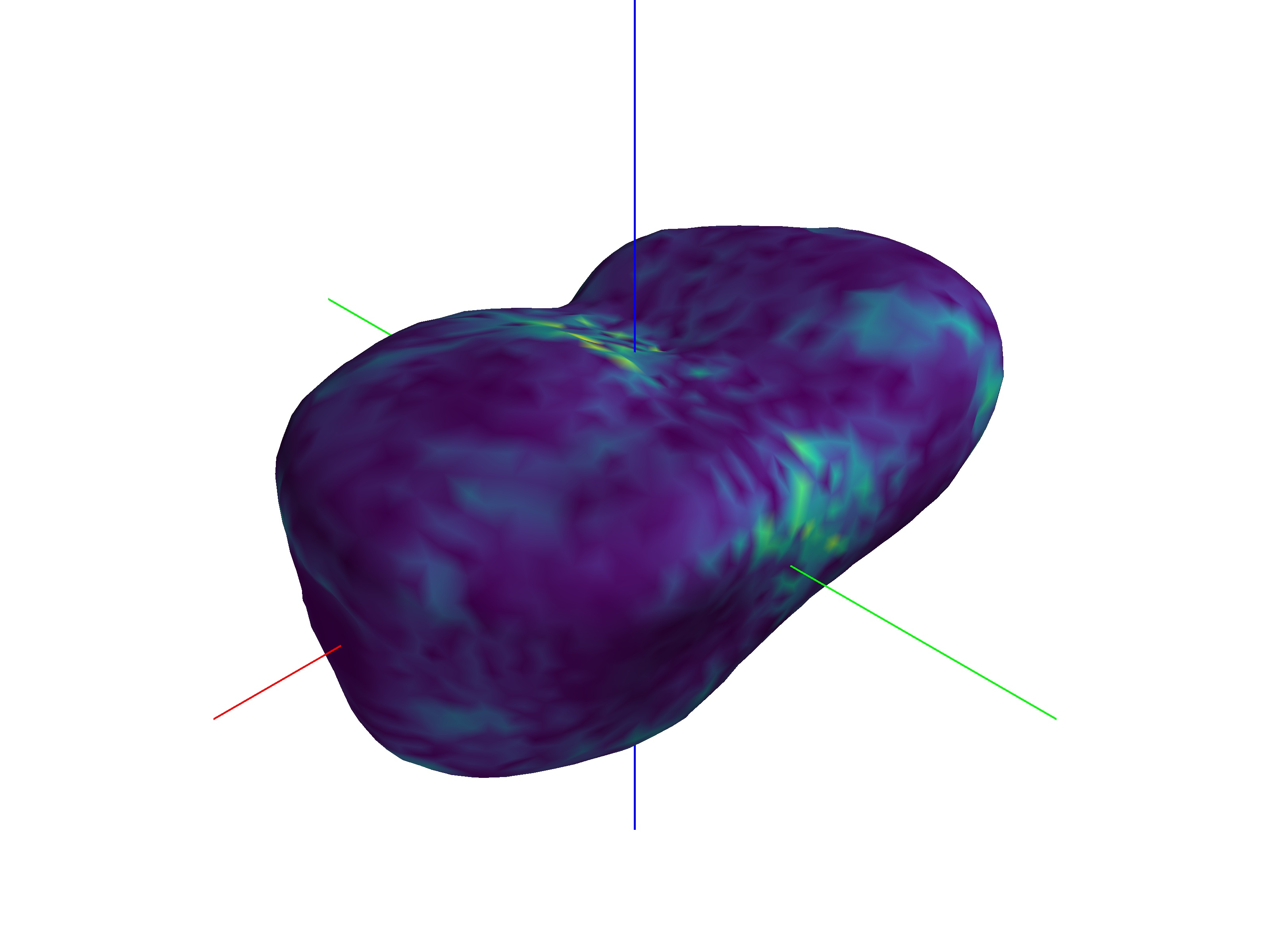}}%
    \subcaptionbox{\SI{100}{\percent} of measurements added\label{fig:castalia_partial_weights_100}}{\includegraphics[trim={15cm 10cm 15cm 10cm},clip,height=0.5\textheight,width=0.3\textwidth,keepaspectratio]{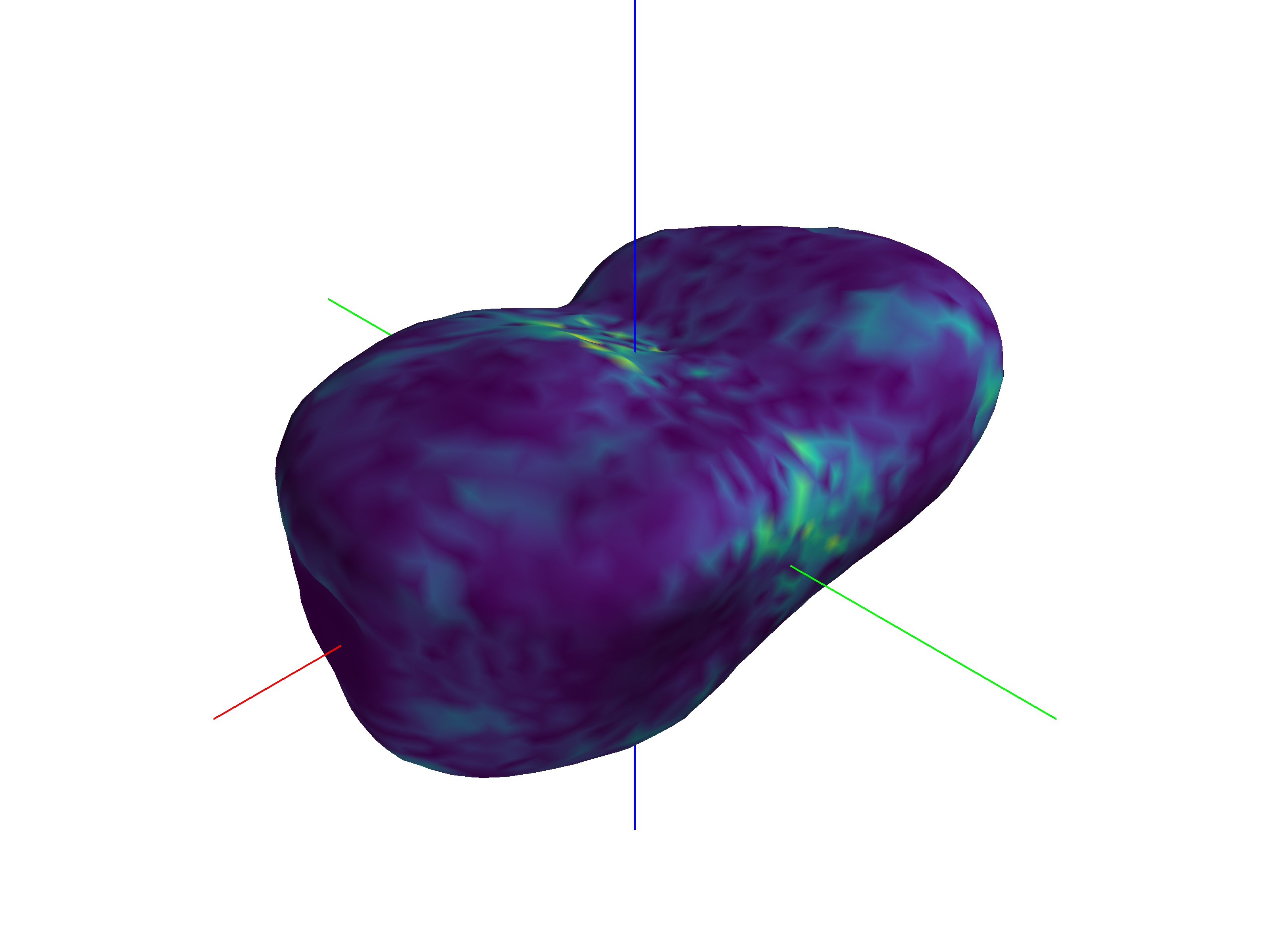}}%
    \subcaptionbox{True Shape Model\label{fig:castalia_weights_truth}}{\includegraphics[trim={15cm 10cm 15cm 10cm},clip,height=0.5\textheight,width=0.3\textwidth,keepaspectratio]{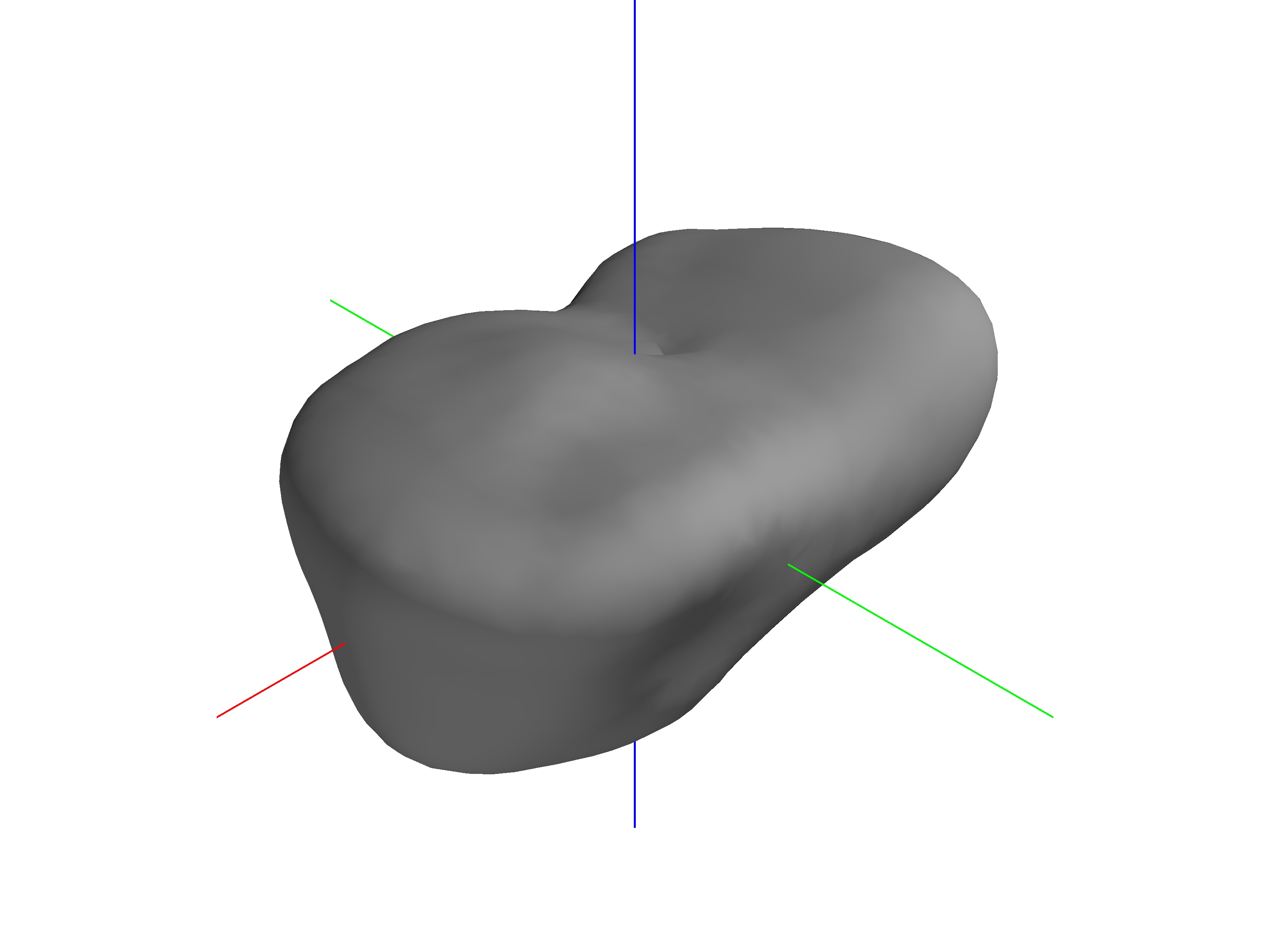}}
    \caption[Asteroid Castalia shape reconstruction with uncertainty]{Incremental reconstruction of asteroid Castalia. The images colored according to the shape uncertainty. Areas of high uncertainty are in yellow while ares of low uncertainty are in purple.
    The guidance scheme continually selects regions of high uncertainty until the total shape uncertainty is sufficiently small.
    ~\label{fig:castalia_weights_reconstruction}}
\end{figure}

Asteroid 4769 Castalia is a small near Earth asteroid of the Apollo group.
In addition, it is classified as a potentially hazardous object with a closed approach distance of less than \SI{0.05}{\astronomicalunit}.
Castalia was discovered in \num{1989} and is the first asteroid to be modeled using radar imagery~\cite{hudson1994}.
Castalia is composed of two distinct lobes suggesting that it is a contact binary of two smaller objects held together by their mutual gravity.
\Cref{fig:castalia_weights_reconstruction} show the shape reconstruction at several discrete points during the simulation.
In addition~\cref{fig:castalia_weights_reconstruction} displays the vertex uncertainty \( w_i \) as a colormap on the surface. 
Areas of high uncertainty are denoted in yellow while areas of low uncertainty are in purple/blue.
Within \SI{50}{\percent} of the simulation span the spacecraft is able to achieve an accurate estimate of the true shape of Castalia.

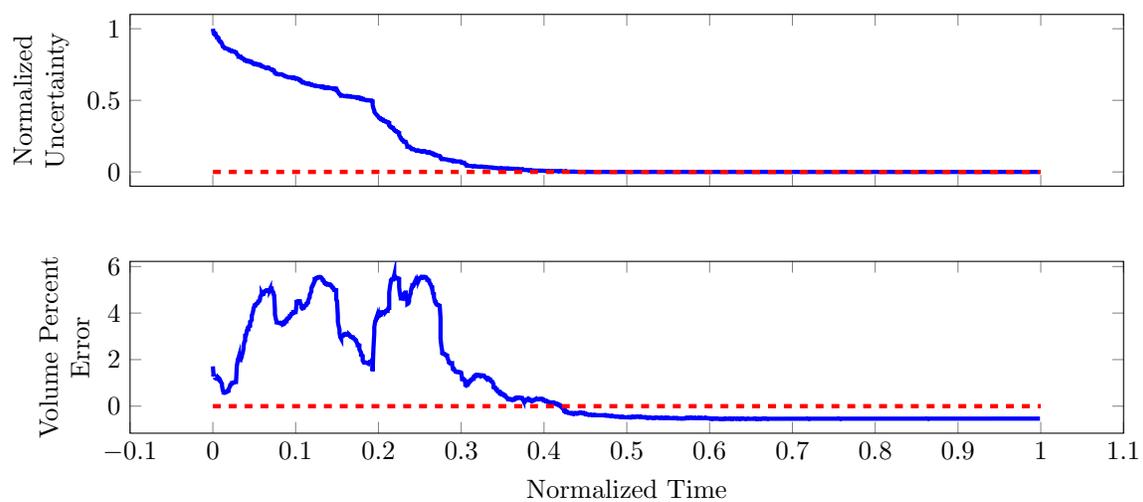
\begin{figure}[htbp]
    \centering
    \tikzsetnextfilename{castalia_metrics}
\begin{tikzpicture}[baseline]
    \begin{groupplot}[
        group style={
            group name={castalia_metrics},
            group size=1 by 2,
            xlabels at=edge bottom,
            ylabels at=edge left,
            xticklabels at=edge bottom,
        },
        xlabel={Normalized Time},
        scale only axis,
        width=0.8\textwidth,
        height=0.1\textheight,
        ylabel style={align=center},
    ]
    \nextgroupplot[ylabel={Normalized\\Uncertainty}]
    \addplot [ultra thick, color=blue, mark=none] table [x=NORMALIZED_TIME, y=NORMALIZED_UNCERTAINTY, col sep=comma] {dynamic_exploration_castalia_uncertainty.csv};
    \addplot [ultra thick,red, mark=none, dashed] coordinates {
        (0.0, 0.0) (1.0, 0.0) 
    };

    \nextgroupplot[ylabel={Volume Percent\\Error}]
    \addplot [ultra thick, blue, mark=none] table [x=NORMALIZED_TIME, y=VOLUME_PERCENT_ERROR, col sep=comma] {dynamic_exploration_castalia_volume.csv};
        \addplot [ultra thick,red, mark=none, dashed] coordinates {
            (0.0, 0.0) (1.0, 0.0) 
        };
\end{groupplot}
\end{tikzpicture}
    \caption{Normalized uncertainty and volume percent error for Castalia\label{fig:castalia_metrics}}
\end{figure}

\Cref{fig:castalia_metrics} shows the total normalized uncertainty and percent error for the volume estimate. 
The plots show that the reconstruction converges to an accurate shape estimate after approximately \SI{8000}{\second}.
In addition, the volume estimate is initially a much larger value but quickly converges to the true value.
The spacecraft autonomously navigates around asteroid Castalia to best minimize the cost function in~\cref{eq:control_cost}.

\section{Multi-resolution Landing Area Refinement}\label{sec:landing_refinement}

In this section, we consider a scenario of autonomous landing on the surface of asteroid, after its shape is constructed as described above.
The shape reconstruction scheme is based on an initial coarse shape estimate that is iteratively updated with range measurements of the surface.
Consequently, the original mesh is uniformly distributed with a relatively large mesh size and many small topological features such as rocks or small craters may not be captured accurately.
However, these small features are critical for surface operations and safe landings.
In addition, it would be computationally prohibitive to have a uniformly high resolution mesh.
In this section, we extend the previous shape update approach to enable a much higher fidelity in a specific location.

\subsection{Landing Site Determination}

The selection of a landing site will typically require a vast quantity of data and weigh a multitude of possible metrics, such as scientific value, hardware constraints, timing and communication limits, or safety considerations. 
In our analysis we consider the surface slope, the distance to the surface, and a fictitious science metric in order to determine the best landing site based on the complete shape estimate.
This approach allows for a spacecraft to autonomously select and to land on small body.

The surface slope is computed according to the method developed in Reference~\cite{scheeres1996}.
Due to the small size, and therefore low gravitational attraction, the force at each point on the surface is a combination of the gravitational attraction and the centripetal acceleration.
At the center of each face, \( f_i = [ f_x,\,  f_y,\,  f_z ] \), we compute a modified surface acceleration as
\begin{align}\label{eq:surface_force}
    U_m = \omega^2 \begin{bmatrix} f_x \\ f_y \\ 0 \end{bmatrix} + \begin{bmatrix} U_x \\ U_y \\ U_z \end{bmatrix},
\end{align}
where \( \omega \in \R^1 \) is the angular velocity of the asteroid and \( U_i \) is computed from~\cref{eq:attraction}.
Then the surface slope can be computed from
\begin{align}\label{eq:surface_slope}
    \cos \parenth{ \pi - \phi } = \frac{\vc{n}_f \cdot U_m}{\norm{U_m}},
\end{align}
where \( \phi \in \R \) is the surface slope defines the angle between the surface normal \( \vc{n}_f \in \R^3 \) and the force vector at the surface.
If \( \phi = \SI{0}{\degree} \) then the force vector and the surface normal are anti-parallel, while \( \phi > \SI{90}{\degree} \) means that a particle on the surface would be thrown off the body as the centripetal force is larger than the gravitational attraction.

Additionally, we compute the distance, using~\cref{eq:geodesic_distance}, between the spacecraft state and each face of the asteroid. 
Finally, we also assign a random science value to the surface in the form of a two dimensional Gaussian.
This can be modified depending on the specific objective of the mission. 

Utilizing these metrics, a landing site is chosen to minimize the surface cost given as
\begin{align}\label{eq:surface_cost}
    \min_{v_i} \mathcal{J}_l(x, v_i) =  \mathcal{J}_{\text{distance}}(x, v_i) - \mathcal{J}_{\text{science}}(v_i),
\end{align}
subject to a hard inequality constraint requiring that the surface slope is less than a threshold, i.e., \( \phi \leq \phi_m \).
The surface cost described here is relatively simple and chosen to be solely a function of the current spacecraft position and shape model.
Additional scientific criteria such as surface material or location on the body, e.g. equatorial vs. polar regions, can be utilized instead of a random science value as presented here.

\subsection{High-Resolution Mapping}\label{sec:refinement}

Mixed resolution surface meshes are routinely used in finite element and geometric modeling applications~\cite{botsch2010}.
As shown in Reference~\cite{mcmahon2017}, utilizing mixed resolution shape models for asteroid missions offers the potential of reduced computational demands.
The computational cost of the polyhedron potential model, given by~\cref{eq:attraction}, is roughly proportional to the number of faces in the shape model.
As a result, a uniformly high resolution shape would quickly become intractable for real time operations.
However, utilizing a mixed resolution approach allows for a high fidelity in a smaller mission critical area, such as a landing site, with a limited impact on the computational cost.

Once a suitable landing site is selected, the surrounding area is isolated and refined by adding new vertices and faces in the specified area.
The goal of refinement, or more generally remeshing, is given a mesh (or a portion of it), compute another mesh whose elements satisfy some quality metrics while suitably approximating the original mesh.
In this work, we utilize the isotropic remeshing algorithm implemented in the Computational Geometry and Algorithms Library (CGAL)~\cite{cgalproject2018}.
This algorithm uses an iterative method which repeatedly splits long edges, collapses short edges, and relocates vertices until all edges are approximately the desired target edge length.

\begin{figure}[htbp]
    \centering
    \subcaptionbox{Original Cube\label{fig:cube_original_mesh}}{\includegraphics[width=0.45\textwidth]{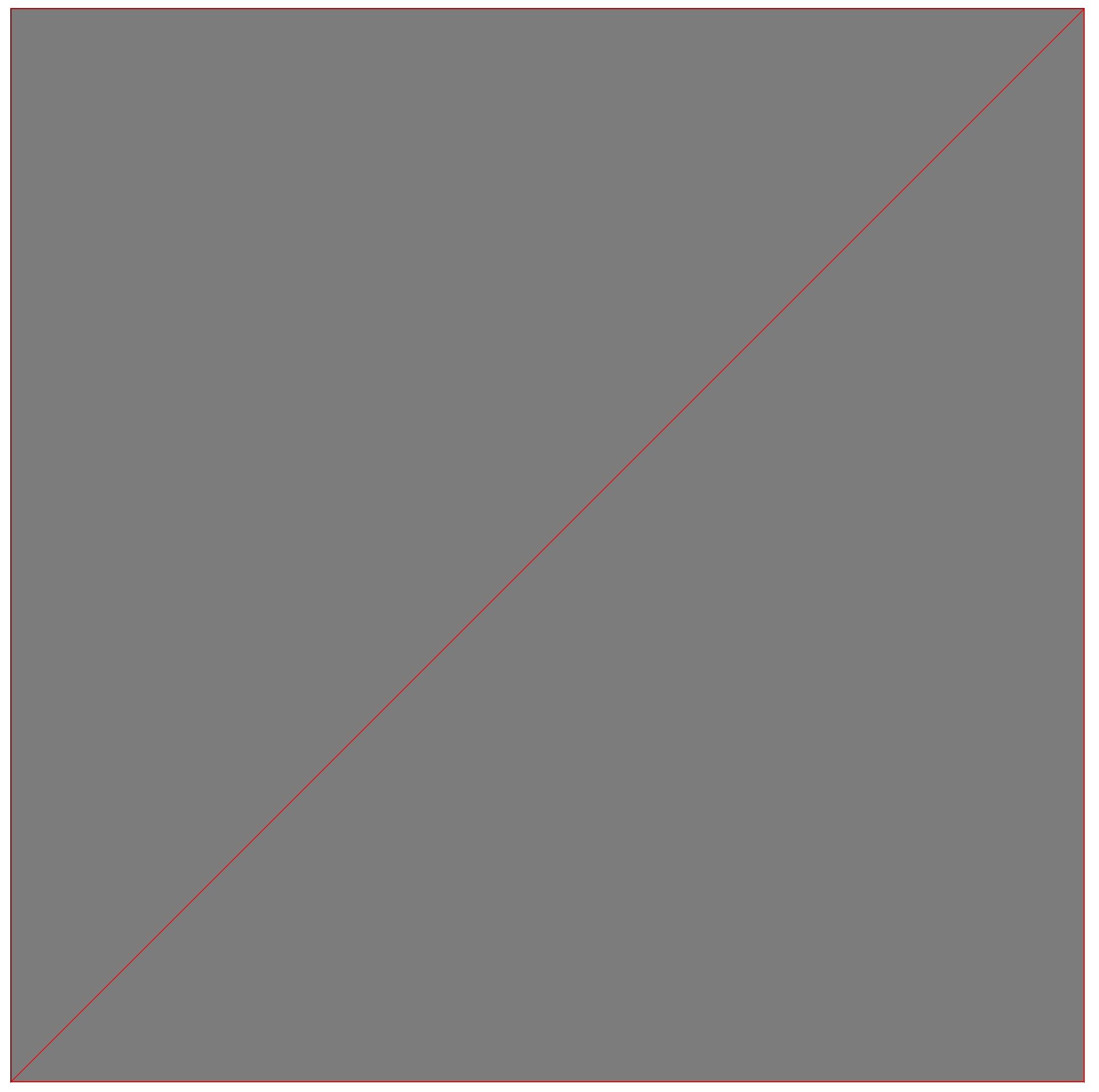}}~
    \subcaptionbox{Remeshed Cube\label{fig:cube_refine_mesh}}{\includegraphics[width=0.45\textwidth]{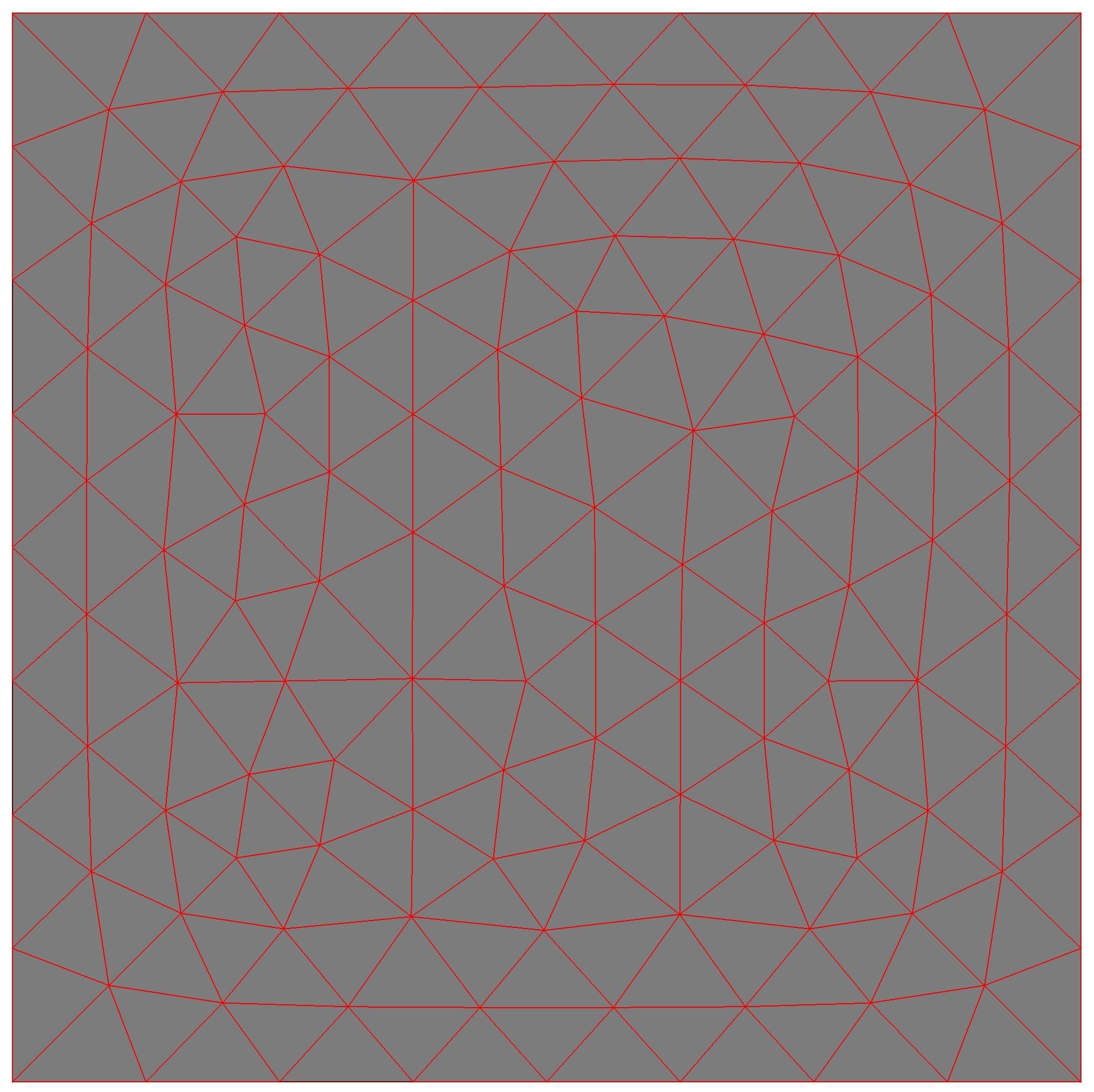}}
    \caption{Example of isotropic remeshing of a face of a cube\label{fig:cube_remesh}}
\end{figure}

For example,~\cref{fig:cube_remesh} shows the isotropic remeshing result for the selected faces of a unit cube.
The original unit cube is composed of \num{8} vertices and \num{12} faces.
The two triangular faces of the visible side of the cube are selected for the isotropic remeshing operation as shown in~\cref{fig:cube_original_mesh}.
A target edge length of \num{0.1} is selected for these faces and used to generate~\cref{fig:cube_refine_mesh}.
The two large triangular faces are divided into a number of smaller triangular faces.
Furthermore, the additional faces are all approximately the same size and preserve the original surface of the cube.
After the isotropic remeshing operation the number of vertices has increased from \num{8} to \num{140}.

Once the area around the landing site is remeshed, the preceding shape reconstruction scheme is applied to those area to develop a high-fidelity topological map. 

\subsection{Numerical Example}

The proposed schemes for landing site selection and refinement are applied to Castalia.



\paragraph{Castalia Landing Site Selection}

With an appropriate shape estimate, the spacecraft can autonomously transition from a shape reconstruction to a landing mode.
Based on the shape estimate we seek to determine the best location to land. 
In reality, any landing site selection would be based on a wide variety of factors and constraints. 
However we highlight a few which can be determined autonomously and from the shape estimate.
The first metric is related to the surface slope which is computed using the completed shape estimate and~\cref{eq:surface_slope}.
Areas which violate the slope constraint of \( \phi < \SI{5}{\degree} \) are excluded from further consideration.
\begin{figure}[htbp]
    \centering
    \subcaptionbox{Surface slope of Castalia\label{fig:surface_slope_castalia}}{\includegraphics[width=\textwidth]{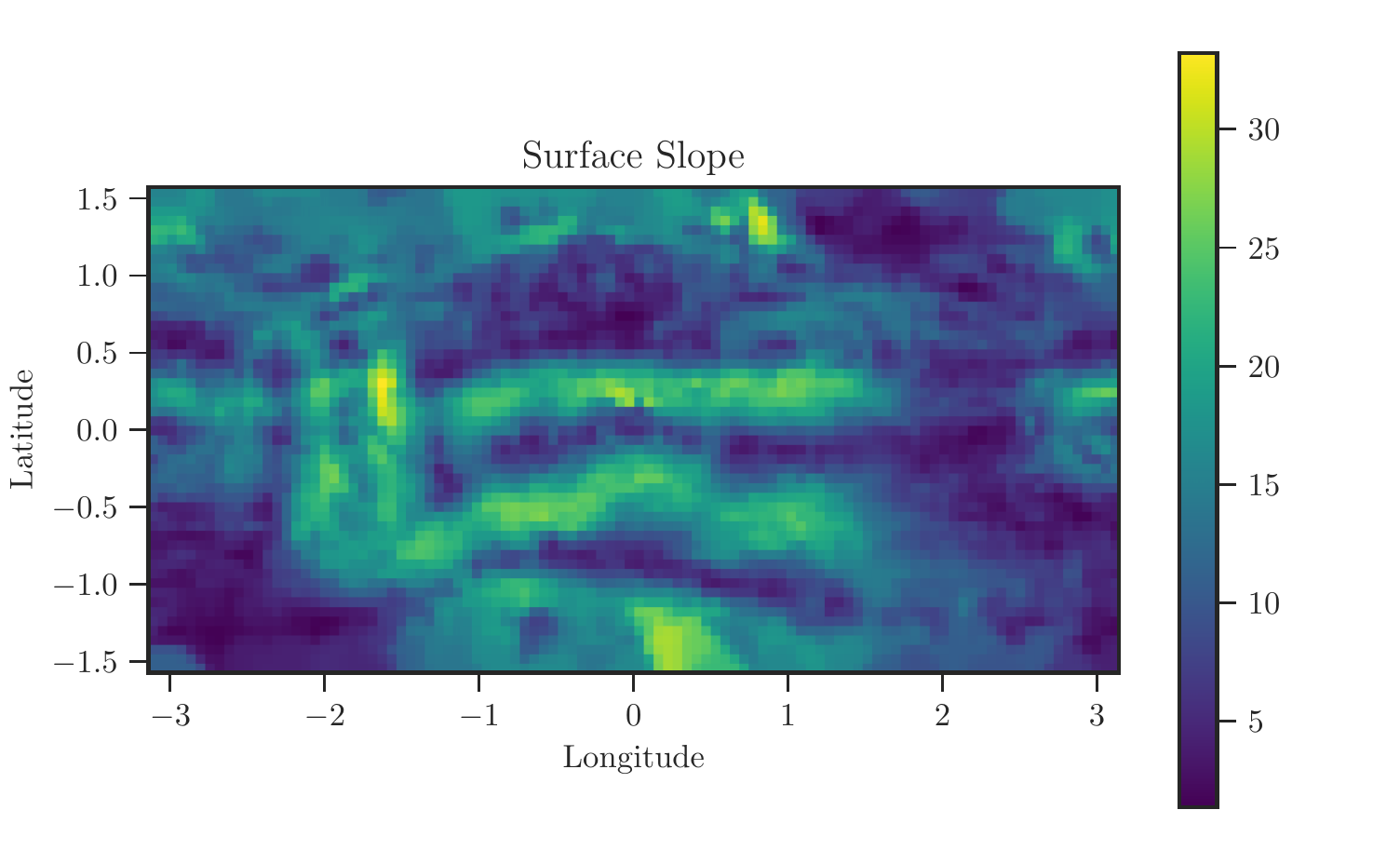}}\\
    \subcaptionbox{Masked surface slope with areas \( \phi > \SI{5}{\degree}\) excluded\label{fig:surface_slope_castalia_masked}}{\includegraphics[width=\textwidth,keepaspectratio]{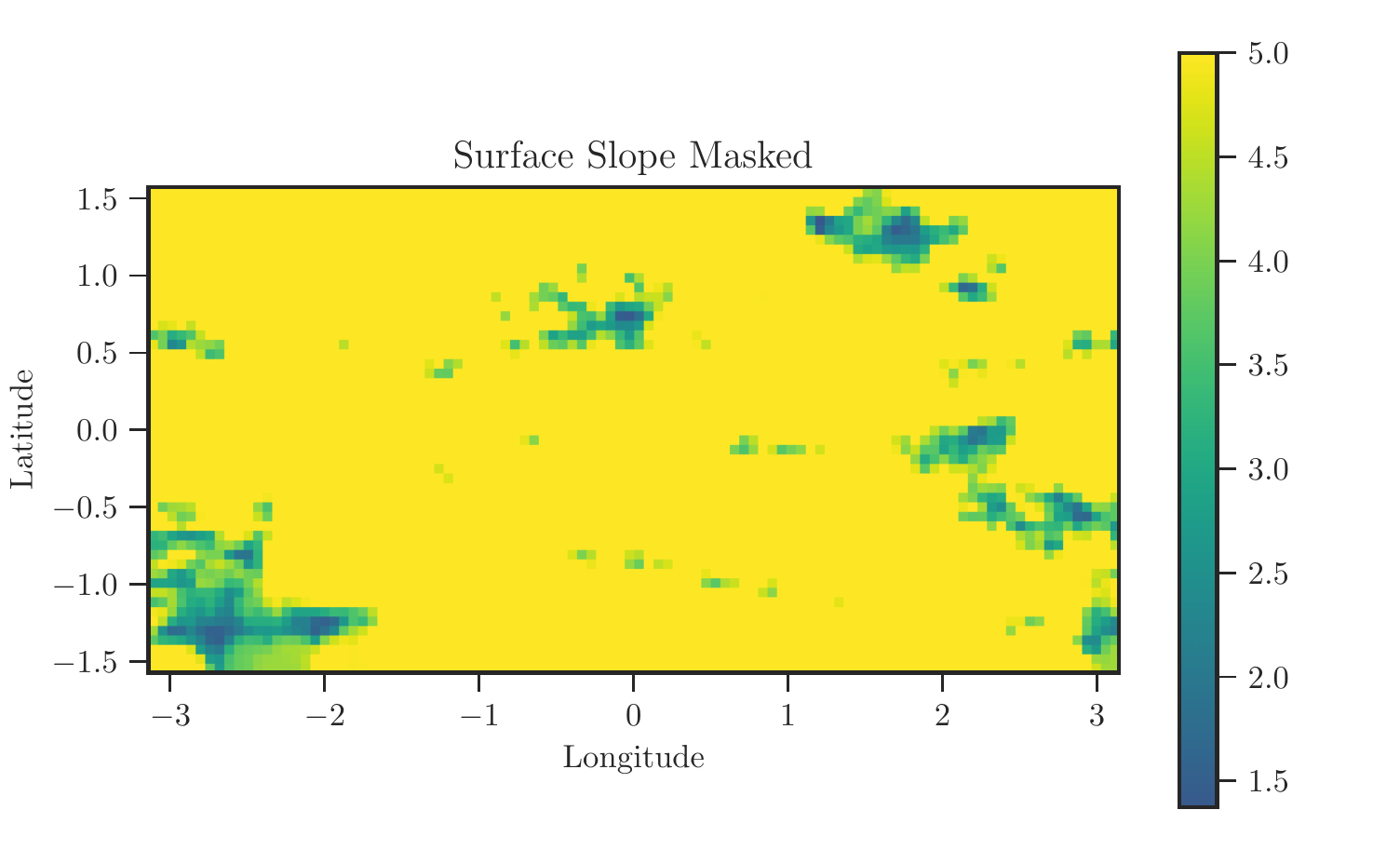}}
    \caption{Surface slope of asteroid Castalia\label{fig:surface_slope_castalia_both}}
\end{figure}

\begin{figure}[htbp]
    \centering
    \subcaptionbox{Surface Distance to surface of Castalia\label{fig:surface_distance_castalia}}{\includegraphics[width=\textwidth]{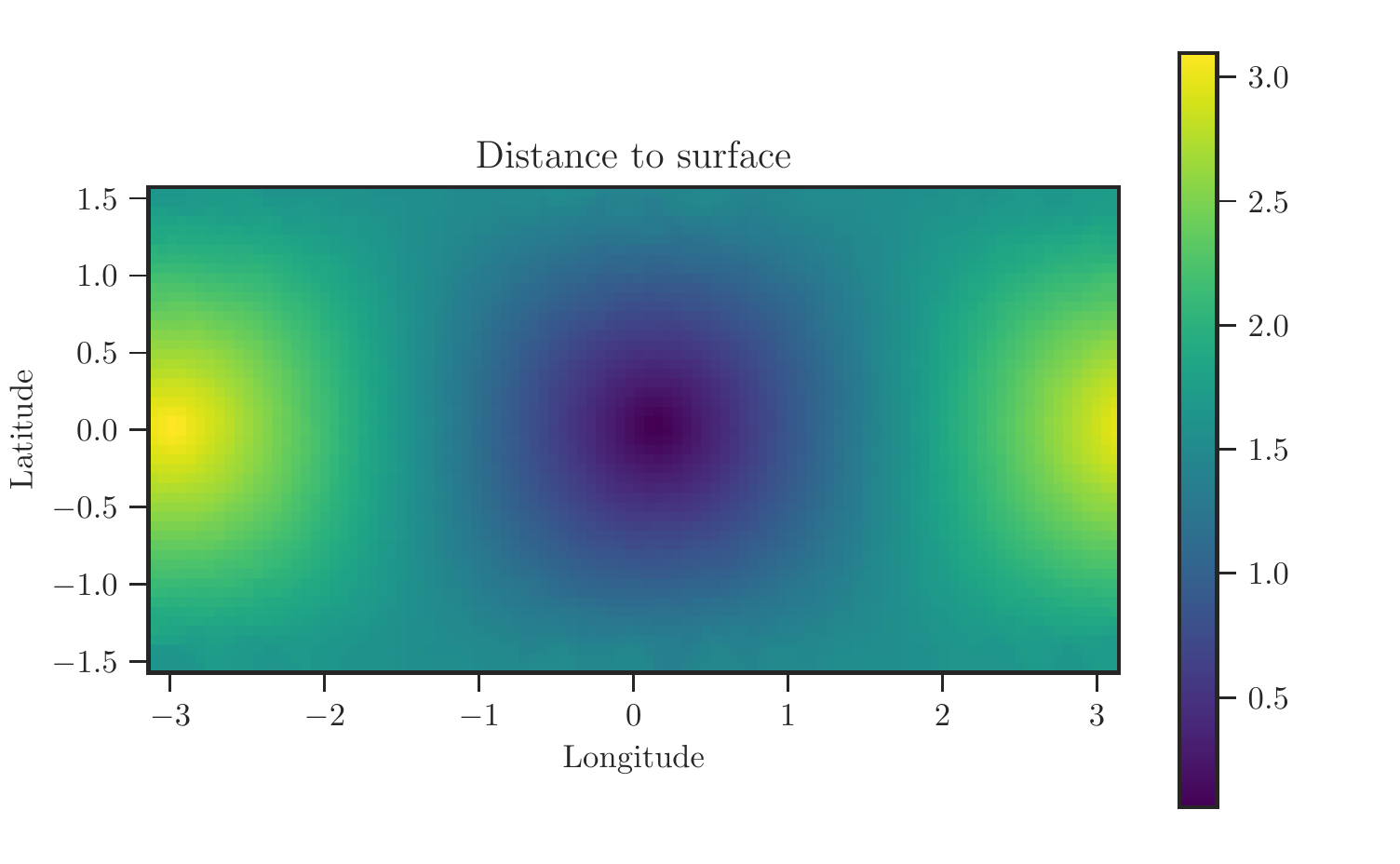}}\\
    \subcaptionbox{Masked distance to surface with areas \( \phi > \SI{5}{\degree}\) excluded\label{fig:surface_distance_castalia_masked}}{\includegraphics[width=\textwidth,keepaspectratio]{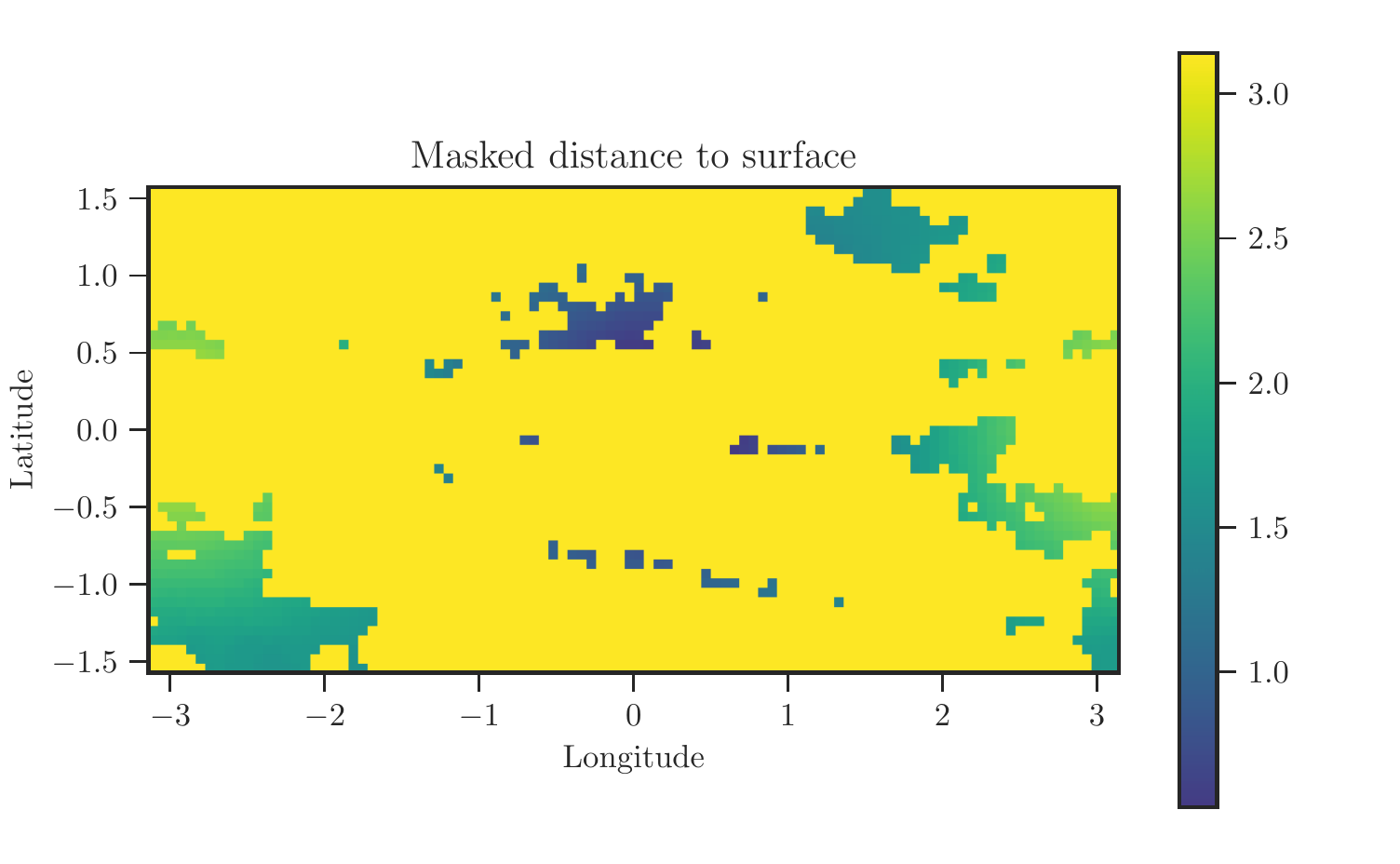}}
    \caption{Distance to surface of asteroid Castalia\label{fig:distance_castalia_both}}
\end{figure}

\begin{figure}[htbp]
    \centering
    \includegraphics[width=\textwidth,keepaspectratio]{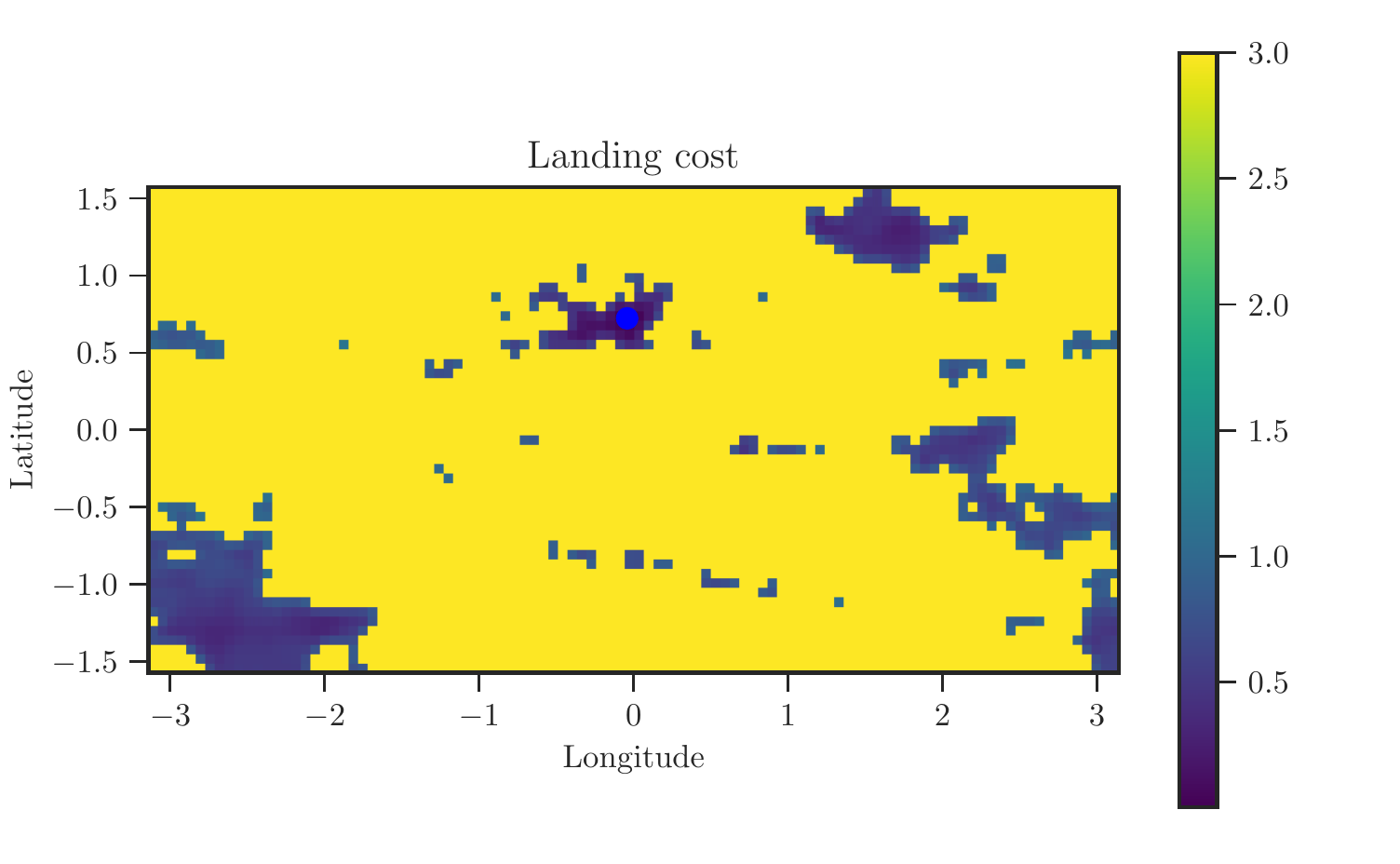}
    \caption{Total cost for surface landing based on surface slope and distance\label{fig:landing_site_cost}}
\end{figure}

The next metric is related to the distance from the current spacecraft position to all candidate landing sites on the surface.
We utilize~\cref{eq:geodesic_distance} to compute a distance metric to the surface.
Landing sites which are closer will be considered preferentially over those at a larger distance.
\Cref{fig:distance_castalia_both} shows a surface plot of the distance to the surface. 
The area immediately beneath the spacecraft has a small cost while those on the opposite side of the asteroid have a much larger cost.
We can combine~\cref{fig:distance_castalia_both,fig:surface_slope_castalia_both} to determine the best landing site. 
The combination of the two is shown in~\cref{fig:landing_site_cost} with the desired landing site shown by the blue marker.
After selecting the appropriate landing site we then prepare for landing by collecting more measurements in the region around the landing area.

\paragraph{Castalia Landing Site Refinement}

\begin{figure}[htbp]
    \centering
    \subcaptionbox{Augmented shape of Castalia with surface features\label{fig:orig_castalia}}{\includegraphics[trim={15cm 0 15cm 0},clip,keepaspectratio,height=0.3\textheight,width=0.5\textwidth]{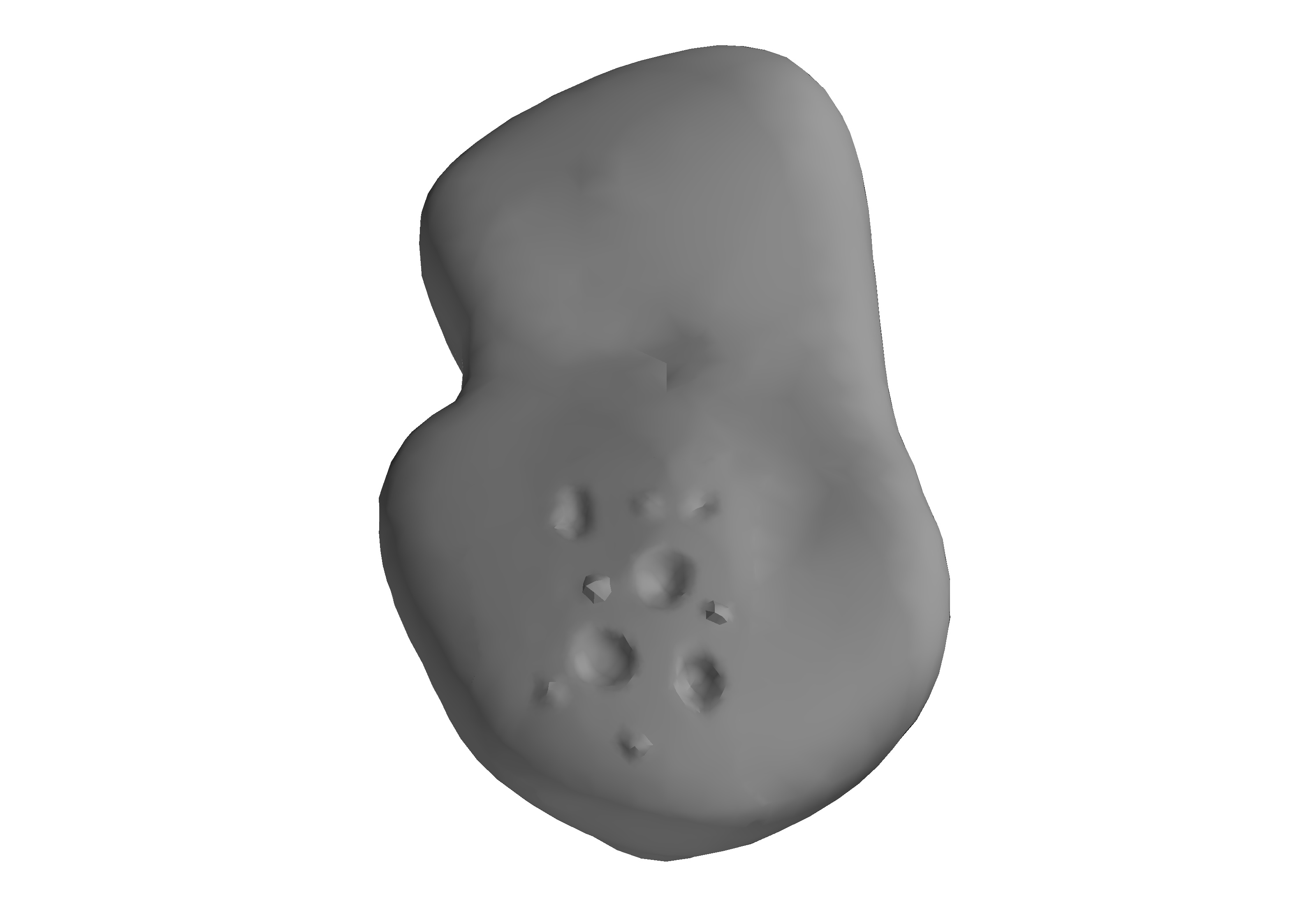}}%
    \subcaptionbox{Estimated shape model after measuring the surface\label{fig:bump_castalia}}{\includegraphics[trim={15cm 0 15cm 0},clip,keepaspectratio,height=0.3\textheight,width=0.5\textwidth]{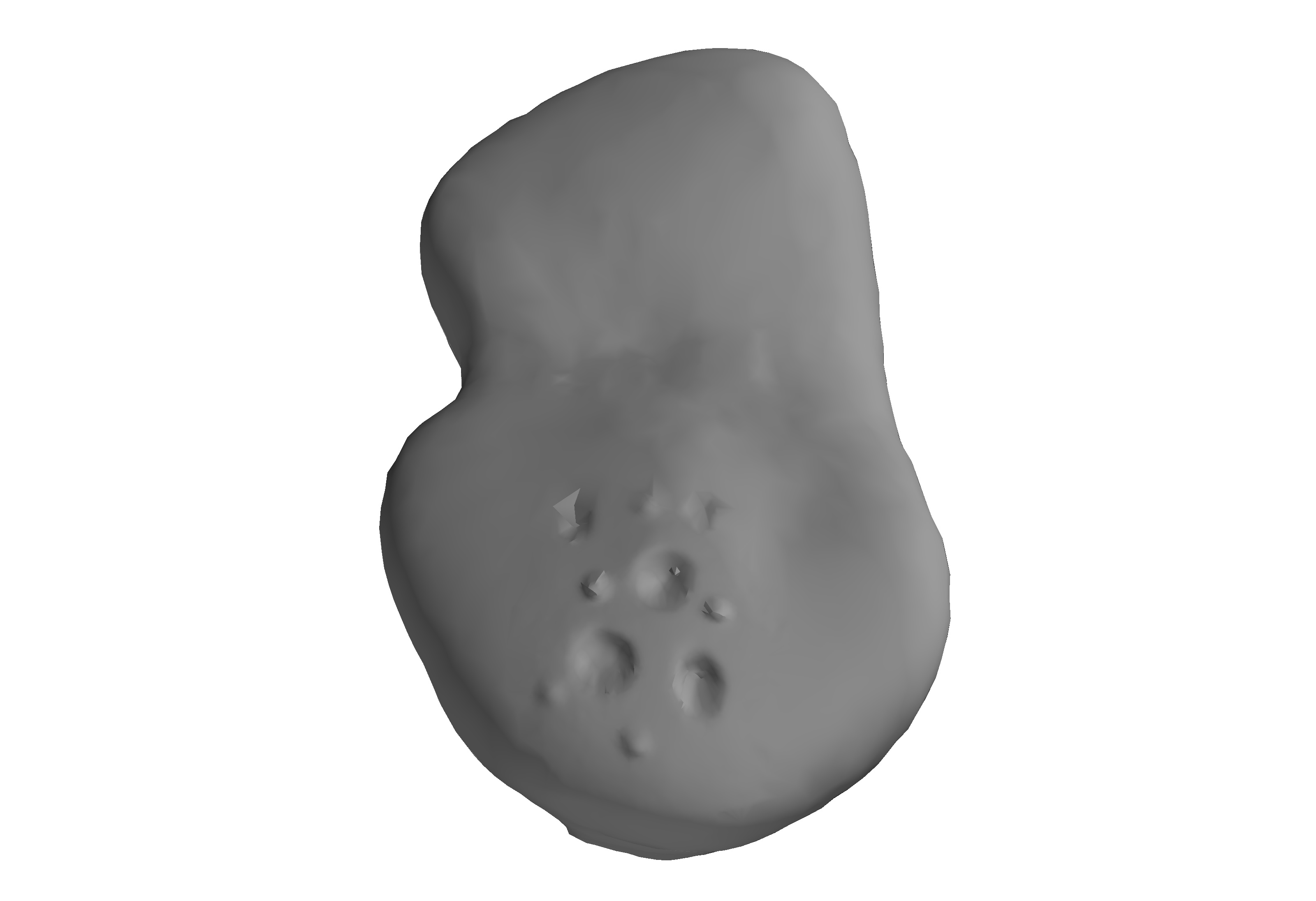}}
    \caption{Asteroid Castalia with augmented with additional surface features~\label{fig:castalia_refinement}}
\end{figure}

One key benefit of an in-situ spacecraft is the ability to measure surface features at a much higher resolution than is possible from the ground. 
This detail provides a much higher fidelity data source than ground based measurements. 
In order to emulate this in the simulation we augment the shape model of Castalia with several small craters and outcroppings as shown in~\cref{fig:castalia_refinement}.
However, these small features would be difficult to capture with the current low resolution shape model of approximately \num{4000} faces.
The lower number of faces is useful for the evaluation of the polyhedron potential model, it is not ideal for the capture of minute surface features. 
As a result, we utilize the isotropic remeshing operation described previously to selectively increase the fidelity in the region about the desired landing site.

\begin{figure}[htbp]
    \centering
    \subcaptionbox{Original vertex density of the initial shape estimate\label{fig:intial_vertex_density}}{\includegraphics[width=\textwidth]{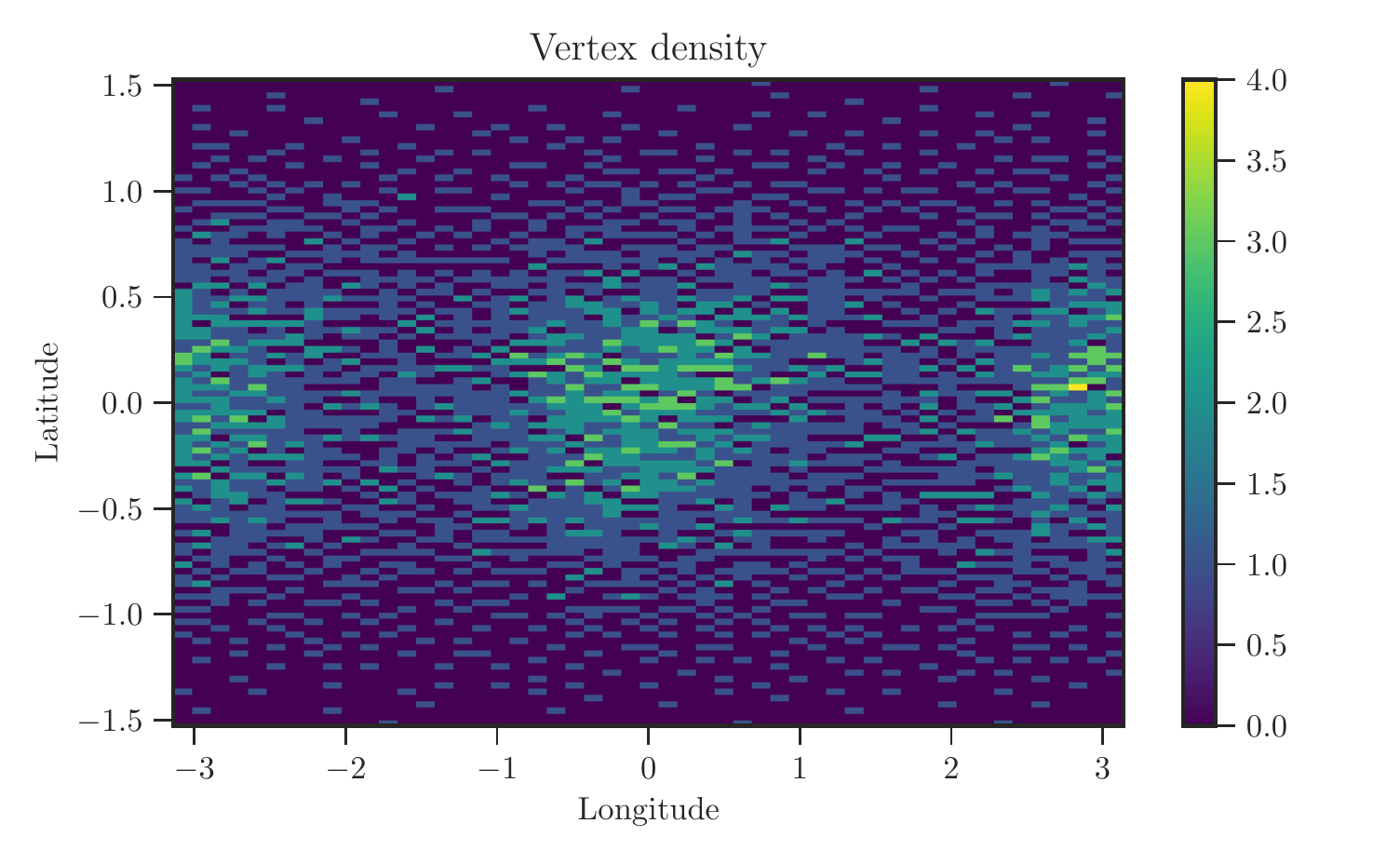}}\\
    \subcaptionbox{Vertex density after refinement  around landing site\label{fig:refine_vertex_density}}{\includegraphics[width=\textwidth]{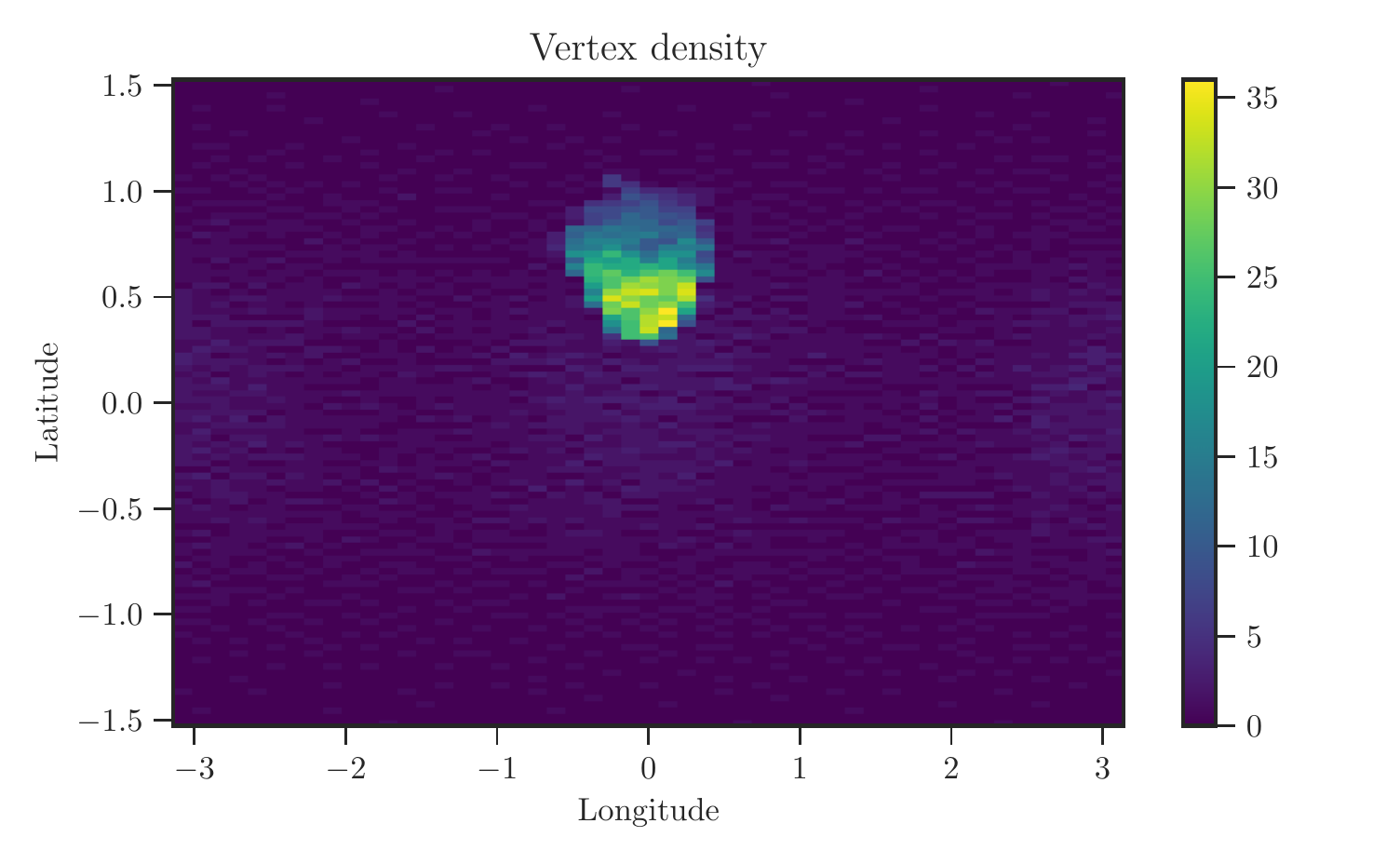}}
    \caption{Vertex density before and after refinement at asteroid Castalia\label{fig:castalia_refine_density}}
\end{figure}

\Cref{fig:castalia_refine_density} shows that the vertex density increases by approximately an order of magnitude in region immediately surround the landing site.

The increased number of faces and vertices in the landing area allows for the capture of the small surface features.
\Cref{fig:castalia_refinement} shows that given LIDAR measurements alone and the shape reconstruction algorithm that the small features are effectively estimated.

\paragraph{Castalia Vertical Descent}

\begin{figure}[htbp]
    \centering
    \includegraphics[width=0.8\textwidth]{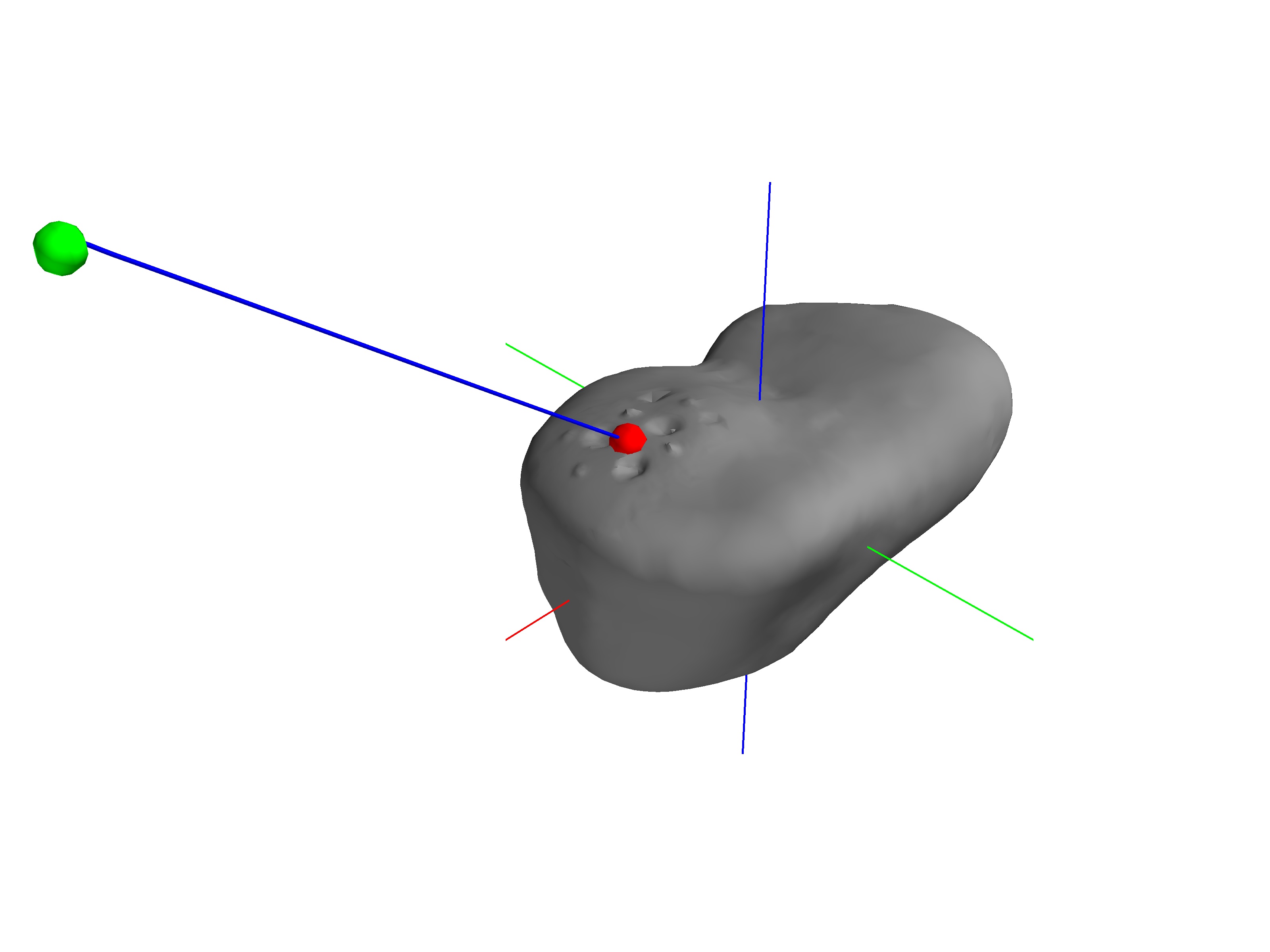}
    \caption{Vertical descent onto 4769 Castalia~\label{fig:castalia_landing}}
\end{figure}

The final phase of the simulation is to utilize the estimated shape model and vertically descend to the desired landing site. 
This is accomplished using the closed loop control of the vehicle and a trajectory which transitions from the home position to the surface over \SI{3600}{\second}.
The landing trajectory, using the estimated shape, is visualized in the asteroid frame in~\cref{fig:castalia_landing}.
During the vertical descent the vehicle is able to accurately track the desired trajectory using the estimated shape to compute the gravitational potential.
A number of videos are available at \url{http://bit.ly/shape_reconstruction} which demonstrate these results.

\section{Conclusions}

This paper developed a Bayesian update scheme to reconstruct the shape of a small body from range measurements. 
The approach allows for a local update operation that is able to reconstruct the shape in real time. 
This is in contrast to standard shape reconstruction algorithms which operate over the entire surface and require significant computational resources.
Next, an optimal guidance scheme is derived which determines the desired state to best update the shape estimate. 
Finally, a mixed resolution shape representation is presented to allow for an increased fidelity in a specific region of the asteroid.
This allows for a much greater shape accuracy in a local area while avoiding the computational costs associated with a uniformly high resolution mesh.
This allows for a spacecraft to autonomously maneuver and reconstruct the shape of a small body without operator intervention.
Several numerical examples were presented which demonstrates the approach on a number of real asteroids.

Throughout this paper, it is considered that the position of the spacecraft relative the asteroid is available. 
Future works include integrating a localization technique with the proposed shape reconstruction scheme such that the relative location and the shape model are estimated simultaneously.

\bibliographystyle{IEEEtran}
\bibliography{library}

\end{document}